\documentclass[11pt,a4paper]{article}

\usepackage[utf8]{inputenc}
\usepackage[T1]{fontenc}
\usepackage[margin=1in]{geometry}
\usepackage{amsmath}
\usepackage{amssymb}
\usepackage{stmaryrd}
\usepackage{booktabs}
\usepackage{array}
\usepackage{tabularx}
\usepackage{authblk}
\usepackage{xcolor}
\usepackage{graphicx}
\usepackage[export]{adjustbox}
\usepackage{longtable}
\usepackage{calc}
\usepackage{newunicodechar}
\usepackage{listings}
\usepackage[colorlinks=true,linkcolor=blue!40!black,citecolor=blue!40!black,urlcolor=blue!40!black]{hyperref}

\let\origtextunderscore\_
\renewcommand{\_}{\origtextunderscore\allowbreak}

\lstdefinestyle{leanbox}{
  basicstyle=\footnotesize\ttfamily,
  columns=fullflexible,
  keepspaces=true,
  breaklines=true,
  breakatwhitespace=false,
  showstringspaces=false,
  upquote=true,
  extendedchars=true,
  inputencoding=utf8,
  literate=
    {§}{{\S}}1
    {¬}{{\ensuremath{\neg}}}1
    {·}{{\ensuremath{\cdot}}}1
    {×}{{\ensuremath{\times}}}1
    {Γ}{{\ensuremath{\Gamma}}}1
    {Δ}{{\ensuremath{\Delta}}}1
    {Η}{{\ensuremath{H}}}1
    {Θ}{{\ensuremath{\Theta}}}1
    {Κ}{{\ensuremath{K}}}1
    {Σ}{{\ensuremath{\Sigma}}}1
    {α}{{\ensuremath{\alpha}}}1
    {β}{{\ensuremath{\beta}}}1
    {λ}{{\ensuremath{\lambda}}}1
    {σ}{{\ensuremath{\sigma}}}1
    {τ}{{\ensuremath{\tau}}}1
    {φ}{{\ensuremath{\varphi}}}1
    {χ}{{\ensuremath{\chi}}}1
    {ψ}{{\ensuremath{\psi}}}1
    {ᵢ}{{\textsubscript{i}}}1
    {ⱼ}{{\textsubscript{j}}}1
    {—}{{\textemdash}}1
    {–}{{\textendash}}1
    {…}{{\ldots}}1
    {′}{{\ensuremath{^{\prime}}}}1
    {⁺}{{\textsuperscript{+}}}1
    {⁻}{{\textsuperscript{\ensuremath{-}}}}1
    {ⁱ}{{\textsuperscript{i}}}1
    {¹}{{\textsuperscript{1}}}1
    {₀}{{\textsubscript{0}}}1
    {₁}{{\textsubscript{1}}}1
    {₂}{{\textsubscript{2}}}1
    {₃}{{\textsubscript{3}}}1
    {₊}{{\textsubscript{+}}}1
    {ₙ}{{\textsubscript{n}}}1
    {ₚ}{{\textsubscript{p}}}1
    {ₖ}{{\textsubscript{k}}}1
    {ₘ}{{\textsubscript{m}}}1
    {ₜ}{{\textsubscript{t}}}1
    {ι}{{\ensuremath{\iota}}}1
    {ω}{{\ensuremath{\omega}}}1
    {ḡ}{{\={g}}}1
    {ē}{{\={e}}}1
    {ń}{{\'{n}}}1
    {ˢ}{{\textsuperscript{s}}}1
    {⦃}{{\textbraceleft\textbraceleft}}2
    {⦄}{{\textbraceright\textbraceright}}2
    {𝒵}{{\ensuremath{\mathcal{Z}}}}1
    {𝓕}{{\ensuremath{\mathcal{F}}}}1
    {ℕ}{{\ensuremath{\mathbb{N}}}}1
    {←}{{\ensuremath{\leftarrow}}}1
    {↑}{{\ensuremath{\uparrow}}}1
    {→}{{\ensuremath{\rightarrow}}}1
    {↓}{{\ensuremath{\downarrow}}}1
    {↔}{{\ensuremath{\leftrightarrow}}}1
    {↥}{{\ensuremath{\Uparrow}}}1
    {↦}{{\ensuremath{\mapsto}}}1
    {↪}{{\ensuremath{\hookrightarrow}}}1
    {↟}{{\ensuremath{\Uparrow}}}1
    {⇄}{{\ensuremath{\rightleftarrows}}}1
    {⇑}{{\ensuremath{\Uparrow}}}1
    {⇒}{{\ensuremath{\Rightarrow}}}1
    {⨆}{{\ensuremath{\bigcup}}}1
    {𝕆}{{\ensuremath{\mathbb{O}}}}1
    {∀}{{\ensuremath{\forall}}}1
    {∃}{{\ensuremath{\exists}}}1
    {∅}{{\ensuremath{\emptyset}}}1
    {∈}{{\ensuremath{\in}}}1
    {∉}{{\ensuremath{\notin}}}1
    {∘}{{\ensuremath{\circ}}}1
    {∋}{{\ensuremath{\ni}}}1
    {∩}{{\ensuremath{\cap}}}1
    {∏}{{\ensuremath{\prod}}}1
    {∞}{{\ensuremath{\infty}}}1
    {⋂}{{\ensuremath{\bigcap}}}1
    {⋃}{{\ensuremath{\bigcup}}}1
    {∧}{{\ensuremath{\wedge}}}1
    {∨}{{\ensuremath{\vee}}}1
    {∪}{{\ensuremath{\cup}}}1
    {∼}{{\ensuremath{\sim}}}1
    {≠}{{\ensuremath{\neq}}}1
    {≤}{{\ensuremath{\leq}}}1
    {≥}{{\ensuremath{\geq}}}1
    {≪}{{\ensuremath{\ll}}}1
    {≃}{{\ensuremath{\simeq}}}1
    {≅}{{\ensuremath{\cong}}}1
    {⊆}{{\ensuremath{\subseteq}}}1
    {⊑}{{\ensuremath{\sqsubseteq}}}1
    {⊔}{{\ensuremath{\sqcup}}}1
    {⊓}{{\ensuremath{\sqcap}}}1
    {⊢}{{\ensuremath{\vdash}}}1
    {⊣}{{\ensuremath{\dashv}}}1
    {⊤}{{\ensuremath{\top}}}1
    {⊥}{{\ensuremath{\bot}}}1
    {⊨}{{\ensuremath{\vDash}}}1
    {⊬}{{\ensuremath{\nvdash}}}1
    {⊭}{{\ensuremath{\nvDash}}}1
    {⋀}{{\ensuremath{\bigwedge}}}1
    {⋁}{{\ensuremath{\bigvee}}}1
    {─}{{\ensuremath{-}}}1
    {│}{{\ensuremath{|}}}1
    {└}{{+}}1
    {├}{{+}}1
    {□}{{\ensuremath{\square}}}1
    {▸}{{\ensuremath{\blacktriangleright}}}1
    {◇}{{\ensuremath{\lozenge}}}1
    {⟨}{{\ensuremath{\langle}}}1
    {⟩}{{\ensuremath{\rangle}}}1
    {⟶}{{\ensuremath{\longrightarrow}}}2
    {⟷}{{\ensuremath{\longleftrightarrow}}}2
    {⟹}{{\ensuremath{\Longrightarrow}}}2
}
\newunicodechar{α}{\ensuremath{\alpha}}
\newunicodechar{β}{\ensuremath{\beta}}
\newunicodechar{σ}{\ensuremath{\sigma}}
\newunicodechar{φ}{\ensuremath{\varphi}}
\newunicodechar{χ}{\ensuremath{\chi}}
\newunicodechar{ψ}{\ensuremath{\psi}}
\newunicodechar{Γ}{\ensuremath{\Gamma}}
\newunicodechar{Δ}{\ensuremath{\Delta}}
\newunicodechar{ℕ}{\ensuremath{\mathbb{N}}}
\newunicodechar{∀}{\ensuremath{\forall}}
\newunicodechar{∃}{\ensuremath{\exists}}
\newunicodechar{¬}{\ensuremath{\neg}}
\newunicodechar{∧}{\ensuremath{\wedge}}
\newunicodechar{□}{\ensuremath{\square}}
\newunicodechar{◇}{\ensuremath{\lozenge}}
\newunicodechar{∼}{\ensuremath{\sim}}
\newunicodechar{→}{\ensuremath{\rightarrow}}
\newunicodechar{←}{\ensuremath{\leftarrow}}
\newunicodechar{↔}{\ensuremath{\leftrightarrow}}
\newunicodechar{↦}{\ensuremath{\mapsto}}
\newunicodechar{⇒}{\ensuremath{\Rightarrow}}
\newunicodechar{⟶}{\ensuremath{\longrightarrow}}
\newunicodechar{⟷}{\ensuremath{\longleftrightarrow}}
\newunicodechar{⟹}{\ensuremath{\Longrightarrow}}
\newunicodechar{⟺}{\ensuremath{\Longleftrightarrow}}
\newunicodechar{∈}{\ensuremath{\in}}
\newunicodechar{⊢}{\ensuremath{\vdash}}
\newunicodechar{⊨}{\ensuremath{\vDash}}
\newunicodechar{≠}{\ensuremath{\neq}}
\newunicodechar{≈}{\ensuremath{\approx}}
\newunicodechar{≅}{\ensuremath{\cong}}
\newunicodechar{≪}{\ensuremath{\ll}}
\newunicodechar{⊆}{\ensuremath{\subseteq}}
\newunicodechar{⊇}{\ensuremath{\supseteq}}
\newunicodechar{⊋}{\ensuremath{\supsetneq}}
\newunicodechar{⊑}{\ensuremath{\sqsubseteq}}
\newunicodechar{⊔}{\ensuremath{\sqcup}}
\newunicodechar{⊓}{\ensuremath{\sqcap}}
\newunicodechar{⊤}{\ensuremath{\top}}
\newunicodechar{⊥}{\ensuremath{\bot}}
\newunicodechar{⊣}{\ensuremath{\dashv}}
\newunicodechar{↪}{\ensuremath{\hookrightarrow}}
\newunicodechar{↟}{\ensuremath{\Uparrow}}
\newunicodechar{⇄}{\ensuremath{\rightleftarrows}}
\newunicodechar{⇑}{\ensuremath{\Uparrow}}
\newunicodechar{⨆}{\ensuremath{\bigcup}}
\newunicodechar{𝕆}{\ensuremath{\mathbb{O}}}
\newunicodechar{∪}{\ensuremath{\cup}}
\newunicodechar{⊕}{\ensuremath{\oplus}}
\newunicodechar{×}{\ensuremath{\times}}
\newunicodechar{⟨}{\ensuremath{\langle}}
\newunicodechar{⟩}{\ensuremath{\rangle}}
\newunicodechar{₀}{\textsubscript{0}}
\newunicodechar{₁}{\textsubscript{1}}
\newunicodechar{ₙ}{\textsubscript{n}}
\newunicodechar{⁺}{\textsuperscript{+}}
\newunicodechar{′}{\ensuremath{^{\prime}}}
\newunicodechar{¹}{\textsuperscript{1}}
\newunicodechar{ⁱ}{\textsuperscript{i}}
\newunicodechar{⁻}{\textsuperscript{-}}
\newunicodechar{ⁿ}{\textsuperscript{n}}
\newunicodechar{₂}{\textsubscript{2}}
\newunicodechar{₃}{\textsubscript{3}}
\newunicodechar{ₖ}{\textsubscript{k}}
\newunicodechar{ₘ}{\textsubscript{m}}
\newunicodechar{ₚ}{\textsubscript{p}}
\newunicodechar{ₜ}{\textsubscript{t}}
\newunicodechar{ᵢ}{\textsubscript{i}}
\newunicodechar{ⱼ}{\textsubscript{j}}
\newunicodechar{₊}{\textsubscript{+}}
\newunicodechar{Η}{\ensuremath{\mathrm{H}}}
\newunicodechar{Θ}{\ensuremath{\Theta}}
\newunicodechar{Κ}{\ensuremath{\mathrm{K}}}
\newunicodechar{Σ}{\ensuremath{\Sigma}}
\newunicodechar{ι}{\ensuremath{\iota}}
\newunicodechar{λ}{\ensuremath{\lambda}}
\newunicodechar{τ}{\ensuremath{\tau}}
\newunicodechar{ω}{\ensuremath{\omega}}
\newunicodechar{≤}{\ensuremath{\leq}}
\newunicodechar{≥}{\ensuremath{\geq}}
\newunicodechar{≃}{\ensuremath{\simeq}}
\newunicodechar{∉}{\ensuremath{\notin}}
\newunicodechar{∋}{\ensuremath{\ni}}
\newunicodechar{∅}{\ensuremath{\emptyset}}
\newunicodechar{∘}{\ensuremath{\circ}}
\newunicodechar{∞}{\ensuremath{\infty}}
\newunicodechar{∨}{\ensuremath{\vee}}
\newunicodechar{∩}{\ensuremath{\cap}}
\newunicodechar{∏}{\ensuremath{\prod}}
\newunicodechar{⋀}{\ensuremath{\bigwedge}}
\newunicodechar{⋁}{\ensuremath{\bigvee}}
\newunicodechar{⋂}{\ensuremath{\bigcap}}
\newunicodechar{⊬}{\ensuremath{\nvdash}}
\newunicodechar{⊭}{\ensuremath{\nvDash}}
\newunicodechar{↑}{\ensuremath{\uparrow}}
\newunicodechar{↓}{\ensuremath{\downarrow}}
\newunicodechar{↥}{\ensuremath{\Uparrow}}
\newunicodechar{⟸}{\ensuremath{\Longleftarrow}}
\newunicodechar{⨅}{\ensuremath{\bigcap}}
\newunicodechar{◁}{\ensuremath{\triangleleft}}
\newunicodechar{▸}{\ensuremath{\blacktriangleright}}
\newunicodechar{𝓕}{\ensuremath{\mathcal{F}}}
\newunicodechar{─}{\-}
\newunicodechar{└}{\textasciigrave{}|}
\newunicodechar{├}{\textbar{}}
\newunicodechar{ē}{\={e}}
\newunicodechar{ń}{\'{n}}
\newunicodechar{ḡ}{\={g}}
\newunicodechar{·}{\textperiodcentered{}}
\newunicodechar{—}{\textemdash{}}
\newunicodechar{–}{\textendash{}}
\newunicodechar{…}{\ldots}
\newunicodechar{§}{\S}
\newunicodechar{│}{\ensuremath{\mid}}
\newunicodechar{✓}{\ensuremath{\checkmark}}
\newunicodechar{æ}{\ae}
\newunicodechar{ö}{\"o}

\title{\textbf{A Lean 4 Formalization of Scott's \emph{Continuous Lattices} (1972)}}

        \author[1]{\textbf{Lars Warren Ericson}}
        \affil[1]{ORCID: 0000-0001-8299-9361}
        \affil[1]{Catskills Research Company}
        \affil[1]{\texttt{lars.ericson@catskillsresearch.com}}

        \date{\today}

\begin{document}

        \maketitle

        \begin{center}
          \small
          \textbf{Github:} \url{https://github.com/catskillsresearch/scott1972} \\
          \textbf{Palomar Registration:} \url{https://palomar-registry.org/entry?id=PALOMAR-2026-08-24-000003&version=2} \\
          \textbf{Primary Category:} cs.LO (Logic in Computer Science) \\
          \textbf{Secondary Category:} math.LO (Logic)
        \end{center}

        \begin{abstract}
        We present a complete machine-checked formalization of Dana Scott's landmark 1972 paper \emph{Continuous Lattices} \textbf{[Sco72]}, carried out in Lean 4 against mathlib and including the March 1972 Milner correction in \textbf{[Sco72]} (pp.~135--136).

Scott's paper develops a model for \(\lambda\)-calculus from a topological starting point. He defines \emph{injective} \(T_0\)-spaces---those with a strong extension property for continuous maps---and shows that they are exactly the \emph{continuous lattices}: complete lattices whose Scott topology is determined by the order via the way-below relation (\(\ll\)). On this foundation he studies projections, retractions, products, function spaces, and inverse limits. The capstone (Theorem 4.4) constructs an inverse limit \(D_\infty\) of function-space approximants and proves that \(D_\infty\) is homeomorphic to its own function space \([D_\infty \to D_\infty]\), yielding a purely mathematical model for Church's untyped \(\lambda\)-calculus.

Our development formalizes \textbf{43 numbered results} from Scott's Sections 1--4 (Propositions, Corollaries, Lemmas, and Theorems), each as a sorry-free Lean theorem, together with supporting infrastructure (step functions, the \(\Uparrow a\) basis of Scott opens, Milner's coarser-than-Scott hypothesis, the function-space tower, and the \(i_\infty\)/\(j_\infty\) pair). The formalization is \textbf{classical} (uses \texttt{Classical.choice} transitively) and follows Scott's proof dependency order. Where the Lean proof required choices not visible in the original---or where dead ends were encountered---we record detailed notes in Section 5. All proofs check with the standard footprint \(\texttt{[propext, Classical.choice, Quot.sound]}\).
        \end{abstract}

\hypertarget{introduction}{%
\section{Introduction}\label{introduction}}

Scott's 1972 paper \emph{Continuous Lattices} is a fully detailed, peer-reviewed publication of the \(D_\infty\) model for the semantics of Church's untyped \(\lambda\)-calculus---but with one crucial historical nuance: \textbf{the model was a complete accident}. While the 1972 paper is the landmark account, Scott had been trying to prove that such a model was mathematically \emph{impossible}.

Three factors frame the breakthrough:

\hypertarget{the-goal-was-types-not-the-untyped-calculus}{%
\subsection{The goal was types, not the untyped calculus}\label{the-goal-was-types-not-the-untyped-calculus}}

In the late 1960s Dana Scott worked alongside Christopher Strachey at Oxford. Scott was a skeptic of Alonzo Church's untyped \(\lambda\)-calculus. He believed programming languages should be strictly typed and argued that the untyped calculus lacked a legitimate mathematical foundation. He began developing a model for \(\lambda\)-calculus specifically to provide a denotational semantics for \emph{typed} languages.

\hypertarget{the-discovery-of-d_infty-1969}{%
\subsection{\texorpdfstring{The discovery of \(D_\infty\) (1969)}{The discovery of D\_\textbackslash infty (1969)}}\label{the-discovery-of-d_infty-1969}}

To model the untyped \(\lambda\)-calculus, a space \(D\) must be isomorphic to its own function space (\(D \cong [D \to D]\)). In standard set theory, Cantor's theorem makes this size-wise impossible: the power of a function space strictly exceeds the size of the base set.

In November 1969, while attempting to formalize why this restriction made untyped models impossible, Scott realized that if one restricts to Scott-continuous functions (those preserving directed suprema) rather than all arbitrary functions, the space does not explode in size. By constructing an inverse limit of algebraic lattices (\(D_0 \to D_1 \to D_2 \to \cdots\)), he built \(D_\infty\), a landmark non-degenerate mathematical model of the untyped \(\lambda\)-calculus. This account does not establish comparative priority over other early models.

\hypertarget{chronology-of-the-papers}{%
\subsection{Chronology of the papers}\label{chronology-of-the-papers}}

Although the definitive mathematical breakdown appeared in 1972, Scott's ``first attempts'' span a few tightly knit manuscripts:

\begin{itemize}
\item
  \textbf{1969 (unpublished manuscript):} \textbf{{[}Sco69{]}} \emph{Lattice-theoretic models for the \(\lambda\)-calculus}---the earliest write-up in the Scott manuscript sequence discussed here, distributed among colleagues after the November discovery. No broader historical priority claim is intended.
\item
  \textbf{1970 (conference paper):} \textbf{{[}Sco70{]}} \emph{Outline of a mathematical theory of computation}---a brief, high-level introductory account.
\item
  \textbf{1972 (published paper):} \textbf{{[}Sco72{]}} \emph{Continuous Lattices}---prepared as a technical report in 1971; recognized as the landmark paper that gave the mathematical and computer science communities the \(D_\infty\) model.
\end{itemize}

Scott's paper opens by arguing that \(T_0\)-spaces, long treated as a mere exercise in separation axioms, are natural once one cares about function spaces and extension properties rather than geometry. That shift---from typed skepticism to the accidental \(D_\infty\) model---is the backdrop for the formalization in §3--§5.

The development documented below follows \textbf{{[}Sco72{]}} through the March 1972 Milner correction. §2 summarizes Scott's original paper; §3--§4 describe the Lean development and catalog the formalized theorems; §5 records proof notes where the mechanization adds detail beyond the published text.

\begin{center}\rule{0.5\linewidth}{0.5pt}\end{center}

\hypertarget{scotts-continuous-lattices}{%
\section{\texorpdfstring{Scott's \emph{Continuous Lattices}}{Scott's Continuous Lattices}}\label{scotts-continuous-lattices}}

\textbf{{[}Sco72{]}} develops a model for \(\lambda\)-calculus from injective \(T_0\)-spaces. Scott's own abstract states the arc of the paper: starting topologically, he introduces spaces with a strong extension property for continuous maps; shows they are exactly the continuous lattices---complete lattices whose topology is the Scott topology determined by the order; studies projections, subspaces, embeddings, products, and function spaces; and proves the main result that one can embed every space in a continuous lattice \(D_\infty\) whose product/subspace inverse-limit topology is homeomorphic to the pointwise Pi topology on its own function space \([D_\infty \to D_\infty]\), yielding models for the Church--Curry \(\lambda\)-calculus.

Scott organizes the paper in four technical sections (following an introductory §0):

\begin{longtable}[]{@{}
  >{\raggedright\arraybackslash}p{(\columnwidth - 4\tabcolsep) * \real{0.3333}}
  >{\raggedright\arraybackslash}p{(\columnwidth - 4\tabcolsep) * \real{0.3333}}
  >{\raggedright\arraybackslash}p{(\columnwidth - 4\tabcolsep) * \real{0.3333}}@{}}
\toprule\noalign{}
\begin{minipage}[b]{\linewidth}\raggedright
Scott §
\end{minipage} & \begin{minipage}[b]{\linewidth}\raggedright
Title
\end{minipage} & \begin{minipage}[b]{\linewidth}\raggedright
Main content
\end{minipage} \\
\midrule\noalign{}
\endhead
\bottomrule\noalign{}
\endlastfoot
§1 & \textbf{Injective spaces} & Definition of injectivity; \(\mathbb{O}\) and its powers; retract characterization (Cor. 1.6--1.7) \\
§2 & \textbf{Continuous lattices} & Way-below (\(\ll\)), Scott topology, Scott-continuous maps; products and retractions; injectivity ⟺ continuous lattice (Thm. 2.12) \\
§3 & \textbf{Function spaces} & \([D \to D']\) as a continuous lattice (Thm. 3.3); \(\lambda\)-abstraction, evaluation, projections, fixed points \\
§4 & \textbf{Inverse limits} & \(D_\infty\) as inverse/direct limit; capstone \(D_\infty\) homeomorphic to \([D_\infty \to D_\infty]\) (Thm. 4.4) \\
\end{longtable}

Our working source text is \href{sources/ScottContinLatt1972.md}{\texttt{sources/ScottContinLatt1972.md}}: a plain-text OCR transcription of \textbf{{[}Sco72{]}} through the Milner correction (pp.~135--136). Image-based PDFs are poor inputs for mechanized proof development; this file gives the formalization a reliable, searchable text to quote against. (An earlier filename suffix \texttt{\_vision} reflected the OCR toolchain only and has been dropped to avoid confusion with the published paper.)

\begin{center}\rule{0.5\linewidth}{0.5pt}\end{center}

\hypertarget{the-lean-formalization}{%
\section{The Lean formalization}\label{the-lean-formalization}}

\hypertarget{scope-and-methodology}{%
\subsection{Scope and methodology}\label{scope-and-methodology}}

The Lean development lives under \texttt{Scott1972/ContinuousLattice/} (root import \texttt{Scott1972.lean}). We track Scott's numbered statements: each row in §4 corresponds to a named theorem in the repository, proved sorry-free. Results not separately numbered by Scott but required for later steps---such as \texttt{wayBelow\_interpolate}, \texttt{exists\_wayBelow\_subset}, and the Milner infrastructure---appear as supporting lemmas in the module map below.

\textbf{Milner infrastructure} (March 1972 correction): \texttt{CoarserThanScottTopology}, \texttt{scottOpen\_of\_coarserThanScott}, \texttt{scottLowerSubbasisSet}, \texttt{scottPrincipalUpSet} in \texttt{MilnerCorrection.lean}.

\textbf{Notation:} \texttt{⊔S′} denotes the ambient join in \(D′\) (\texttt{ambientSSup}); \texttt{⊔S} the subspace join; \texttt{j(⊔S′)\ =\ ⊔S} is \texttt{retr\_ambientSSup\_eq\_sSup}.

\hypertarget{constructivity}{%
\subsection{Constructivity}\label{constructivity}}

The formalization is \textbf{classical}. It uses mathlib topology, \texttt{Classical.choice} (transitively), embedding into Sierpiński powers, and order-theoretic arguments that have not been audited for constructivity. Every completed proof reports \texttt{\#print\ axioms} as \texttt{{[}propext,\ Classical.choice,\ Quot.sound{]}}.

\hypertarget{module-map}{%
\subsection{Module map}\label{module-map}}

\begin{longtable}[]{@{}
  >{\raggedright\arraybackslash}p{(\columnwidth - 4\tabcolsep) * \real{0.3333}}
  >{\raggedright\arraybackslash}p{(\columnwidth - 4\tabcolsep) * \real{0.3333}}
  >{\raggedright\arraybackslash}p{(\columnwidth - 4\tabcolsep) * \real{0.3333}}@{}}
\toprule\noalign{}
\begin{minipage}[b]{\linewidth}\raggedright
Scott §
\end{minipage} & \begin{minipage}[b]{\linewidth}\raggedright
Title
\end{minipage} & \begin{minipage}[b]{\linewidth}\raggedright
Lean modules
\end{minipage} \\
\midrule\noalign{}
\endhead
\bottomrule\noalign{}
\endlastfoot
§1 & \textbf{Injective spaces} & \texttt{Injective.lean} \\
§2 & \textbf{Continuous lattices} & \texttt{WayBelow.lean}, \texttt{Specialization.lean}, \texttt{ScottMaps.lean}, \texttt{Constructions.lean}, \texttt{MilnerCorrection.lean} \\
§3 & \textbf{Function spaces} & \texttt{FunctionSpaces.lean} \\
§4 & \textbf{Inverse limits} & \texttt{InverseLimits.lean}, \texttt{FunctionSpaceTower.lean} \\
\end{longtable}

\begin{center}\rule{0.5\linewidth}{0.5pt}\end{center}

\hypertarget{catalog-of-formalized-results}{%
\section{Catalog of formalized results}\label{catalog-of-formalized-results}}

We formalize \textbf{43} numbered results from Scott §1--§4. Each entry is a full statement matching Scott's numbering, proved without \texttt{sorry}. Theorem 4.4 is split into four subgoals \textbf{(a)--(d)} in both Scott and Lean.

\textbf{Supporting keystones} (not separately numbered by Scott): \texttt{directedOn\_wayBelow}, \texttt{wayBelow\_interpolate} (interpolation for \(\ll\), \textbf{axiom-free}), and \texttt{exists\_wayBelow\_subset} (the \(\uparrow a\) basis of the Scott topology) in \texttt{WayBelow.lean}; these underpin Proposition 2.11.

\begin{longtable}[]{@{}
  >{\raggedright\arraybackslash}p{(\columnwidth - 6\tabcolsep) * \real{0.2500}}
  >{\raggedright\arraybackslash}p{(\columnwidth - 6\tabcolsep) * \real{0.2500}}
  >{\raggedright\arraybackslash}p{(\columnwidth - 6\tabcolsep) * \real{0.2500}}
  >{\raggedright\arraybackslash}p{(\columnwidth - 6\tabcolsep) * \real{0.2500}}@{}}
\toprule\noalign{}
\begin{minipage}[b]{\linewidth}\raggedright
§
\end{minipage} & \begin{minipage}[b]{\linewidth}\raggedright
Scott
\end{minipage} & \begin{minipage}[b]{\linewidth}\raggedright
Lean identifier(s)
\end{minipage} & \begin{minipage}[b]{\linewidth}\raggedright
Module
\end{minipage} \\
\midrule\noalign{}
\endhead
\bottomrule\noalign{}
\endlastfoot
1 & Prop 1.2 & \texttt{proposition\_1\_2} & \texttt{Injective.lean} \\
1 & Prop 1.3 & \texttt{proposition\_1\_3} & \texttt{Injective.lean} \\
1 & Prop 1.4 & \texttt{proposition\_1\_4} & \texttt{Injective.lean} \\
1 & Prop 1.5 & \texttt{proposition\_1\_5} & \texttt{Injective.lean} \\
1 & Cor 1.6 & \texttt{corollary\_1\_6} & \texttt{Injective.lean} \\
1 & Cor 1.7 & \texttt{corollary\_1\_7} & \texttt{Injective.lean} \\
2 & Prop 2.1 & \texttt{proposition\_2\_1} & \texttt{Specialization.lean} \\
2 & Prop 2.2 & \texttt{bot\_wayBelow}, \texttt{WayBelow.sup}, \texttt{WayBelow.trans\_le}, \texttt{WayBelow.le\_trans}, \texttt{wayBelow\_self\_iff\_scottOpen\_Ici}, \texttt{wayBelow\_sSup\_iff} & \texttt{WayBelow.lean} \\
2 & Prop 2.4 & \texttt{isContinuousLattice\_iff\_isLUB\_sInf\_nhds} & \texttt{WayBelow.lean} \\
2 & Prop 2.5 & \texttt{proposition\_2\_5} & \texttt{ScottMaps.lean} \\
2 & Prop 2.6 & \texttt{proposition\_2\_6} & \texttt{ScottMaps.lean} \\
2 & Prop 2.8 & \texttt{proposition\_2\_8} & \texttt{Constructions.lean} \\
2 & Prop 2.9(a) & \texttt{proposition\_2\_9\_a} & \texttt{Constructions.lean} \\
2 & Prop 2.9(b) & \texttt{proposition\_2\_9\_b}, \texttt{proposition\_2\_9} & \texttt{Constructions.lean} \\
2 & Prop 2.10(a) & \texttt{proposition\_2\_10\_a} & \texttt{FunctionSpaces.lean} \\
2 & Prop 2.10(b) & \texttt{proposition\_2\_10\_b}, \texttt{proposition\_2\_10} & \texttt{FunctionSpaces.lean} \\
2 & Prop 2.11 & \texttt{proposition\_2\_11} & \texttt{Constructions.lean} \\
2 & Thm 2.12 & \texttt{theorem\_2\_12}, \texttt{theorem\_2\_12\_backward}, \texttt{theorem\_2\_12\_forward} & \texttt{Theorem212.lean} \\
3 & Prop 3.2 & \texttt{proposition\_3\_2} & \texttt{FunctionSpaces.lean} \\
3 & Thm 3.3(a) & \texttt{theorem\_3\_3\_isContinuousLattice}, \texttt{ScottMap.instCompleteLattice}, \texttt{stepMap}, \texttt{stepMap\_wayBelow}, \texttt{stepMap\_pointwise\_sSup} & \texttt{FunctionSpaces.lean} \\
3 & Thm 3.3(b) & \texttt{theorem\_3\_3\_topology}, \texttt{theorem\_3\_3}, \texttt{wayBelow\_le\_finset\_sup\_step}, \texttt{pointwiseSubbasic\_scottOpen} & \texttt{FunctionSpaces.lean} \\
3 & Cor 3.4 & \texttt{corollary\_3\_4\_jointly\_continuous}, \texttt{corollary\_3\_4\_preservesDirectedSup}, \texttt{corollary\_3\_4} & \texttt{FunctionSpaces.lean} \\
3 & Prop 3.5 & \texttt{proposition\_3\_5}, \texttt{scottLambda}, \texttt{curry\_left/right\_preservesDirectedSup}, \texttt{lambda\_outer\_preservesDirectedSup} & \texttt{FunctionSpaces.lean} \\
3 & Prop 3.7 & \texttt{proposition\_3\_7\_retraction}, \texttt{proposition\_3\_7\_projection} & \texttt{FunctionSpaces.lean} \\
3 & Prop 3.8 & \texttt{proposition\_3\_8}, \texttt{scottExtend\_maximal}, \texttt{continuous\_eq\_sSup\_openInfs} & \texttt{Constructions.lean} \\
3 & Lemma 3.9 & \texttt{lemma\_3\_9}, \texttt{scottExtend\_maximal\_le} & \texttt{Theorem212.lean} \\
3 & Prop 3.10 & \texttt{incl\_sSup}, \texttt{incl\_injective}, \texttt{incl\_wayBelow}, \texttt{proposition\_3\_10\_converse}, \texttt{retr\_eq\_sSup} & \texttt{FunctionSpaces.lean} \\
3 & Prop 3.12 & \texttt{proposition\_3\_12}, \texttt{IsProjection}, \texttt{isProjection\_sSup}, \texttt{Projections.instCompleteLattice} & \texttt{FunctionSpaces.lean} \\
3 & Prop 3.13 & \texttt{proposition\_3\_13}, \texttt{Proposition313.projection} & \texttt{FunctionSpaces.lean} \\
3 & Prop 3.14 & \texttt{proposition\_3\_14}, \texttt{Proposition314.fixMap}, \texttt{fix\_eq}, \texttt{fix\_le}, \texttt{fix\_unique} & \texttt{FunctionSpaces.lean} \\
4 & Prop 4.1 & \texttt{proposition\_4\_1}, \texttt{InverseLimit}, \texttt{inverseLimitTopology}, \texttt{inverseLimit\_isInjectiveSpace} & \texttt{InverseLimits.lean} \\
4 & Prop 4.2 & \texttt{proposition\_4\_2}, \texttt{embInf}, \texttt{projInf}, \texttt{iComp}, \texttt{embInf\_succ}, \texttt{inverseLimit\_eq\_iSup} & \texttt{InverseLimits.lean} \\
4 & Cor 4.3 & \texttt{corollary\_4\_3}, \texttt{coconeInf}, \texttt{coconeInf\_comp\_embInf} & \texttt{InverseLimits.lean} \\
4 & Lemma 4.5 & \texttt{lemma\_4\_5}, \texttt{idInf\_eq\_iSup} & \texttt{InverseLimits.lean} \\
4 & Thm 4.4(a) & \texttt{embInfInf}, \texttt{projInfInf}, \texttt{iInfTerm}, \texttt{jInfTerm}, \texttt{*\_apply}, \texttt{*\_preservesDirectedSup} & \texttt{FunctionSpaceTower.lean} \\
4 & Thm 4.4(b) & \texttt{projInfInf\_comp\_embInfInf} & \texttt{FunctionSpaceTower.lean} \\
4 & Thm 4.4(c) & \texttt{embInfInf\_comp\_projInfInf} & \texttt{FunctionSpaceTower.lean} \\
4 & Thm 4.4(d) & \texttt{theorem\_4\_4}, \texttt{theorem\_4\_4\_inverses}, \texttt{theorem\_4\_4\_sourceTopology\_homeomorph}, \texttt{theorem\_4\_4\_orderIso} & \texttt{FunctionSpaceTower.lean} \\
\end{longtable}

\hypertarget{proof-dependency-structure}{%
\subsection{Proof dependency structure}\label{proof-dependency-structure}}

Scott §1--§4 are not independent modules; the Lean import graph follows Scott's exposition order. Note that Propositions 2.10(a)--(b) live in \texttt{FunctionSpaces.lean} (§3 module) even though Scott numbers them in §2; \texttt{Theorem212.lean} bridges §2 and §3 for Theorem 2.12 and Lemma 3.9.

\begin{center}
\includegraphics[max width=\linewidth,max totalheight=0.85\textheight,keepaspectratio]{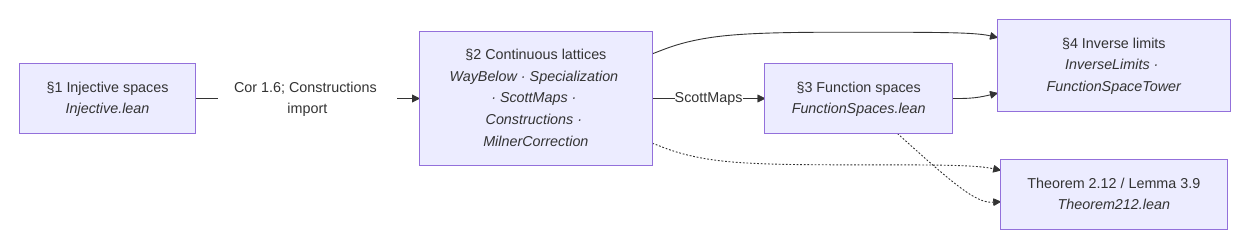}
\end{center}

\hypertarget{injective-spaces-result-hierarchy}{%
\subsection{§1 Injective spaces --- result hierarchy}\label{injective-spaces-result-hierarchy}}

The six results of §1 form a short chain from the Sierpiński space \(\mathbb{O}\) to the retract characterization of injectivity.

\begin{center}
\includegraphics[max width=\linewidth,max totalheight=0.85\textheight,keepaspectratio]{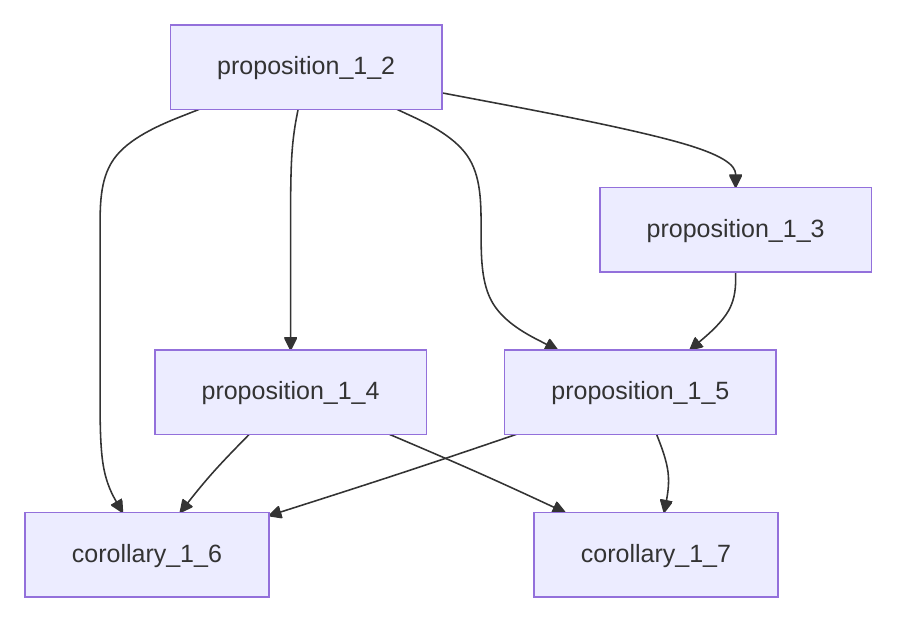}
\end{center}

\hypertarget{continuous-lattices-result-hierarchy}{%
\subsection{§2 Continuous lattices --- result hierarchy}\label{continuous-lattices-result-hierarchy}}

\begin{center}
\includegraphics[max width=\linewidth,max totalheight=0.85\textheight,keepaspectratio]{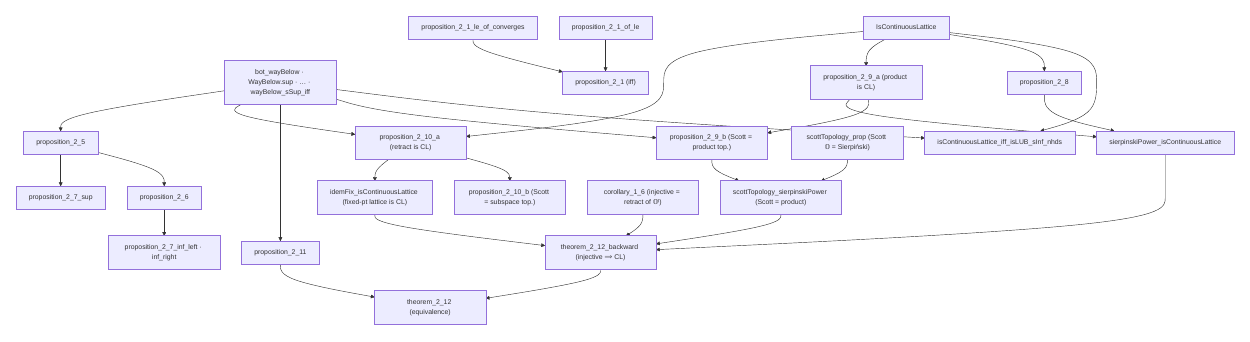}
\end{center}

\hypertarget{function-spaces-result-hierarchy}{%
\subsection{§3 Function spaces --- result hierarchy}\label{function-spaces-result-hierarchy}}

Proposition 2.7 is proved in \texttt{ScottMaps.lean} but not tracked separately in §4; it feeds Proposition 2.6's infrastructure. Propositions 2.10(a)--(b) are proved in \texttt{FunctionSpaces.lean}. Proposition 3.2 is formalized but is not on the critical path to later results.

\begin{center}
\includegraphics[max width=\linewidth,max totalheight=0.85\textheight,keepaspectratio]{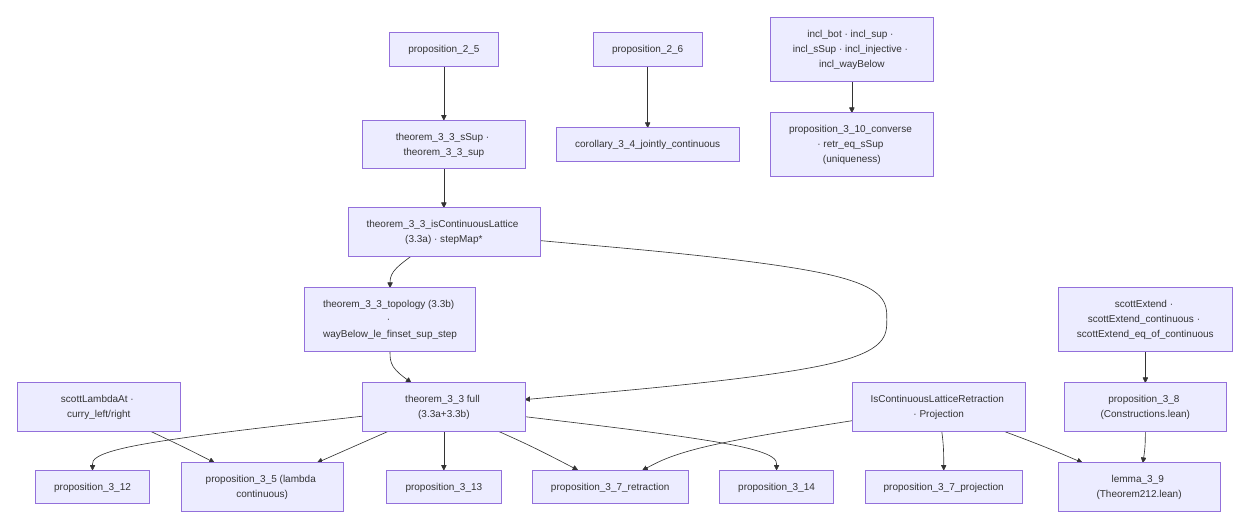}
\end{center}

\hypertarget{inverse-limits-result-hierarchy}{%
\subsection{§4 Inverse limits --- result hierarchy}\label{inverse-limits-result-hierarchy}}

Scott §4 is complete in Lean: Propositions 4.1--4.2, Corollary 4.3, Lemma 4.5, and Theorem 4.4 \textbf{(a)--(d)}. See §5.3 for proof notes on the capstone.

The Lean proof of Proposition 4.1 follows Scott's source route: Proposition 3.8 constructs maximal coordinate extensions, Lemma 3.9 makes them compatible, and injectivity plus Theorem 2.12 yields the continuous lattice and identifies its Scott topology with the product/subspace inverse-limit topology. Lemma 4.5 follows Scott's induction and enters Theorem 4.4(c).

\begin{center}
\includegraphics[max width=\linewidth,max totalheight=0.85\textheight,keepaspectratio]{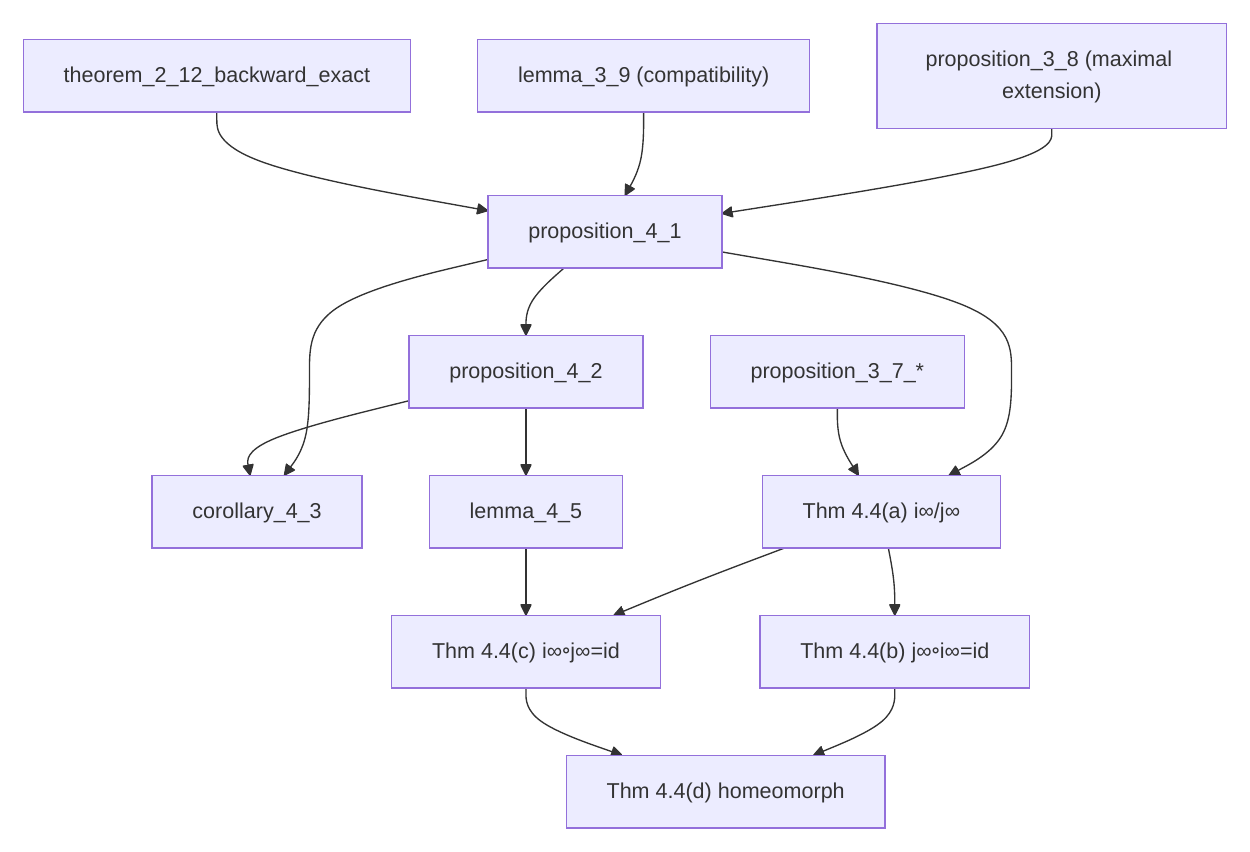}
\end{center}

\begin{center}\rule{0.5\linewidth}{0.5pt}\end{center}

\hypertarget{notes-on-the-formalization}{%
\section{Notes on the formalization}\label{notes-on-the-formalization}}

This section records proof strategy, Lean engineering choices, and lessons learned where the mechanization goes beyond Scott's published text. Results in §1 follow Scott's short arguments directly and are not discussed here. Proposition 2.1 is split in Lean as \texttt{proposition\_2\_1\_of\_le} and \texttt{proposition\_2\_1\_le\_of\_converges}, then bundled as \texttt{proposition\_2\_1}.

\hypertarget{continuous-lattices-scott-2}{%
\subsection{Continuous lattices (Scott §2)}\label{continuous-lattices-scott-2}}

Section §2 is where most of the order-topology alignment work lives: products (2.9), retractions (2.10), injectivity (2.11), and the equivalence theorem (2.12) all required careful handling of Scott topologies without registering competing instances.

\hypertarget{proposition-2.6-joint-separate-continuity-proposition_2_6}{%
\subsubsection{\texorpdfstring{Proposition 2.6 (joint ↔ separate continuity) --- \texttt{proposition\_2\_6}}{Proposition 2.6 (joint ↔ separate continuity) --- proposition\_2\_6}}\label{proposition-2.6-joint-separate-continuity-proposition_2_6}}

Scott states that a function of several variables between complete lattices is continuous jointly if and only if it is continuous in each variable separately. We formalize the two-variable case \texttt{f\ :\ D\ ×\ D\textquotesingle{}\ →\ D\textquotesingle{}\textquotesingle{}}, with continuity phrased as \texttt{PreservesDirectedSup} (justified by Prop 2.5), and the product \texttt{D\ ×\ D\textquotesingle{}} carrying the componentwise complete-lattice structure (whose induced topology is the product topology). The proof follows Scott's directed-net argument:

\begin{itemize}
\item
  \textbf{Joint ⟹ separate.} Precompose \texttt{f} with the slice map \texttt{x\ ↦\ (x,\ y)}. The image of a directed \texttt{S\ ⊆\ D} under this map is directed in \texttt{D\ ×\ D\textquotesingle{}} with least upper bound \texttt{(⊔S,\ y)} (computed componentwise via \texttt{Prod.fst\_sSup} / \texttt{Prod.snd\_sSup}, using \texttt{S} nonempty for the constant second coordinate). Joint preservation of that supremum therefore yields preservation in the first variable; the second variable is symmetric.
\item
  \textbf{Separate ⟹ joint} (the substance). For directed \texttt{S*\ ⊆\ D\ ×\ D\textquotesingle{}}, project to the directed sets \texttt{S\ =\ fst\ \textquotesingle{}\textquotesingle{}\ S*} and \texttt{S\textquotesingle{}\ =\ snd\ \textquotesingle{}\textquotesingle{}\ S*} (directedness via \texttt{DirectedOn.fst} / \texttt{DirectedOn.snd}), so that \texttt{⊔S*\ =\ (⊔S,\ ⊔S\textquotesingle{})}. Then:

  \begin{itemize}
    \item
    \texttt{⊔(f\ \textquotesingle{}\textquotesingle{}\ S*)\ ≤\ f(⊔S*)} is immediate from monotonicity of \texttt{f} (assembled from the separate monotonicities \texttt{hmono1}, \texttt{hmono2}).
  \item
    \texttt{f(⊔S*)\ ≤\ ⊔(f\ \textquotesingle{}\textquotesingle{}\ S*)}: unfolding separate continuity twice gives \texttt{f(⊔S*)\ =\ ⊔\_\{x∈S\}\ ⊔\_\{y∈S\textquotesingle{}\}\ f(x,\ y)}; for each pair \texttt{x\ ∈\ S}, \texttt{y\ ∈\ S\textquotesingle{}} there exist witnesses \texttt{(x,\ b),\ (a,\ y)\ ∈\ S*}, and \textbf{directedness of \texttt{S*}} supplies \texttt{r\ ∈\ S*} above both, so \texttt{(x,\ y)\ ≤\ r} and \texttt{f(x,\ y)\ ≤\ f(r)\ ≤\ ⊔(f\ \textquotesingle{}\textquotesingle{}\ S*)} by monotonicity. This is exactly Scott's ``monotonicity + directedness'' step.
  \end{itemize}
\end{itemize}

Sorry-free; \texttt{\#print\ axioms} gives \texttt{{[}propext,\ Classical.choice,\ Quot.sound{]}} (the standard classical footprint throughout this development).

\hypertarget{proposition-2.8-finite-lattices-are-continuous-proposition_2_8}{%
\subsubsection{\texorpdfstring{Proposition 2.8 (finite lattices are continuous) --- \texttt{proposition\_2\_8}}{Proposition 2.8 (finite lattices are continuous) --- proposition\_2\_8}}\label{proposition-2.8-finite-lattices-are-continuous-proposition_2_8}}

Scott states this as a one-line example. The Lean proof isolates the genuinely finite step in a reusable lemma \texttt{directedOn\_finite\_sSup\_mem}: \emph{a non-empty finite directed set attains its supremum} (\texttt{⊔S\ ∈\ S}). A maximal element \texttt{m\ ∈\ S} exists by \texttt{Set.Finite.exists\_maximal}; by directedness any \texttt{s\ ∈\ S} and \texttt{m} have an upper bound \texttt{c\ ∈\ S}, and maximality forces \texttt{c\ ≤\ m}, so \texttt{s\ ≤\ m}. Hence \texttt{m} is the greatest element, \texttt{IsLUB\ S\ m}, and \texttt{⊔S\ =\ m\ ∈\ S}. With this, every principal up-set \texttt{Set.Ici\ y} is Scott-open (a directed \texttt{S} with \texttt{y\ ≤\ ⊔S} has \texttt{⊔S\ ∈\ S}), so \texttt{y\ ≪\ y} via \texttt{wayBelow\_self\_iff\_scottOpen\_Ici}, and \texttt{y} is trivially the supremum of \texttt{\{x\ \textbar{}\ x\ ≪\ y\}}. \texttt{{[}Finite\ D{]}} suffices (subsets are finite via \texttt{Set.toFinite}).

\hypertarget{proposition-2.9-products-of-continuous-lattices-proposition_2_9_a-proposition_2_9_b}{%
\subsubsection{\texorpdfstring{Proposition 2.9 (products of continuous lattices) --- \texttt{proposition\_2\_9\_a}, \texttt{proposition\_2\_9\_b}}{Proposition 2.9 (products of continuous lattices) --- proposition\_2\_9\_a, proposition\_2\_9\_b}}\label{proposition-2.9-products-of-continuous-lattices-proposition_2_9_a-proposition_2_9_b}}

Scott's Proposition 2.9 is a \textbf{conjunction} of an order-theoretic and a topological claim, so we split it: \texttt{proposition\_2\_9\_a} (the product is a continuous lattice), \texttt{proposition\_2\_9\_b} (the Scott topology of the product equals the product of the Scott topologies), and the bundled \texttt{proposition\_2\_9\ :=\ ⟨a,\ b⟩}.

\textbf{2.9(a) --- order content (\texttt{proposition\_2\_9\_a}).} A product \texttt{∀\ i,\ Eᵢ} of continuous lattices is a continuous lattice. The construction is the cylinder element: for \texttt{a\ ≪\ yᵢ} in factor \texttt{Eᵢ}, let \texttt{{[}a{]}ⁱ\ :=\ Function.update\ ⊥\ i\ a}. Then \texttt{{[}a{]}ⁱ\ ≪\ y} in the product, witnessed by the preimage \texttt{\{z\ \textbar{}\ zᵢ\ ∈\ U\}} of a Scott-open \texttt{U\ ⊆\ Eᵢ} with \texttt{yᵢ\ ∈\ U\ ⊆\ Ici\ a}: this set is an upper set, and inaccessible because suprema are coordinatewise (\texttt{sSup\_apply\_eq\_sSup\_image}), so a directed \texttt{S} with \texttt{(⊔S)ᵢ\ ∈\ U} already has some \texttt{f\ ∈\ S} with \texttt{fᵢ\ ∈\ U}. Given any upper bound \texttt{b} of \texttt{\{x\ \textbar{}\ x\ ≪\ y\}}, each \texttt{{[}a{]}ⁱ\ ≤\ b} gives \texttt{a\ =\ ({[}a{]}ⁱ)ᵢ\ ≤\ bᵢ}; ranging over \texttt{a\ ≪\ yᵢ} and using continuity of \texttt{Eᵢ} (\texttt{(hE\ i).sSup\_wayBelow}) yields \texttt{yᵢ\ ≤\ bᵢ} for all \texttt{i}, i.e.~\texttt{y\ ≤\ b}.

\textbf{2.9(b) --- topology agreement (\texttt{proposition\_2\_9\_b}).} We prove the \emph{full equality} of topologies \texttt{scottTopologicalSpace\ =\ Pi.topologicalSpace\ (fun\ \_\ =\textgreater{}\ scottTopologicalSpace)} by \texttt{le\_antisymm}; no Milner-style coarseness hypothesis is needed. Working with explicit topology terms (\texttt{Eᵢ} carries no \texttt{TopologicalSpace} instance) keeps us clear of the \texttt{specializationPreorder} diamond, and the mathlib order \texttt{t₁\ ≤\ t₂} unfolds \emph{definitionally} to \texttt{∀\ U,\ IsOpen{[}t₂{]}\ U\ →\ IsOpen{[}t₁{]}\ U}. - \textbf{Product ⊆ Scott} (\texttt{scott\ ≤\ ⨅ᵢ\ induced\ (eval\ i)}): each projection preserves directed suprema (\texttt{sSup\_apply\_eq\_sSup\_image}), hence is Scott-continuous (\texttt{continuous\_of\_preservesDirectedSup}); \texttt{le\_iInf} + \texttt{continuous\_iff\_le\_induced} finish. - \textbf{Scott ⊆ Product}: for a Scott-open \texttt{U\ ∋\ z} the \texttt{↟a} basis (\texttt{exists\_wayBelow\_Ici\_subset}, the \texttt{Ici}-strengthening of \texttt{exists\_wayBelow\_subset}) gives \texttt{a\ ≪\ z} with \texttt{↑a\ ⊆\ U}. Three new structural lemmas about way-below in a product do the rest: \texttt{wayBelow\_proj} (\texttt{a\ ≪\ z\ ⟹\ aᵢ\ ≪\ zᵢ}, via the preimage under \texttt{v\ ↦\ Function.update\ z\ i\ v}, Scott-open by \texttt{update\_preservesDirectedSup}) and \texttt{wayBelow\_finite\_support} (\texttt{a\ ≪\ z} has finite support: the truncations \texttt{Z\ F\ =\ (if\ ·\ ∈\ F\ then\ z·\ else\ ⊥)} are directed with sup \texttt{z}, so \texttt{a\ ≤\ Z\ F} for some finite \texttt{F}). The finite box \texttt{⋂\_\{i∈F\}\ eval\ i\ ⁻¹\textquotesingle{}\ Vᵢ} (with \texttt{Vᵢ\ ∋\ zᵢ} Scott-open inside \texttt{Ici\ aᵢ}) is product-open (\texttt{isOpen\_biInter\_finset} of induced-opens, each \texttt{≥} the product topology by \texttt{iInf\_le}) and lies in \texttt{↑a\ ⊆\ U} (off \texttt{F}, \texttt{aⱼ\ =\ ⊥\ ≤\ wⱼ}; on \texttt{F}, \texttt{aᵢ\ ≤\ wᵢ}).

\texttt{classical} supplies the \texttt{DecidableEq} for \texttt{Function.update}; footprint \texttt{{[}propext,\ Classical.choice,\ Quot.sound{]}} for all of 2.9(a)/(b).

\textbf{Engineering notes / lessons from 2.9(b).} This was the hardest single proof in the development; recording the dead-ends so the next session does not re-pay the cost):

\begin{itemize}
\item
  \emph{Avoid \texttt{letI} for the factor/product topologies.} The tempting move is \texttt{letI\ :\ ∀\ i,\ TopologicalSpace\ (Eᵢ)\ :=\ fun\ \_\ =\textgreater{}\ scottTopologicalSpace} so that mathlib's \texttt{Pi.topologicalSpace}, \texttt{continuous\_apply}, \texttt{isOpen\_biInter\_finset}, \ldots{} resolve by instance. But our imports make \texttt{specializationPreorder} an active instance, so a \texttt{TopologicalSpace\ (Eᵢ)} in scope introduces a \textbf{second \texttt{Preorder\ (Eᵢ)}} that fights the \texttt{CompleteLattice} one --- the same diamond that broke \texttt{scottExtend\_eq\_of\_continuous} earlier. Keeping every topology an \textbf{explicit term} (\texttt{@Pi.topologicalSpace\ …}, \texttt{@IsOpen\ \_\ scottTopologicalSpace\ …}) and never registering an instance is what makes the proof go through. The order reasoning (way-below, \texttt{sSup}, finite support) lives in \emph{instance-free} lemmas (\texttt{wayBelow\_proj}, \texttt{wayBelow\_finite\_support}) precisely so they never see a competing topology.
\item
  \emph{Use the definitional unfolding of the topology order.} \texttt{TopologicalSpace.le\_def} shows \texttt{t₁\ ≤\ t₂} \textbf{is} \texttt{∀\ U,\ IsOpen{[}t₂{]}\ U\ →\ IsOpen{[}t₁{]}\ U} (the partial order's \texttt{le} field), so \texttt{intro\ U\ hU} works directly on a \texttt{P\ ≤\ S} goal and \texttt{iInf\_le\ \_\ i\ \_\ hopen} turns an induced-open into a product-open with no \texttt{le\_def} rewrite or \texttt{IsOpen.mono} lemma. This is the single most useful fact for product/Scott topology bridges.
\item
  \emph{Prefer \texttt{Set.Ici\ a\ ⊆\ U} over \texttt{↟a\ ⊆\ U}.} \texttt{exists\_wayBelow\_subset} actually proves the stronger \texttt{Set.Ici\ a\ ⊆\ U} (the witness \texttt{a} lies in the upper-set \texttt{U}), so the new \texttt{exists\_wayBelow\_Ici\_subset} lets the box-containment step ask only for \texttt{a\ ≤\ w} instead of \texttt{a\ ≪\ w}. This \textbf{eliminates the way-below \texttt{⟸} characterization} (componentwise-\texttt{≪} + finite-support ⟹ product-\texttt{≪}) entirely --- a large, fiddly \texttt{Finset.sup}-of-cylinders argument we would otherwise have needed.
\item
  \emph{Finite support falls out of the truncations, not a separate axiom.} \texttt{a\ ≪\ z} plus the directed family \texttt{Z\ F\ =\ (if\ ·\ ∈\ F\ then\ z·\ else\ ⊥)} (sup \texttt{z}) gives \texttt{a\ ≤\ Z\ F} for some finite \texttt{F} via \texttt{wayBelow\_sSup\_iff}; then \texttt{aⱼ\ ≤\ (Z\ F)ⱼ\ =\ ⊥} off \texttt{F}. No independent ``way-below ⟹ finite support'' theorem is required.
\item
  \emph{\texttt{@}-argument order is worth checking empirically.} \texttt{isOpen\_biInter\_finset} autobinds as \texttt{@isOpen\_biInter\_finset\ X\ α\ {[}inst{]}\ s\ f\ h} (space first, index second); \texttt{isOpen\_induced\_iff} needs the codomain topology, supplied painlessly by the named argument \texttt{(t\ :=\ scottTopologicalSpace)} rather than a positional \texttt{@}. When in doubt, feed one wrong argument and read the ``expected type'' in the error to recover the true order.
\item
  \emph{Beta-reduce before \texttt{rw}.} \texttt{PreservesDirectedSup\ f} unfolds to \texttt{f\ (sSup\ T)\ =\ …} with \texttt{f} a literal lambda, so the goal is \texttt{(fun\ v\ =\textgreater{}\ update\ z\ i\ v)\ (sSup\ T)\ j}; a \texttt{Function.update\_self} rewrite only matches after a \texttt{show} (or \texttt{dsimp\ only}) forces the beta reduction to \texttt{Function.update\ z\ i\ (sSup\ T)}.
\end{itemize}

\hypertarget{proposition-2.10-a-retract-of-a-cl-is-a-cl-proposition_2_10_a-proposition_2_10_b}{%
\subsubsection{\texorpdfstring{Proposition 2.10 (a retract of a CL is a CL) --- \texttt{proposition\_2\_10\_a}, \texttt{proposition\_2\_10\_b}}{Proposition 2.10 (a retract of a CL is a CL) --- proposition\_2\_10\_a, proposition\_2\_10\_b}}\label{proposition-2.10-a-retract-of-a-cl-is-a-cl-proposition_2_10_a-proposition_2_10_b}}

Like 2.9, Scott's 2.10 bundles an order claim and a topology claim; we split it as \texttt{proposition\_2\_10\_a} / \texttt{proposition\_2\_10\_b} with the bundled \texttt{proposition\_2\_10}. A \emph{retract} is the existing \texttt{IsContinuousLatticeRetraction\ D\ D\textquotesingle{}}: Scott maps \texttt{i\ :\ D\ →\ D\textquotesingle{}}, \texttt{j\ :\ D\textquotesingle{}\ →\ D} with \texttt{j\ ∘\ i\ =\ id}. We take \texttt{D\textquotesingle{}} continuous and conclude both halves for \texttt{D}.

The single engine is \texttt{retr\_wayBelow\_of\_wayBelow\_incl}: \textbf{\texttt{x\textquotesingle{}\ ≪\ i(d)} in \texttt{D\textquotesingle{}} ⟹ \texttt{j(x\textquotesingle{})\ ≪\ d} in \texttt{D}}. Witness the \texttt{D}-way-below by \texttt{i⁻¹V\textquotesingle{}} for an ambient Scott-open witness \texttt{V\textquotesingle{}} of \texttt{x\textquotesingle{}\ ≪\ i(d)} (\texttt{i⁻¹V\textquotesingle{}} is Scott-open since \texttt{i} preserves directed sups, \texttt{scottOpen\_preimage}); for \texttt{z\ ∈\ i⁻¹V\textquotesingle{}}, \texttt{x\textquotesingle{}\ ⊑\ i(z)} gives \texttt{j(x\textquotesingle{})\ ⊑\ j(i(z))\ =\ z}. With \texttt{sSup\_image\_retr\_wayBelow} (\texttt{d\ =\ ⊔\_D\ \{j(x\textquotesingle{})\ :\ x\textquotesingle{}\ ≪\ i(d)\}}, from \texttt{j(⊔\textquotesingle{}S′)\ =\ ⊔S} + continuity of \texttt{D\textquotesingle{}}): - \textbf{2.10(a).} Any upper bound \texttt{b} of \texttt{\{x\ \textbar{}\ x\ ≪\ d\}} dominates every \texttt{j(x\textquotesingle{})}, hence the supremum \texttt{d}. (\texttt{IsLUB} is immediate.) - \textbf{2.10(b).} \texttt{scott\ =\ induced\ i\ scott\textquotesingle{}}. The easy \texttt{scott\ ≤\ induced} is \texttt{scottOpen\_preimage} again. The hard \texttt{induced\ ≤\ scott} (Milner) shows the family \texttt{\{i⁻¹(↟x\textquotesingle{})\ :\ x\textquotesingle{}\ ∈\ D\textquotesingle{}\}} is a \textbf{basis} of \texttt{D}'s Scott topology: given Scott-open \texttt{U\ ∋\ d}, the directed family \texttt{\{j(x\textquotesingle{})\ :\ x\textquotesingle{}\ ≪\ i(d)\}} (sup \texttt{d}) meets \texttt{U} at some \texttt{j(x\textquotesingle{})}, and \texttt{i⁻¹(↟x\textquotesingle{})\ ⊆\ U} because \texttt{x\textquotesingle{}\ ≪\ i(z)\ ⟹\ j(x\textquotesingle{})\ ⊑\ z} and \texttt{U} is upper. Each \texttt{i⁻¹(↟x\textquotesingle{})} is induced-open by construction, so every Scott-open is a union of induced-opens, i.e.~induced-open.

\textbf{Engineering notes / lessons from 2.10:}

\begin{itemize}
\item
  \emph{No projection, no Milner hypothesis needed.} Scott proves 2.10 for general retractions and only needs \emph{projections} later (for the function-space 3.7/3.9). The whole proof goes through with the bare \texttt{IsContinuousLatticeRetraction} (Scott maps + \texttt{j\ ∘\ i\ =\ id}); \texttt{incl\_retr\_le} is never used. And, as with 2.9(b), the topology agreement is a genuine equality --- \texttt{CoarserThanScottTopology} does not appear. The Milner subtlety (``lubs in the subspace are \emph{larger}, so a relativised open need not be lattice-open'') is dissolved by the retraction: \texttt{j(⊔S′)\ =\ ⊔S} realigns the inequality.
\item
  \emph{Reuse the abstract structure instead of building a \texttt{CompleteLattice} on a subtype.} The tempting faithful reading --- fixed points \texttt{\{x\ //\ j\ x\ =\ x\}} of an idempotent Scott map, with transported joins \texttt{sSup\_K\ S\ =\ j(⊔\textquotesingle{}\ i\textquotesingle{}\textquotesingle{}S)} --- forces a hand-built \texttt{CompleteLattice} instance (every axiom, then continuity, then topology) and is several hundred lines. Phrasing the retract as \emph{its own} lattice \texttt{D} with Scott maps to/from \texttt{D\textquotesingle{}} captures exactly the same content (\texttt{i} preserving directed sups \textbf{is} the statement that \texttt{D}-joins are \texttt{j} of ambient joins) at a fraction of the cost.
\item
  \emph{\texttt{isOpen\_induced\_iff} needs the codomain topology pinned.} \texttt{Eᵢ}/\texttt{D\textquotesingle{}} carry no \texttt{TopologicalSpace} instance, so \texttt{rw\ {[}isOpen\_induced\_iff{]}} fails instance synthesis; supply \texttt{(t\ :=\ scottTopologicalSpace)} (same trick as 2.9(b)).
\item
  \emph{\texttt{scottOpen\_preimage} is the workhorse.} ``Preimage of a Scott-open under a Scott map is Scott-open'' appears three times here (the way-below witness, and both topology inclusions). Packaging \texttt{incl\_preservesDirectedSup\ :\ PreservesDirectedSup\ ⇑i} once keeps the call sites clean.
\end{itemize}

This unblocks the \textbf{backward half of Theorem 2.12} (injective ⟹ CL) at the \emph{retract} level; the embedding of an injective space into a power of \texttt{𝕆} (1.6) supplies the rest, and 2.12 is now \textbf{complete} (see the Theorem 2.12 note below).

\hypertarget{keystones-for-2.11-interpolation-and-the-a-basis-waybelow.lean}{%
\subsubsection{\texorpdfstring{Keystones for 2.11: interpolation and the \texttt{↟a} basis --- \texttt{WayBelow.lean}}{Keystones for 2.11: interpolation and the ↟a basis --- WayBelow.lean}}\label{keystones-for-2.11-interpolation-and-the-a-basis-waybelow.lean}}

Two standard facts about \texttt{≪} that mathlib does not provide and that the capstone needs:

\begin{itemize}
\item
  \textbf{Interpolation} (\texttt{wayBelow\_interpolate}): in a continuous lattice \texttt{a\ ≪\ c\ ⟹\ ∃\ b,\ a\ ≪\ b\ ≪\ c}. The set \texttt{M\ =\ \{m\ \textbar{}\ ∃\ x,\ m\ ≪\ x\ ∧\ x\ ≪\ c\}} is directed (apply directedness of \texttt{\{·\ ≪\ x\}} twice) with \texttt{⊔M\ =\ c} (continuity twice); then \texttt{a\ ≪\ c\ =\ ⊔M} forces \texttt{a\ ≪\ m\ ≤\ x\ ≪\ c} for some \texttt{m\ ≪\ x\ ≪\ c}, so \texttt{b\ :=\ x}. Notably this is \textbf{axiom-free} (\texttt{\#print\ axioms} reports none).
\item
  \textbf{\texttt{↟a} basis} (\texttt{exists\_wayBelow\_subset}): every Scott-open \texttt{U\ ∋\ z} contains a basic neighbourhood \texttt{↟a\ =\ \{w\ \textbar{}\ a\ ≪\ w\}} with \texttt{a\ ≪\ z}. Since \texttt{z\ =\ ⊔\{a\ \textbar{}\ a\ ≪\ z\}} is a directed sup in the open \texttt{U}, inaccessibility yields \texttt{a\ ≪\ z} with \texttt{a\ ∈\ U}, and \texttt{↟a\ ⊆\ ↑a\ ⊆\ U}.
\end{itemize}

\hypertarget{proposition-2.11-continuous-lattices-are-injective-proposition_2_11}{%
\subsubsection{\texorpdfstring{Proposition 2.11 (continuous lattices are injective) --- \texttt{proposition\_2\_11}}{Proposition 2.11 (continuous lattices are injective) --- proposition\_2\_11}}\label{proposition-2.11-continuous-lattices-are-injective-proposition_2_11}}

The substantial half of Theorem 2.12. The witness is an explicit operator \texttt{scottExtend\ e\ f\ y\ =\ ⊔\ \{\ ⊓\ f\textquotesingle{}\textquotesingle{}(e⁻¹V)\ :\ V\ an\ open\ nbhd\ of\ y\ \}} (a standalone \texttt{def}, purely order-theoretic). Two lemmas about it:

\begin{itemize}
\item
  \textbf{Extends \texttt{f}} (\texttt{scottExtend\_eq\_of\_continuous}). The \texttt{≤} bound is immediate (\texttt{f\ x₀} is one of the values met). For \texttt{≥}, continuity of the lattice is essential: for each \texttt{a\ ≪\ f\ x₀}, the Scott-open \texttt{↟a} pulls back along the continuous \texttt{f}, and the \textbf{embedding} turns that into an open \texttt{V\ ⊆\ Y} with \texttt{e⁻¹V\ =\ f⁻¹(↟a)}; on \texttt{e⁻¹V}, \texttt{f\ ≥\ a}, so \texttt{a\ ≤\ ⊓\ f\textquotesingle{}\textquotesingle{}(e⁻¹V)\ ≤\ g(e\ x₀)}. Summing over \texttt{a\ ≪\ f\ x₀} (continuity) gives \texttt{f\ x₀\ ≤\ g(e\ x₀)}.
\item
  \textbf{Continuous} (\texttt{scottExtend\_continuous}). Uses the \texttt{↟a} basis: for Scott-open \texttt{U} and \texttt{g\ y₀\ ∈\ U} pick \texttt{a\ ≪\ g\ y₀} with \texttt{↟a\ ⊆\ U}; as \texttt{g\ y₀} is a directed sup, \texttt{a\ ≪\ ⊓\ f\textquotesingle{}\textquotesingle{}(e⁻¹V)} for some open \texttt{V\ ∋\ y₀}, and that value is \texttt{≤\ g\ y\textquotesingle{}} for all \texttt{y\textquotesingle{}\ ∈\ V}, so \texttt{V\ ⊆\ g⁻¹U}.
\end{itemize}

A Lean-specific wrinkle: \texttt{E} carries no global \texttt{TopologicalSpace} instance (its topology is \texttt{scottTopologicalSpace}), so lemmas like \texttt{IsOpen.preimage} that \emph{synthesize} \texttt{{[}TopologicalSpace\ E{]}} fail. The order-heavy \texttt{scottExtend\_eq\_of\_continuous} uses \texttt{continuous\_def} (whose topology arguments are ordinary implicits, unified from the hypothesis) to avoid both the synthesis failure and the specialization-order diamond a \texttt{letI} would introduce; the purely topological \texttt{scottExtend\_continuous} and \texttt{proposition\_2\_11} use \texttt{letI\ :\ TopologicalSpace\ E\ :=\ scottTopologicalSpace}. Footprint \texttt{{[}propext,\ Classical.choice,\ Quot.sound{]}}.

\hypertarget{theorem-2.12-injective-continuous-lattice-theorem_2_12-theorem_2_12_backward-theorem212.lean}{%
\subsubsection{\texorpdfstring{Theorem 2.12 (injective ⟺ continuous lattice) --- \texttt{theorem\_2\_12}, \texttt{theorem\_2\_12\_backward} (\texttt{Theorem212.lean})}{Theorem 2.12 (injective ⟺ continuous lattice) --- theorem\_2\_12, theorem\_2\_12\_backward (Theorem212.lean)}}\label{theorem-2.12-injective-continuous-lattice-theorem_2_12-theorem_2_12_backward-theorem212.lean}}

Both directions are now closed; \texttt{theorem\_2\_12} is the full biconditional:

\begin{quote}
A \texttt{T₀}-space is injective \textbf{iff} it is homeomorphic to a continuous lattice under its Scott topology.
\end{quote}

\begin{itemize}
\item
  \textbf{Forward} (CL ⟹ injective) is \texttt{theorem\_2\_12\_forward} (= 2.11).
\item
  \textbf{Backward} (injective ⟹ CL) is \texttt{theorem\_2\_12\_backward}. The argument:

  \begin{enumerate}
  \def\labelenumi{\arabic{enumi}.}
    \item
    By Corollary 1.6, an injective \texttt{T₀}-space \texttt{D} is a \emph{retract} of a Sierpiński power \texttt{L\ =\ ι\ →\ 𝕆} (\texttt{𝕆\ =\ Prop}): there are continuous \texttt{s\ :\ D\ →\ L}, \texttt{r\ :\ L\ →\ D} with \texttt{r\ ∘\ s\ =\ id}.
  \item
    \texttt{L} is a continuous lattice (\texttt{sierpinskiPower\_isContinuousLattice}, from 2.8 + 2.9a) whose Scott topology \emph{is} its product topology (\texttt{scottTopology\_sierpinskiPower}, from 2.9b plus \texttt{scottTopology\_prop}: the Scott topology on \texttt{𝕆} is the Sierpiński topology).
  \item
    \texttt{e\ :=\ s\ ∘\ r} is therefore a \textbf{Scott-continuous idempotent} on \texttt{L}. Its fixed-point set \texttt{IdemFix\ e} carries the ambient-supremum-corrected complete-lattice structure (\texttt{IdemFix.completeLattice}), is a continuous lattice by Proposition 2.10 (\texttt{idemFix\_isContinuousLattice}), and \texttt{d\ ↦\ s\ d} is a homeomorphism \texttt{D\ ≃ₜ\ IdemFix\ e}.
  \end{enumerate}
\end{itemize}

\textbf{Engineering notes / lessons from 2.12:}

\begin{itemize}
\item
  \emph{Fixed points of a monotone idempotent are a complete lattice} for free via \texttt{completeLatticeOfSup}: take \texttt{sSup\_K\ S\ =\ e\ (sSup\_L\ (val\ \textquotesingle{}\textquotesingle{}\ S))} and \texttt{sInf} derived. No closure/kernel (\texttt{e\ ≤\ id} or \texttt{e\ ≥\ id}) hypothesis is needed --- only monotone + idempotent --- and Scott-continuity of \texttt{e} is what makes the inclusion/corestriction \texttt{ScottMap}s, so the retract machinery of 2.10 applies verbatim.
\item
  \emph{The subtype-topology trap.} \texttt{IdemFix\ e\ =\ \{x\ :\ L\ //\ e\ x\ =\ x\}} is reducibly a subtype of \texttt{L}, so it \textbf{auto-inherits the subspace \texttt{TopologicalSpace}}, which competes with the Scott topology coming from its (non-instance) \texttt{CompleteLattice}. This breaks \texttt{Continuous.comp}/\texttt{subtype\_mk} (they synthesize the \emph{subspace} instance, not Scott). The fix: build the homeomorphism against the canonical subspace topology (where those lemmas work), then transport across the propositional equality \texttt{scott\ =\ subspace} --- itself \texttt{idemFix\_scottTopology} (= \texttt{induced\ val\ scott\_L}) composed with \texttt{scottTopology\_sierpinskiPower} (\texttt{scott\_L\ =\ product}), closing by \texttt{rfl}.
\item
  \emph{Statement shape.} Endowing an abstract injective space with a lattice is impossible literally, so the faithful statement is ``homeomorphic to a continuous lattice under its Scott topology''; the reverse arrow transfers injectivity across the homeomorphism via \texttt{IsInjectiveSpace.of\_retract}.
\item
  Footprint \texttt{{[}propext,\ Classical.choice,\ Quot.sound{]}}.
\end{itemize}

\hypertarget{function-spaces-scott-3}{%
\subsection{Function spaces (Scott §3)}\label{function-spaces-scott-3}}

Section §3 builds the function-space lattice, proves agreement with pointwise convergence (Theorem 3.3), and develops the projection and fixed-point infrastructure needed for §4.

\hypertarget{theorem-3.3a-d-d-is-a-continuous-lattice-theorem_3_3_iscontinuouslattice-functionspaces.lean}{%
\subsubsection{\texorpdfstring{Theorem 3.3(a) (\texttt{{[}D\ →\ D\textquotesingle{}{]}} is a continuous lattice) --- \texttt{theorem\_3\_3\_isContinuousLattice} (\texttt{FunctionSpaces.lean})}{Theorem 3.3(a) ({[}D → D\textquotesingle{]} is a continuous lattice) --- theorem\_3\_3\_isContinuousLattice (FunctionSpaces.lean)}}\label{theorem-3.3a-d-d-is-a-continuous-lattice-theorem_3_3_iscontinuouslattice-functionspaces.lean}}

Scott's ``pointwise'' argument, in three movements.

\begin{enumerate}
\def\labelenumi{\arabic{enumi}.}
\item
  \textbf{Complete lattice on \texttt{{[}D\ →\ D\textquotesingle{}{]}}.} \texttt{ScottMap\ D\ D\textquotesingle{}} is a genuine \texttt{def} (a subtype of \texttt{D\ →\ D\textquotesingle{}}), so --- unlike the \texttt{IdemFix} subtype trap of 2.12 --- it carries \emph{no} auto-synthesized order/topology to fight. We register \texttt{instPartialOrder} (pointwise \texttt{≤}), \texttt{instSupSet} (\texttt{sSupMaps\ F\ x\ =\ ⊔\{g\ x\ \textbar{}\ g\ ∈\ F\}}, which is itself a \texttt{ScottMap} because pointwise suprema of Scott maps preserve directed sups), prove \texttt{isLUB\_sSup}, and close with \texttt{completeLatticeOfSup}. Crucially \texttt{sSup} here is \emph{pointwise} (\texttt{sSup\_apply} is \texttt{rfl}), matching Scott's observation that \textbf{arbitrary} (not just directed) joins are computed pointwise --- while infima are \emph{not} (derived as \texttt{⊔} of lower bounds by \texttt{completeLatticeOfSup}).
\item
  \textbf{Step functions.} \texttt{ē{[}e,e\textquotesingle{}{]}(x)\ =\ e\textquotesingle{}} if \texttt{e\ ≪\ x} else \texttt{⊥}, encoded as \texttt{⨆\ \_\ :\ e\ ≪\ x,\ e\textquotesingle{}} (\texttt{stepFun}) to dodge any \texttt{Decidable\ (e\ ≪\ x)}. Scott-continuity of \texttt{stepFun} is exactly the Scott-openness of the way-above set \texttt{\{x\ \textbar{}\ e\ ≪\ x\}} (\texttt{scottOpen\_wayBelow}, true in \emph{any} complete lattice): inaccessibility of that open set supplies the member of a directed \texttt{S} realizing the value.
\item
  \textbf{Way-below + reconstruction.} \texttt{e\textquotesingle{}\ ≪\ f\ e\ ⟹\ ē{[}e,e\textquotesingle{}{]}\ ≪\ f}, witnessed by the Scott-open \texttt{\{g\ \textbar{}\ e\textquotesingle{}\ ≪\ g\ e\}} (open because joins are pointwise, so inaccessibility reduces to \texttt{wayBelow\_sSup\_iff} in \texttt{D\textquotesingle{}}); this is the \textbf{topological} way-below of \texttt{WayBelow.lean}, so we never need an order-theoretic ≪-characterization. And \texttt{f\ x\ =\ ⊔\{e\textquotesingle{}\ \textbar{}\ ∃\ e\ ≪\ x,\ e\textquotesingle{}\ ≪\ f\ e\}} (\texttt{stepMap\_pointwise\_sSup}) follows from \texttt{x\ =\ ⊔\{e\ ≪\ x\}} (continuity of \texttt{D}), \texttt{f} preserving that directed sup, and \texttt{f\ x\ =\ ⊔\{w\ ≪\ f\ x\}} (continuity of \texttt{D\textquotesingle{}}) + \texttt{wayBelow\_sSup\_iff}. Continuity of \texttt{{[}D\ →\ D\textquotesingle{}{]}} then drops out: any upper bound \texttt{g} of \texttt{\{h\ ≪\ f\}} dominates every \texttt{ē{[}e,e\textquotesingle{}{]}\ ≪\ f}, hence pointwise \texttt{e\textquotesingle{}\ ≤\ g\ x}, hence \texttt{f\ x\ =\ ⊔\{…\}\ ≤\ g\ x}.
\end{enumerate}

\textbf{Engineering notes / lessons from 3.3(a):}

\begin{itemize}
\item
  \emph{Register the lattice as a real instance.} Because \texttt{ScottMap} is a plain \texttt{def}, a global \texttt{CompleteLattice} instance is safe and makes \texttt{≪}, \texttt{ScottOpen}, and \texttt{IsContinuousLattice} typecheck on the function space with no \texttt{@}/\texttt{letI} gymnastics --- the opposite experience to the \texttt{IdemFix} subtype.
\item
  \emph{\texttt{⨆\ \_\ :\ p,\ a} is the clean ``indicator''.} It is \texttt{a} when \texttt{p} holds (\texttt{iSup\_pos}) and \texttt{⊥} otherwise (\texttt{iSup\_neg}), needs no \texttt{Decidable}, and commutes with the proofs cleanly.
\item
  \emph{Topological ≪ is an asset, not a burden here.} Proving \texttt{ē{[}e,e\textquotesingle{}{]}\ ≪\ f} by exhibiting one Scott-open set is shorter than any directed-set argument; the lattice's pointwise \texttt{sSup} makes its inaccessibility immediate.
\item
  Footprint \texttt{{[}propext,\ Classical.choice,\ Quot.sound{]}}.
\end{itemize}

\hypertarget{theorem-3.3b-lattice-topology-pointwise-convergence-topology-theorem_3_3_topology-functionspaces.lean}{%
\subsubsection{\texorpdfstring{Theorem 3.3(b) (lattice topology = pointwise-convergence topology) --- \texttt{theorem\_3\_3\_topology} (\texttt{FunctionSpaces.lean})}{Theorem 3.3(b) (lattice topology = pointwise-convergence topology) --- theorem\_3\_3\_topology (FunctionSpaces.lean)}}\label{theorem-3.3b-lattice-topology-pointwise-convergence-topology-theorem_3_3_topology-functionspaces.lean}}

The function space carries two topologies: the Scott topology of the continuous lattice \texttt{{[}D\ →\ D\textquotesingle{}{]}} (from \texttt{ScottMap.instCompleteLattice}) and the product/pointwise-convergence topology \texttt{scottMapPointwiseTopology} generated by \texttt{\{f\ \textbar{}\ f\ x\ ∈\ U\}} (\texttt{U} Scott-open in \texttt{D\textquotesingle{}}). They are equal.

\begin{itemize}
\item
  \textbf{pointwise ⊆ Scott} (\texttt{le\_generateFrom\_iff\_subset\_isOpen}): each subbasic \texttt{\{f\ \textbar{}\ f\ x\ ∈\ U\}} is Scott-open in the lattice (\texttt{pointwiseSubbasic\_scottOpen}). Inaccessibility is immediate because the lattice's \texttt{sSup} is \emph{pointwise} (\texttt{ScottMap.sSup\_apply}), reducing to inaccessibility of \texttt{U} in \texttt{D\textquotesingle{}}. (This is the half Scott notes is automatic on p.~136: lubs are pointwise, so \textbf{no Milner hypothesis is needed} --- unlike 2.9--2.10.)
\item
  \textbf{Scott ⊆ pointwise} is the substance, via the \texttt{↟φ}-basis of a continuous lattice (\texttt{exists\_wayBelow\_subset}, using 3.3(a)): given a Scott-open \texttt{U\ ∋\ g}, pick \texttt{φ\ ≪\ g} with \texttt{↟φ\ ⊆\ U}. The key lemma \texttt{wayBelow\_le\_finset\_sup\_step} then shows \texttt{φ\ ≪\ g} forces \texttt{φ\ ≤\ ⊔ᵢ\ ē{[}eᵢ,eᵢ\textquotesingle{}{]}} for \textbf{finitely many} pairs with \texttt{eᵢ\textquotesingle{}\ ≪\ g\ eᵢ}: the finite joins of step functions below \texttt{g} form a \emph{directed} family (\texttt{Finset.sup} over \texttt{F₁\ ∪\ F₂}) with supremum \texttt{g} (pointwise reconstruction again), so \texttt{wayBelow\_sSup\_iff} lands \texttt{φ} below one of them. The finite intersection \texttt{W\ =\ ⋂ᵢ\ \{h\ \textbar{}\ eᵢ\textquotesingle{}\ ≪\ h\ eᵢ\}} is then a pointwise-open (\texttt{isOpen\_biInter\_finset}) neighbourhood of \texttt{g} with \texttt{W\ ⊆\ ↟φ\ ⊆\ U}: any \texttt{h\ ∈\ W} has every \texttt{ē{[}eᵢ,eᵢ\textquotesingle{}{]}\ ≪\ h} (\texttt{stepMap\_wayBelow}), hence \texttt{⊔ᵢ\ ē{[}eᵢ,eᵢ\textquotesingle{}{]}\ ≪\ h} (\texttt{wayBelow\_finset\_sup}) and \texttt{φ\ ≪\ h}.
\end{itemize}

\textbf{Engineering notes / lessons from 3.3(b):}

\begin{itemize}
\item
  \emph{Finiteness enters exactly once.} The only place finiteness of the step-function decomposition is needed is to keep \texttt{W} a \emph{finite} intersection (hence open). It is delivered by realizing \texttt{g} as a directed sup of \texttt{Finset.sup}s of step functions and invoking \texttt{wayBelow\_sSup\_iff} --- the standard ``compact basis'' move, here done concretely with \texttt{Finset\ (D\ ×\ D\textquotesingle{})}.
\item
  \emph{No ambient instance on \texttt{ScottMap}.} Since the two topologies are competing non-instances, the proof passes them explicitly (\texttt{@isOpen\_iff\_forall\_mem\_open}, \texttt{@isOpen\_biInter\_finset}); this is painless precisely because \texttt{ScottMap} carries no auto-synthesized \texttt{TopologicalSpace}.
\item
  \emph{Beware ascription into \texttt{sSup}.} \texttt{(sSup\ Sg\ :\ D\ →\ D\textquotesingle{})} makes Lean elaborate \texttt{sSup} at type \texttt{D\ →\ D\textquotesingle{}} (wrong \texttt{SupSet}); write \texttt{((sSup\ Sg\ :\ ScottMap\ D\ D\textquotesingle{})\ :\ D\ →\ D\textquotesingle{})} to keep the lattice join.
\item
  This closes \textbf{3.3 in full} (\texttt{theorem\_3\_3}), with no Milner hypothesis, contrary to the earlier expectation recorded for 2.9--2.10.
\item
  Footprint \texttt{{[}propext,\ Classical.choice,\ Quot.sound{]}}.
\end{itemize}

\hypertarget{corollary-3.4-joint-continuity-of-evaluation-corollary_3_4_jointly_continuous-functionspaces.lean}{%
\subsubsection{\texorpdfstring{Corollary 3.4 (joint continuity of evaluation) --- \texttt{corollary\_3\_4\_jointly\_continuous} (\texttt{FunctionSpaces.lean})}{Corollary 3.4 (joint continuity of evaluation) --- corollary\_3\_4\_jointly\_continuous (FunctionSpaces.lean)}}\label{corollary-3.4-joint-continuity-of-evaluation-corollary_3_4_jointly_continuous-functionspaces.lean}}

\texttt{eval\ :\ {[}D\ →\ D\textquotesingle{}{]}\ ×\ D\ →\ D\textquotesingle{}}, \texttt{(f,\ x)\ ↦\ f\ x}, is jointly Scott-continuous. The proof is a clean application of \textbf{Proposition 2.6} (joint ↔ separate Scott-continuity on a product lattice): reduce \texttt{PreservesDirectedSup\ eval} to the two separate slots. In \texttt{x} (fixed \texttt{f}) it is exactly \texttt{f}'s own Scott-continuity (\texttt{proposition\_2\_5} + \texttt{ScottMap.continuous}); in \texttt{f} (fixed \texttt{x}) it is the pointwise formula for suprema in \texttt{{[}D\ →\ D\textquotesingle{}{]}} (\texttt{ScottMap.sSup\_apply}: \texttt{(⊔F)\ x\ =\ ⊔\ \{g\ x\ \textbar{}\ g\ ∈\ F\}}). Then \texttt{continuous\_of\_preservesDirectedSup} upgrades to topological continuity. Via Theorem 3.3(b) (and 2.9(b)) the Scott topology of the product lattice is the product of the pointwise topology on \texttt{{[}D\ →\ D\textquotesingle{}{]}} and the Scott topology on \texttt{D}, so this is joint continuity for Scott's product topology. Footprint \texttt{{[}propext,\ Classical.choice,\ Quot.sound{]}}.

\hypertarget{proposition-3.5-functional-abstraction-proposition_3_5-functionspaces.lean}{%
\subsubsection{\texorpdfstring{Proposition 3.5 (functional abstraction) --- \texttt{proposition\_3\_5} (\texttt{FunctionSpaces.lean})}{Proposition 3.5 (functional abstraction) --- proposition\_3\_5 (FunctionSpaces.lean)}}\label{proposition-3.5-functional-abstraction-proposition_3_5-functionspaces.lean}}

\texttt{lambda\ :\ {[}{[}D\ ×\ D\textquotesingle{}{]}\ →\ D\textquotesingle{}\textquotesingle{}{]}\ →\ {[}D\ →\ {[}D\textquotesingle{}\ →\ D\textquotesingle{}\textquotesingle{}{]}{]}}, \texttt{lambda\ f\ x\ y\ =\ f\ (x,\ y)}, is Scott-continuous --- note this \emph{uses 3.3} twice, since the codomain \texttt{{[}D\ →\ {[}D\textquotesingle{}\ →\ D\textquotesingle{}\textquotesingle{}{]}{]}} must itself be a continuous lattice (hence a legitimate target). Two layers:

\begin{itemize}
\item
  \emph{\texttt{lambda\ f} is a Scott map} (\texttt{lambda\_outer\_preservesDirectedSup}): equality in \texttt{{[}D\textquotesingle{}\ →\ D\textquotesingle{}\textquotesingle{}{]}} is pointwise, so it reduces to \textbf{left}-currying \texttt{x\ ↦\ f\ (x,\ y)} being Scott-continuous (\texttt{curry\_left\_preservesDirectedSup}, mirror of the existing right-currying), itself a one-line consequence of \texttt{f}'s joint continuity and \texttt{sSup\ \{(x,\ y)\ \textbar{}\ x\ ∈\ S\}\ =\ (⊔S,\ y)}.
\item
  \emph{\texttt{lambda} is a Scott map} (\texttt{proposition\_3\_5\_preservesDirectedSup}): evaluating both sides pointwise at \texttt{(x,\ y)} and unfolding the (three nested!) pointwise \texttt{ScottMap.sSup\_apply}, both collapse to \texttt{⊔\ \{f\ (x,\ y)\ \textbar{}\ f\ ∈\ 𝓕\}}; \texttt{@{[}simp{]}\ scottLambda\_apply} (definitional) closes the resulting image-set equality with a bare \texttt{congr\ 1}.
\end{itemize}

The pleasant outcome: once \texttt{{[}D\ →\ D\textquotesingle{}{]}} is a genuine \texttt{CompleteLattice} instance with \emph{pointwise} \texttt{sSup} (\texttt{ScottMap.sSup\_apply} is \texttt{rfl}), all of §3's continuity facts (3.4, 3.5) are short pointwise computations. Footprint \texttt{{[}propext,\ Classical.choice,\ Quot.sound{]}}.

\hypertarget{proposition-3.8-maximal-extension-along-a-subspace-embedding-proposition_3_8-constructions.lean}{%
\subsubsection{\texorpdfstring{Proposition 3.8 (maximal extension along a subspace embedding) --- \texttt{proposition\_3\_8} (\texttt{Constructions.lean})}{Proposition 3.8 (maximal extension along a subspace embedding) --- proposition\_3\_8 (Constructions.lean)}}\label{proposition-3.8-maximal-extension-along-a-subspace-embedding-proposition_3_8-constructions.lean}}

For \texttt{E} a continuous lattice and \texttt{e\ :\ X\ →\ Y} a subspace embedding, Scott's explicit formula \texttt{scottExtend\ e\ f\ y\ =\ ⊔\ \{\ ⊓\ f\textquotesingle{}\textquotesingle{}(e⁻¹V)\ :\ V\ an\ open\ nbhd\ of\ y\ \}} is \emph{the maximal extension} of a continuous \texttt{f\ :\ X\ →\ E} to \texttt{{[}Y\ →\ E{]}}. The full statement bundles three clauses:

\begin{itemize}
\item
  \textbf{Continuous} and \textbf{extends \texttt{f}}: reused verbatim from the 2.11 injectivity machinery (\texttt{scottExtend\_continuous}, \texttt{scottExtend\_eq\_of\_continuous}) --- the \emph{same} operator \texttt{scottExtend} serves both 2.11 and 3.8, so 3.8 is essentially 2.11 plus a maximality clause.
\item
  \textbf{Maximal} (\texttt{scottExtend\_maximal}): for any continuous solution \texttt{f\textquotesingle{}} of \texttt{f\textquotesingle{}\ ∘\ e\ =\ f}, expand \texttt{f\textquotesingle{}} itself via \texttt{continuous\_eq\_sSup\_openInfs} (the order-theoretic identity \texttt{f\textquotesingle{}\ y\ =\ ⊔\ \{\ ⊓\ f\textquotesingle{}\textquotesingle{}(U)\ :\ U\ open\ nbhd\ of\ y\ \}}, proved by interpolating from below with \texttt{f\textquotesingle{}\ y\ =\ ⊔\ \{a\ ≪\ f\textquotesingle{}\ y\}} and openness of each \texttt{f\textquotesingle{}⁻¹(↟a)}). Restricting each meet from the open \texttt{U} to the embedded subspace \texttt{e(X)\ ∩\ U} only \emph{enlarges} the meet and lands it on a defining term of \texttt{scottExtend}, giving \texttt{f\textquotesingle{}\ y\ ≤\ scottExtend\ e\ f\ y} --- exactly Scott's two-line chain on p.116.
\end{itemize}

\textbf{Engineering notes / lessons from 3.8:} the partial \texttt{FunctionSpaces.scottSubspaceExtend} (renamed \texttt{scottSubspaceExtend\_maximal}) had ranged \texttt{U} over the \emph{Scott} topology of \texttt{Y} (forcing a spurious \texttt{CompleteLattice\ Y}), which is unfaithful to Scott (where \texttt{Y} is an arbitrary \texttt{T₀} space). The faithful route was to retarget the whole proposition onto the already-continuous \texttt{scottExtend} from 2.11, which ranges \texttt{U} over \texttt{Y}'s \emph{given} topology --- turning ``Stuck (one-sided bound)'' into a clean \textbf{Pass} that simply repackages existing lemmas. Footprint \texttt{{[}propext,\ Classical.choice,\ Quot.sound{]}}.

\hypertarget{proposition-3.10-characterization-of-projection-inclusions-proposition_3_10_converse-retr_eq_ssup-functionspaces.lean}{%
\subsubsection{\texorpdfstring{Proposition 3.10 (characterization of projection inclusions) --- \texttt{proposition\_3\_10\_converse}, \texttt{retr\_eq\_sSup} (\texttt{FunctionSpaces.lean})}{Proposition 3.10 (characterization of projection inclusions) --- proposition\_3\_10\_converse, retr\_eq\_sSup (FunctionSpaces.lean)}}\label{proposition-3.10-characterization-of-projection-inclusions-proposition_3_10_converse-retr_eq_ssup-functionspaces.lean}}

A map \texttt{i\ :\ D\ →\ D\textquotesingle{}} between continuous lattices is the inclusion of a projection \textbf{iff} it (i) preserves arbitrary suprema, (ii) is injective, and (iii) preserves \texttt{≪}. The \textbf{forward} direction was already in place (\texttt{incl\_sSup}, \texttt{incl\_injective}, \texttt{incl\_wayBelow}); this completes the \textbf{converse} and the \textbf{uniqueness} of Scott's formula (iv) \texttt{j(x\textquotesingle{})\ =\ ⊔\ \{\ x\ \textbar{}\ i(x)\ ⊑\ x\textquotesingle{}\ \}}.

\begin{itemize}
\item
  \emph{Order-reflection from (i)+(ii)} (\texttt{le\_of\_incl\_le}): condition (i) on the two-element set gives \texttt{i(x\ ⊔\ y)\ =\ i\ x\ ⊔\ i\ y} (\texttt{incl\_sup\_of\_preservesSSup}); then \texttt{i\ x\ ⊑\ i\ y\ ⟹\ i(x⊔y)=i\ y\ ⟹\ x⊔y=y} (injectivity) \texttt{⟹\ x\ ⊑\ y}. This is exactly Scott's ``\texttt{x\ ⊑\ y\ ⟺\ x\ ⊔\ y\ =\ y}'' remark, and it makes \texttt{i} an order-embedding.
\item
  \emph{\texttt{j\ ∘\ i\ =\ id}} (\texttt{converseRetr\_incl}): order-reflection collapses \texttt{\{x\ \textbar{}\ i\ x\ ⊑\ i\ y\}} to \texttt{Iic\ y}, whose join is \texttt{y}.
\item
  \emph{\texttt{i\ ∘\ j\ ⊑\ id}} (\texttt{incl\_converseRetr\_le}): immediate from (i), since \texttt{i(⊔\{x\ \textbar{}\ i\ x\ ⊑\ x\textquotesingle{}\})\ =\ \ \ ⊔\{i\ x\ \textbar{}\ i\ x\ ⊑\ x\textquotesingle{}\}\ ⊑\ x\textquotesingle{}}.
\item
  \emph{\texttt{j} continuous} (\texttt{converseRetr\_preservesDirectedSup}): the one place (iii) is needed. For a directed \texttt{S\textquotesingle{}} and \texttt{i\ x\ ⊑\ ⊔S\textquotesingle{}}, interpolate \texttt{x\ =\ ⊔\{z\ ≪\ x\}} (continuity of \texttt{D}); each \texttt{z\ ≪\ x} gives \texttt{i\ z\ ≪\ i\ x\ ⊑\ ⊔S\textquotesingle{}}, so \texttt{i\ z\ ⊑\ x\textquotesingle{}} for some \texttt{x\textquotesingle{}\ ∈\ S\textquotesingle{}} (directed \texttt{wayBelow\_sSup\_iff}), whence \texttt{z\ ⊑\ j\ x\textquotesingle{}\ ⊑\ ⊔\ j\textquotesingle{}\textquotesingle{}S\textquotesingle{}}.
\item
  \emph{Uniqueness} (\texttt{retr\_eq\_sSup}): any projection's \texttt{j} satisfies \texttt{j\ x\textquotesingle{}\ =\ ⊔\{x\ \textbar{}\ i\ x\ ⊑\ x\textquotesingle{}\}} --- \texttt{≤} since \texttt{i(j\ x\textquotesingle{})\ ⊑\ x\textquotesingle{}} makes \texttt{j\ x\textquotesingle{}} a member; \texttt{≥} since each member \texttt{x} has \texttt{x\ =\ j(i\ x)\ ⊑\ j\ x\textquotesingle{}}.
\end{itemize}

\textbf{Engineering notes / lessons from 3.10:} condition (i) is stated for \emph{arbitrary} \texttt{S}, so it trivially supplies \texttt{PreservesDirectedSup\ i} (whence \texttt{i} is a legitimate \texttt{ScottMap}) with a one-line \texttt{fun\ \_\ \_\ \_\ =\textgreater{}\ hi\ \_} --- no need to separately assume continuity of \texttt{i}. Set-membership in \texttt{\{x\ \textbar{}\ i\ x\ ⊑\ x\textquotesingle{}\}} is \emph{definitionally} the predicate, so \texttt{le\_sSup}/\texttt{sSup\_le} chains go through with bare \texttt{.le} coercions and \texttt{show} re-statements rather than \texttt{Set.mem\_setOf} rewrites. Footprint \texttt{{[}propext,\ Classical.choice,\ Quot.sound{]}}.

\hypertarget{lemma-3.9-extensions-commute-with-a-range-projection-lemma_3_9-theorem212.lean}{%
\subsubsection{\texorpdfstring{Lemma 3.9 (extensions commute with a range projection) --- \texttt{lemma\_3\_9} (\texttt{Theorem212.lean})}{Lemma 3.9 (extensions commute with a range projection) --- lemma\_3\_9 (Theorem212.lean)}}\label{lemma-3.9-extensions-commute-with-a-range-projection-lemma_3_9-theorem212.lean}}

With \texttt{e\ :\ X\ →\ Y} a subspace embedding and \texttt{i,\ j\ :\ D\ ⇄\ D\textquotesingle{}} a projection on the \emph{range}, if continuous \texttt{f\ :\ X\ →\ D} and \texttt{g\ :\ X\ →\ D\textquotesingle{}} satisfy \texttt{f\ =\ j\ ∘\ g}, then their maximal extensions (3.8) satisfy \(\bar{f} = j \circ \bar{g}\). This is the key compatibility used to build inverse limits (§4: \(\bar{f}_n = j_n \circ \bar{f}_{n+1}\)). The proof is a clean two-inequality sandwich, exactly Scott's:

\begin{itemize}
\item
  \(j \circ \bar{g} \sqsubseteq \bar{f}\): \(j \circ \bar{g}\) is continuous and \((j \circ \bar{g}) \circ e = j \circ g = f\), so the \emph{equality} maximality of \(\bar{f}\) (\texttt{scottExtend\_maximal}) applies.
\item
  \(i \circ \bar{f} \sqsubseteq \bar{g}\): \((i \circ \bar{f}) \circ e = i \circ f = i \circ j \circ g \sqsubseteq g\) (using \(i \circ j \sqsubseteq \mathrm{id}\)), so the \emph{sub-solution} maximality \texttt{scottExtend\_maximal\_le} (the remark after 3.8, added here as the \texttt{≤}-analogue of \texttt{scottExtend\_maximal} --- identical proof, final \texttt{=} weakened to \texttt{≤}) applies.
\item
  combine: \(\bar{f} = j \circ i \circ \bar{f} \sqsubseteq j \circ \bar{g} \sqsubseteq \bar{f}\) (apply monotone \(j\) to the second bound, and \(j \circ i = \mathrm{id}\)).
\end{itemize}

\textbf{Engineering notes / lessons from 3.9:} the lemma lives in \texttt{Theorem212.lean} because it is the only module importing \emph{both} \texttt{scottExtend} (Constructions) and \texttt{IsContinuousLatticeProjection} (FunctionSpaces). The one real friction was composition continuity: the Scott topology is a bare \texttt{def}, not an \texttt{instance}, so \texttt{Continuous.comp} cannot synthesize \texttt{TopologicalSpace\ D}. Registering it with \texttt{letI} works, but \textbf{only if scoped inside the \texttt{have} for the composite} --- registering it at the top of the proof makes the lattice \texttt{≤} ambiguous (it gets re-resolved through the topology's \texttt{specializationPreorder}), which silently breaks every later \texttt{le\_antisymm}/\texttt{calc}. The older inf-level partials \texttt{lemma\_3\_9\_incl\_inf}/\texttt{lemma\_3\_9\_retr\_inf} are now superseded auxiliaries. Footprint \texttt{{[}propext,\ Classical.choice,\ Quot.sound{]}}.

\hypertarget{proposition-3.12-the-lattice-of-projections-j_d-proposition_3_12-functionspaces.lean}{%
\subsubsection{\texorpdfstring{Proposition 3.12 (the lattice of projections \texttt{J\_D}) --- \texttt{proposition\_3\_12} (\texttt{FunctionSpaces.lean})}{Proposition 3.12 (the lattice of projections J\_D) --- proposition\_3\_12 (FunctionSpaces.lean)}}\label{proposition-3.12-the-lattice-of-projections-j_d-proposition_3_12-functionspaces.lean}}

\texttt{J\_D\ =\ \{\ j\ ∈\ {[}D\ →\ D{]}\ :\ j\ =\ j\ ∘\ j\ ⊑\ id\ \}} (\texttt{IsProjection}) is a complete lattice realized as a \texttt{⊔}-closed subspace of \texttt{{[}D\ →\ D{]}}. The whole proof reduces, via the pointwise characterization \texttt{isProjection\_iff} (idempotent \textbf{and} deflationary), to closure of \texttt{J\_D} under arbitrary \texttt{sSup} (\texttt{isProjection\_sSup}); a \texttt{⊔}-closed subset of a complete lattice is a complete lattice (\texttt{completeLatticeOfSup} on the subtype \texttt{Projections\ D}).

\begin{itemize}
\item
  \emph{binary} (\texttt{isProjection\_sup}): since \texttt{j\ x\ ⊔\ k\ x\ ⊑\ x}, monotonicity + idempotency pin \texttt{j\ (j\ x\ ⊔\ k\ x)\ =\ j\ x} (and symmetrically for \texttt{k}), so \texttt{(j\ ⊔\ k)\ ∘\ (j\ ⊔\ k)\ =\ j\ ⊔\ k}. This is the one spot needing \texttt{sup\_apply} --- the new lemma that the \texttt{completeLatticeOfSup}-derived binary join of Scott maps is computed \emph{pointwise} (\texttt{(f\ ⊔\ g)\ x\ =\ f\ x\ ⊔\ g\ x}, since \texttt{⊔\ =\ sSup\ \{·,·\}} and \texttt{sSup} is pointwise).
\item
  \emph{directed} (\texttt{isProjection\_directedSup}): continuity of each \texttt{k\ ∈\ S} distributes \texttt{k\ ((⊔S)\ x)\ =\ ⊔ⱼ\ k\ (j\ x)} over the directed family \texttt{\{\ j\ x\ \}}, and directedness + idempotency collapse the double sup \texttt{\{\ k\ (j\ x)\ \}} back to \texttt{(⊔S)\ x}. (Continuity of \texttt{D} itself is \emph{not} used --- this works for any complete lattice \texttt{D}.)
\item
  \emph{arbitrary} (\texttt{isProjection\_sSup}): reuse \texttt{finsetSupOf} (every \texttt{sSup} is the directed sup of finite sub-joins), with \texttt{isProjection\_finsetSup} via \texttt{Finset.sup\_induction} on \texttt{⊥}/binary.
\end{itemize}

\textbf{Engineering notes / lessons from 3.12:} the identity map is named \texttt{ScottMap.idMap}, \textbf{not} \texttt{id}, to avoid shadowing \texttt{\_root\_.id} (which \texttt{finsetSupOf}'s \texttt{Finset.sup\ id} relies on). The \texttt{Projections\ D} subtype must be an \texttt{abbrev} (not \texttt{def}) so the ambient \texttt{Subtype.partialOrder}/\texttt{SupSet} instances are found by typeclass resolution --- the same reducibility lesson as \texttt{IdemFix} in 2.12. Footprint \texttt{{[}propext,\ Classical.choice,\ Quot.sound{]}}.

\hypertarget{proposition-3.13-d-is-a-projection-of-d-d-proposition_3_13-functionspaces.lean}{%
\subsubsection{\texorpdfstring{Proposition 3.13 (\texttt{D} is a projection of \texttt{{[}D\ →\ D{]}}) --- \texttt{proposition\_3\_13} (\texttt{FunctionSpaces.lean})}{Proposition 3.13 (D is a projection of {[}D → D{]}) --- proposition\_3\_13 (FunctionSpaces.lean)}}\label{proposition-3.13-d-is-a-projection-of-d-d-proposition_3_13-functionspaces.lean}}

Scott's \texttt{con\ :\ D\ →\ {[}D\ →\ D{]}}, \texttt{con\ x\ =\ (const\ x)}, and \texttt{min\ :\ {[}D\ →\ D{]}\ →\ D}, \texttt{min\ f\ =\ f(⊥)}, form a projection: \texttt{min\ (con\ x)\ =\ (const\ x)(⊥)\ =\ x} (so \texttt{min\ ∘\ con\ =\ id}, \texttt{rfl}), and \texttt{con\ (min\ f)\ =\ const\ (f\ ⊥)\ ⊑\ f} pointwise because \texttt{f(⊥)\ ⊑\ f(y)} by monotonicity (so \texttt{con\ ∘\ min\ ⊑\ id}). Both maps are Scott-continuous: \texttt{con} because suprema in \texttt{{[}D\ →\ D{]}} are pointwise (\texttt{con\ (⊔S)\ =\ const\ (⊔S)} and \texttt{⊔ⱼ\ const(j)\ =\ const(⊔S)}), and \texttt{min} because it is evaluation at \texttt{⊥}, which reads off the pointwise supremum (\texttt{ScottMap.sSup\_apply}). The result packages as a term of the existing \texttt{IsContinuousLatticeProjection\ D\ {[}D\ →\ D{]}}, so it immediately feeds Proposition 3.10's machinery. (Continuity of \texttt{D} is again unused; included only to match Scott's hypothesis.) Footprint \texttt{{[}propext,\ Classical.choice,\ Quot.sound{]}}.

\hypertarget{proposition-3.14-the-fixed-point-operator-proposition_3_14-functionspaces.lean}{%
\subsubsection{\texorpdfstring{Proposition 3.14 (the fixed-point operator) --- \texttt{proposition\_3\_14} (\texttt{FunctionSpaces.lean})}{Proposition 3.14 (the fixed-point operator) --- proposition\_3\_14 (FunctionSpaces.lean)}}\label{proposition-3.14-the-fixed-point-operator-proposition_3_14-functionspaces.lean}}

\texttt{fix\ :\ {[}D\ →\ D{]}\ →\ D} is Scott's least-fixed-point combinator: \texttt{f\ (fix\ f)\ =\ fix\ f} and \texttt{f\ x\ ⊑\ x\ ⟹\ fix\ f\ ⊑\ x}, and it is the \emph{unique} operator with these two properties. The \textbf{order content} is mathlib's \texttt{OrderHom.lfp} (\texttt{fix\ f\ :=\ (⟨f,\ f.monotone⟩\ :\ D\ →o\ D).lfp}), giving \texttt{fix\_eq} (\texttt{map\_lfp}), \texttt{fix\_le} (\texttt{lfp\_le}), and \texttt{fix\_unique} (least element of the fixed-point set is unique) for free.

The \textbf{continuity} of \texttt{fix} (Scott's actual claim) is the work. Scott argues via Kleene's \texttt{fix\ f\ =\ ⊔ₙ\ fⁿ(⊥)} (``pointwise lub of continuous functions''); we give a \textbf{direct lattice proof that avoids iteration entirely} (\texttt{fix\_preservesDirectedSup}). For directed \texttt{S\ ⊆\ {[}D\ →\ D{]}}, set \texttt{g\ =\ ⊔S} and \texttt{a\ =\ ⊔\{fix\ f\ :\ f\ ∈\ S\}}:

\begin{itemize}
\item
  \texttt{a\ ⊑\ fix\ g} is just \texttt{fix}-monotonicity (\texttt{fix\_mono}, itself a two-line \texttt{fix\_le}).
\item
  \texttt{fix\ g\ ⊑\ a}: by \texttt{fix\_le} it suffices that \texttt{a} is a pre-fixed point, \texttt{g\ a\ ⊑\ a}. Pointwise sups give \texttt{g\ a\ =\ ⊔\_\{f∈S\}\ f\ a}, and continuity of each \texttt{f} on the \textbf{directed} family \texttt{\{fix\ f\textquotesingle{}\ :\ f\textquotesingle{}\ ∈\ S\}} gives \texttt{f\ a\ =\ ⊔\_\{f\textquotesingle{}∈S\}\ f\ (fix\ f\textquotesingle{})}. For any \texttt{f,\ f\textquotesingle{}\ ∈\ S} choose (directedness) \texttt{h\ ∈\ S} above both: \texttt{f\ (fix\ f\textquotesingle{})\ ⊑\ h\ (fix\ f\textquotesingle{})\ ⊑\ h\ (fix\ h)\ =\ fix\ h\ ⊑\ a}. Hence \texttt{g\ a\ ⊑\ a}.
\end{itemize}

\textbf{Engineering notes / lessons from 3.14:} the direct argument is far shorter than building Kleene's theorem and only needs three ingredients already in hand --- \texttt{OrderHom.lfp} monotonicity facts, \texttt{ScottMap.sSup\_apply} (pointwise sups in \texttt{{[}D\ →\ D{]}}), and \texttt{preservesDirectedSup\_coe}. Two small Lean traps: (1) \texttt{sSup\_le} leaves the bound element as an un-β-reduced \texttt{(fun\ f\ =\textgreater{}\ ↑f\ (sSup\ T))\ f}, so a \texttt{show\ (f\ :\ D\ →\ D)\ (sSup\ T)\ ≤\ sSup\ T} is needed before the \texttt{rw}; (2) in the uniqueness clause an \emph{unannotated} binder \texttt{∀\ f,\ (f\ :\ D\ →\ D)\ …} makes the ascription \textbf{fix the binder type to \texttt{D\ →\ D}} rather than coerce --- the binders must be written \texttt{∀\ f\ :\ ScottMap\ D\ D}. Continuity of \texttt{D} is unused (works for any complete lattice). Footprint \texttt{{[}propext,\ Classical.choice,\ Quot.sound{]}}.

\hypertarget{inverse-limits-scott-4}{%
\subsection{Inverse limits (Scott §4)}\label{inverse-limits-scott-4}}

Section §4 constructs \(D_\infty\) and proves Theorem 4.4. Proposition 4.1 follows Scott's Proposition-3.8/Lemma-3.9 injectivity route, while the explicit topology bridges and the function-space tower scaffolding for 4.4 are the main engineering contributions.

\hypertarget{proposition-4.1-inverse-limit-of-projections-is-a-continuous-lattice-proposition_4_1-inverselimits.lean}{%
\subsubsection{\texorpdfstring{Proposition 4.1 (inverse limit of projections is a continuous lattice) --- \texttt{proposition\_4\_1} (\texttt{InverseLimits.lean})}{Proposition 4.1 (inverse limit of projections is a continuous lattice) --- proposition\_4\_1 (InverseLimits.lean)}}\label{proposition-4.1-inverse-limit-of-projections-is-a-continuous-lattice-proposition_4_1-inverselimits.lean}}

\texttt{D∞\ =\ \{\ x\ :\ ∀n,\ Dₙ\ //\ ∀n,\ jₙ(xₙ₊₁)\ =\ xₙ\ \}} for an ω-system of continuous lattices with projection bonding maps \texttt{jₙ\ :\ D\_\{n+1\}\ →\ Dₙ}. Its named topology \texttt{inverseLimitTopology} is exactly the subspace topology induced from the Pi product of the stage Scott topologies. Each projection is also an adjunction, so its upper adjoint preserves arbitrary infima; this establishes the componentwise complete-lattice structure on compatible sequences.

The proof of Proposition 4.1 itself follows Scott's topological argument. Given a continuous \texttt{f\ :\ X\ →\ D∞} and a subspace embedding \texttt{e\ :\ X\ →\ Y}, define the \texttt{n}-th coordinate extension \texttt{\textbackslash{}bar\ f\_n\ :\ Y\ →\ D\_n} by Proposition 3.8's maximal extension. Lemma 3.9 gives \texttt{j\_n\ ∘\ \textbackslash{}bar\ f\_\{n+1\}\ =\ \textbackslash{}bar\ f\_n}, so the extensions assemble into a continuous map \texttt{Y\ →\ D∞}; continuity is coordinatewise for the Pi-induced subspace topology. Thus \texttt{D∞} is an injective \texttt{T₀}-space. \texttt{theorem\_2\_12\_backward\_exact}, together with the proof that specialization in the subspace topology is componentwise order, yields both \texttt{IsContinuousLattice\ D∞} and \texttt{scottTopologicalSpace\ =\ inverseLimitTopology}.

\textbf{Engineering notes / lessons from 4.1:} the complete lattice remains the componentwise one, while the injectivity proof deliberately installs the named product/subspace topology only locally. Explicitly preserving the specialization order is what lets Theorem 2.12 return the topology identification needed by Theorem 4.4, rather than merely the order-theoretic lattice result. Footprint \texttt{{[}propext,\ Classical.choice,\ Quot.sound{]}}.

\hypertarget{proposition-4.2-the-maps-j_n-are-projections-proposition_4_2-inverselimits.lean}{%
\subsubsection{\texorpdfstring{Proposition 4.2 (the maps \texttt{j\_\{∞n\}} are projections) --- \texttt{proposition\_4\_2} (\texttt{InverseLimits.lean})}{Proposition 4.2 (the maps j\_\{∞n\} are projections) --- proposition\_4\_2 (InverseLimits.lean)}}\label{proposition-4.2-the-maps-j_n-are-projections-proposition_4_2-inverselimits.lean}}

\texttt{j\_\{∞n\}\ :\ D∞\ →\ Dₙ} is evaluation \texttt{x\ ↦\ xₙ}. Scott constructs the inverse embedding \texttt{i\_\{n∞\}\ :\ Dₙ\ →\ D∞} componentwise: \texttt{i\_\{n∞\}(x)\_m\ =\ x} at \texttt{m\ =\ n}, climbs by \texttt{iₖ\ =\ (P\ k).incl} for \texttt{m\ \textgreater{}\ n}, and descends by \texttt{jₖ\ =\ (P\ k).retr} for \texttt{m\ \textless{}\ n}. We realize this with two \texttt{Nat.leRecOn} towers:

\begin{itemize}
\item
  \texttt{embLE\ (h\ :\ n\ ≤\ m)\ :\ Dₙ\ →\ D\_m} (climb, \texttt{=\ i\_\{m-1\}∘…∘iₙ}) and \texttt{projLE\ (h\ :\ m\ ≤\ n)\ :\ D\_n\ →\ Dₘ} (descend, \texttt{=\ j\_m∘…∘j\_\{n-1\}}), with the computation lemmas \texttt{embLE\_self/\_succ/\_succ\_left}, \texttt{projLE\_self/\_succ} reading off \texttt{Nat.leRecOn\_self/\_succ/\_succ\_left};
\item
  \texttt{iComp\ n\ x\ m\ =\ if\ n\ ≤\ m\ then\ embLE\ …\ else\ projLE\ …} is the component map; \texttt{iComp\_compatible} (case split on \texttt{n\ ≤\ m}, \texttt{n\ =\ m+1}, \texttt{m+1\ ≤\ n}, the middle hinge being \texttt{projLE\_retr}) shows the sequence is a genuine point of \texttt{D∞}, and \texttt{iComp\_self} gives \texttt{j\_\{∞n\}∘i\_\{n∞\}\ =\ id}.
\end{itemize}

Both towers are Scott-continuous (\texttt{embLE/projLE\_preservesDirectedSup}, by \texttt{Nat.le\_induction} + \texttt{ScottMap.preservesDirectedSup\_comp}), hence each component \texttt{iComp\ n\ ·\ m} is (\texttt{iComp\_preservesDirectedSup}); since directed sups in \texttt{D∞} are pointwise (\texttt{coe\_sSup\_of\_directed}), the bundled \texttt{embInf\ n\ :\ ScottMap\ Dₙ\ D∞} and \texttt{projInf\ n\ :\ ScottMap\ D∞\ Dₙ} are continuous. \texttt{proposition\_4\_2} packages \texttt{⟨embInf,\ projInf⟩} as an \texttt{IsContinuousLatticeProjection}: \texttt{retr\_incl\ =\ iComp\_self}, and \texttt{incl\_retr\_le} reduces coordinatewise (\texttt{Subtype.coe\_le\_coe}) to \texttt{iComp\_incl\_le} --- for \texttt{m\ ≥\ n} climbing \texttt{yₙ} stays below \texttt{yₘ} (\texttt{embLE\_le}, using \texttt{incl∘retr\ ⊑\ id} and compatibility), for \texttt{m\ \textless{}\ n} it equals \texttt{yₘ} (\texttt{projLE\_compatible}).

Also formalized: the recursion equation \texttt{i\_\{n∞\}\ =\ i\_\{(n+1)∞\}∘iₙ} (\texttt{embInf\_succ}) and the monotone-lub identity \texttt{x\ =\ ⨆ₙ\ i\_\{n∞\}(xₙ)} (\texttt{inverseLimit\_eq\_iSup}); the family is monotone via \texttt{embInf\_succ} + \texttt{incl\_retr\_le} (\texttt{embInf\_le\_succ}), so its range is directed and the lub is computed pointwise, where \texttt{iComp\_self} pins the \texttt{m}-th coordinate to \texttt{xₘ}.

\textbf{Engineering notes / lessons from 4.2:} \texttt{Nat.leRecOn} (and \texttt{Nat.le\_induction}) is the clean way to build/induct on the two dependently-typed towers without \texttt{Nat}-subtraction casts; the descend tower uses the \emph{function} motive \texttt{C\ k\ :=\ D\ k\ →\ Dₘ}. \texttt{Nat.leRecOn} is \texttt{@{[}elab\_as\_elim{]}}, so its computation lemmas must be applied after unfolding the wrapper (\texttt{simp\ only\ {[}embLE{]}} / \texttt{simp\ only\ {[}projLE{]}}) --- a bare term-mode \texttt{:=\ Nat.leRecOn\_self\ x} fails with ``failed to elaborate eliminator''. Lean 4's definitional proof irrelevance means the towers do not depend on \emph{which} \texttt{≤} proof is supplied, so the \texttt{rw} chains match across \texttt{le\_refl}/\texttt{Nat.le\_succ\_of\_le}/\texttt{Nat.le\_of\_succ\_le} freely. The eliminator is invoked as \texttt{induction\ n,\ h\ using\ Nat.le\_induction}. Footprint \texttt{{[}propext,\ Classical.choice,\ Quot.sound{]}}.

\hypertarget{corollary-4.3-d-is-also-the-direct-limit-corollary_4_3-inverselimits.lean}{%
\subsubsection{\texorpdfstring{Corollary 4.3 (\texttt{D∞} is also the \emph{direct} limit) --- \texttt{corollary\_4\_3} (\texttt{InverseLimits.lean})}{Corollary 4.3 (D∞ is also the direct limit) --- corollary\_4\_3 (InverseLimits.lean)}}\label{corollary-4.3-d-is-also-the-direct-limit-corollary_4_3-inverselimits.lean}}

Where Prop 4.2 makes \texttt{D∞} the \emph{inverse} (projective) limit, 4.3 is the dual universal property: it is the \emph{direct} (injective) limit along the embeddings \texttt{iₙ}. Given any complete lattice \texttt{D\textquotesingle{}} and a \textbf{compatible cocone} of Scott maps \texttt{fₙ\ :\ Dₙ\ →\ D\textquotesingle{}} with \texttt{fₙ\ =\ f\_\{n+1\}∘iₙ} (\texttt{hf}), the mediating map is \texttt{coconeInf\ f\ x\ =\ f∞(x)\ =\ ⨆ₙ\ fₙ(xₙ)}. We prove there is a \textbf{unique} continuous \texttt{f∞} with \texttt{fₙ\ =\ f∞∘i\_\{n∞\}} (an \texttt{∃!} over \texttt{ScottMap\ (InverseLimit\ D\ P)\ D\textquotesingle{}}).

\begin{itemize}
\item
  \emph{Factorization} \texttt{coconeInf\_comp\_embInf}: \texttt{f∞(i\_\{n∞\}(x))\ =\ ⨆ₘ\ f\_m(iComp\ n\ x\ m)\ =\ fₙ(x)} by \texttt{le\_antisymm}. The \texttt{≥} direction is \texttt{iComp\_self} at \texttt{m\ =\ n}. For \texttt{≤}, the family \texttt{m\ ↦\ f\_m(iComp\ n\ x\ m)} is dominated by \texttt{fₙ(x)}: above \texttt{n} it is \emph{constant} \texttt{=\ fₙ(x)} (\texttt{coconeInf\_climb}, \texttt{Nat.le\_induction} collapsing \texttt{f\_\{m+1\}∘iₘ\ =\ f\_m}), and below \texttt{n} it only decreases (\texttt{coconeInf\_descend}: peel \texttt{projLE} via \texttt{projLE\_succ}, then \texttt{fₘ∘jₘ\ =\ f\_\{m+1\}∘iₘ∘jₘ\ ⊑\ f\_\{m+1\}} by \texttt{incl\_retr\_le} + monotonicity).
\item
  \emph{Continuity} \texttt{coconeInf\_preservesDirectedSup}: needs \textbf{no} \texttt{hf}. For directed \texttt{S}, push the sup through each coordinate (\texttt{eval\_preservesDirectedSup}) and through each continuous \texttt{fₙ} (\texttt{preservesDirectedSup\_coe}, image of \texttt{S} directed under evaluation), then commute the resulting double sup over \texttt{ℕ\ ×\ S} with \texttt{iSup\_comm} (rewriting images as subtype sups with \texttt{sSup\_image\textquotesingle{}}).
\item
  \emph{Uniqueness}: any continuous \texttt{g} with \texttt{fₙ\ =\ g∘i\_\{n∞\}} satisfies \texttt{g(x)\ =\ g(⨆ₙ\ i\_\{n∞\}(xₙ))\ =\ \ \ ⨆ₙ\ g(i\_\{n∞\}(xₙ))\ =\ ⨆ₙ\ fₙ(xₙ)\ =\ f∞(x)}, using \texttt{inverseLimit\_eq\_iSup} (4.2), continuity of \texttt{g} on the directed family (\texttt{embInf\_family\_directed}), and \texttt{ScottMap.ext}.
\end{itemize}

Footprint \texttt{{[}propext,\ Classical.choice,\ Quot.sound{]}}.

\hypertarget{lemma-4.5-and-the-functional-equation-lemma_4_5-idinf_eq_isup-inverselimits.lean}{%
\subsubsection{\texorpdfstring{Lemma 4.5 and the functional equation --- \texttt{lemma\_4\_5}, \texttt{idInf\_eq\_iSup} (\texttt{InverseLimits.lean})}{Lemma 4.5 and the functional equation --- lemma\_4\_5, idInf\_eq\_iSup (InverseLimits.lean)}}\label{lemma-4.5-and-the-functional-equation-lemma_4_5-idinf_eq_isup-inverselimits.lean}}

\texttt{idInf\_eq\_iSup} records Scott's ``remark following 4.2'': as Scott maps \texttt{D\_∞\ →\ D\_∞}, \texttt{id\ =\ ⨆ₙ\ (i\_\{n∞\}\ ∘\ j\_\{∞n\})}. Pointwise, \texttt{(⨆ₙ\ i\_\{n∞\}∘j\_\{∞n\})(x)\ =\ ⨆ₙ\ i\_\{n∞\}(xₙ)\ =\ x} (\texttt{ScottMap.sSup\_apply} to push the sup of maps through evaluation, then \texttt{inverseLimit\_eq\_iSup}).

\texttt{lemma\_4\_5} is Scott's tool for \emph{recognizing projections from limits}: if \texttt{u\ :\ ∀\ n,\ D\_\{n+1\}} obeys the shifted recursion \texttt{j\_\{n+1\}(u\_\{n+2\})\ =\ u\_\{n+1\}}, then \texttt{u\_∞\ =\ ⨆ₙ\ i\_\{(n+1)∞\}(uₙ)} has \texttt{j\_\{∞(n+1)\}(u\_∞)\ =\ uₙ}. The proof follows Scott's induction. First the defining family is monotone, so continuity moves \texttt{j\_\{∞(n+1)\}} through its directed supremum. For every \texttt{m\ ≥\ n}, induction on \texttt{m\ -\ n} proves \texttt{j\_\{∞(n+1)\}(i\_\{(m+1)∞\}(u\_m))\ =\ u\_n}: peel one projection using \texttt{projLE\_succ}, then apply the recursion hypothesis. Terms with \texttt{m\ \textless{}\ n} are bounded by the diagonal term, and the projected supremum is therefore exactly \texttt{u\_n}.

\hypertarget{theorem-4.4-scaffolding-functionspacetower.lean}{%
\subsubsection{\texorpdfstring{Theorem 4.4 scaffolding --- \texttt{FunctionSpaceTower.lean}}{Theorem 4.4 scaffolding --- FunctionSpaceTower.lean}}\label{theorem-4.4-scaffolding-functionspacetower.lean}}

The capstone needs the \emph{concrete} recursion \texttt{D\_\{n+1\}\ =\ {[}Dₙ\ →\ Dₙ{]}}, \texttt{j\_\{n+1\}\ =\ {[}jₙ\ →\ jₙ{]}} --- the first place in §4 where the levels are genuine function spaces. Because the type at level \texttt{n+1} depends on the \emph{lattice structure} at level \texttt{n}, we bundle carrier + instance in \texttt{CLat} and recurse (\texttt{towerCLat}); \texttt{towerType}/\texttt{towerCompleteLattice} project out the type and its \texttt{CompleteLattice}, and crucially \texttt{towerType\_succ\ :\ D\_\{n+1\}\ =\ {[}Dₙ→Dₙ{]}} holds by \texttt{rfl}, with a \texttt{CoeFun} (\texttt{towerCoeFun}) letting us apply a \texttt{D\_\{n+1\}} element directly as a function \texttt{Dₙ\ →\ Dₙ}.

The bonding maps are exactly Proposition 3.7 specialized diagonally. \texttt{conjMap\ post\ pre} (\texttt{f\ ↦\ post∘f∘pre}) is Scott-continuous because directed sups in \texttt{{[}Y→Y{]}} are pointwise, and supplies the outer continuity needed to bundle Proposition 3.7's canonical plain maps. Iterating from a chosen base \texttt{j₀\ :\ {[}D₀→D₀{]}\ ◁\ D₀} (Proposition 3.13 supplies one) gives the projection tower \texttt{towerProj}. The Scott recursion/algebra laws are then definitional: \texttt{towerProj\_succ\_incl\_apply} (\texttt{i\_\{n+1\}(x)=iₙ∘x∘jₙ}), \texttt{towerProj\_succ\_retr\_apply} (\texttt{j\_\{n+1\}=jₙ∘·∘iₙ}), and \texttt{towerProj\_incl\_apply} (\texttt{iₙ(f(x))=i\_\{n+1\}(f)(iₙ(x))}, application preserved one level up).

\textbf{Thm 4.4(a) --- \texttt{embInfInf} / \texttt{projInfInf}.} With \texttt{DInf\ :=\ InverseLimit\ (towerType\ D₀)\ (towerProj\ D₀\ j₀)} (a continuous lattice by Proposition 4.1) and \texttt{DInfFn\ :=\ {[}D∞\ →\ D∞{]}}, Scott's limit pair is written down directly:

\begin{lstlisting}[style=leanbox]
i inf (x) =  bigcup _n (i_{n inf }  comp  x_{n+1}  comp  j_{ inf n})       : D inf   ->  [D inf   ->  D inf ]
j inf (f) =  bigcup _n i_{(n+1) inf }(j_{ inf n}  comp  f  comp  i_{n inf })   : [D inf   ->  D inf ]  ->  D inf 
\end{lstlisting}

The engineering payoff: \textbf{each summand is already a \texttt{ScottMap}.} The \texttt{n}-th summand of \texttt{i∞}, \texttt{iInfTerm\ n}, is the composite \texttt{conjMap\ (i\_\{n∞\},\ j\_\{∞n\})\ ∘\ j\_\{∞(n+1)\}} (conjugation by the Prop 4.2 projection pair, precomposed with the component projection \texttt{j\_\{∞(n+1)\}\ :\ D∞\ →\ D\_\{n+1\}} reading off \texttt{x\_\{n+1\}}); the \texttt{n}-th summand of \texttt{j∞}, \texttt{jInfTerm\ n}, is \texttt{i\_\{(n+1)∞\}\ ∘\ conjMap\ (j\_\{∞n\},\ i\_\{n∞\})}. Both are honest Scott maps because \texttt{conjMap}, \texttt{embInf}, \texttt{projInf}, and \texttt{.comp} are. Consequently \texttt{i∞} and \texttt{j∞} are \emph{suprema of Scott maps} --- \texttt{⨆ₙ\ iInfTerm\ n}, \texttt{⨆ₙ\ jInfTerm\ n} --- taken in the complete lattices \texttt{{[}D∞\ →\ {[}D∞→D∞{]}{]}} and \texttt{{[}{[}D∞→D∞{]}\ →\ D∞{]}} (Theorem 3.3), so they are Scott-continuous \emph{for free}: no bespoke directed-sup/\texttt{iSup\_comm} argument is needed (contrast the \texttt{coconeInf} template). The pointwise unfolding \texttt{embInfInf\_apply\ :\ i∞(x)\ =\ ⨆ₙ\ iInfTerm\ n\ x} (and \texttt{projInfInf\_apply}) follows from \texttt{ScottMap.sSup\_apply} + \texttt{Set.range\_comp}, and the \texttt{*\_apply} reductions of the summands hold by \texttt{rfl} (riding on \texttt{towerType\_succ} defeq). \texttt{*\_preservesDirectedSup} is then immediate from \texttt{.continuous} via Proposition 2.5. Footprint \texttt{{[}propext,\ Classical.choice,\ Quot.sound{]}}. Theorem 4.4 subgoals \textbf{(a)--(d)} are all complete:

\begin{itemize}
\item
  \textbf{(a)} \texttt{embInfInf} / \texttt{projInfInf}: define \(i_\infty\)/\(j_\infty\) as Scott maps (suprema of Scott maps).
\item
  \textbf{(b)} \texttt{projInfInf\_comp\_embInfInf}: \(j_\infty \circ i_\infty = \mathrm{id}\) on \(D_\infty\).
\item
  \textbf{(c)} \texttt{embInfInf\_comp\_projInfInf}: \(i_\infty \circ j_\infty = \mathrm{id}\) on \([D_\infty \to D_\infty]\).
\item
  \textbf{(d)} \texttt{theorem\_4\_4}: Scott's statement --- \(D_\infty\) is a continuous lattice whose product/subspace inverse-limit topology is homeomorphic to the pointwise Pi topology on \([D_\infty \to D_\infty]\); \texttt{theorem\_4\_4\_inverses} is the inverse pair in the proof; \texttt{theorem\_4\_4\_sourceTopology\_homeomorph} packages that pair with the source topologies; \texttt{theorem\_4\_4\_orderIso} is the companion lattice isomorphism from Scott's abstract.
\end{itemize}

\textbf{Thm 4.4(b) --- \texttt{projInfInf\_comp\_embInfInf}.} Goal: \(j_\infty \circ i_\infty = \mathrm{id}\) on \(D_\infty\). Following Scott's calculation, expand \texttt{j∞(i∞(x))\ =\ ⨆ₙ\ jInfTerm\ n\ (i∞\ x)}. Pushing the two conjugations through the inner/outer suprema (\texttt{conjMap\_iSup}, \texttt{embInf\_succ\_iSup} --- each just \emph{preservation of directed sups} by the relevant \texttt{ScottMap}, since the summand families are monotone in \texttt{m}) rewrites the \texttt{n}-th term as \texttt{⨆ₘ\ g\ n\ m} with \texttt{g\ n\ m\ =\ i\_\{(n+1)∞\}(conjMap\ (j\_\{∞n\},\ i\_\{n∞\})(iInfTerm\ m\ x))}. The double sup \texttt{⨆ₙ\ ⨆ₘ\ g\ n\ m} collapses to the diagonal \texttt{⨆ₙ\ g\ n\ n} (\texttt{iSup₂\_monotone\_eq\_diagonal}); monotonicity in \texttt{m} is routine, and monotonicity in \texttt{n} is the one piece of real content --- \texttt{conjMap\_incl\_le\_conjMap\_succ}, the inequality \texttt{i\_\{n+1\}(conjMap\ (j\_\{∞n\},\ i\_\{n∞\})\ f)\ ⊑\ conjMap\ (j\_\{∞(n+1)\},\ i\_\{(n+1)∞\})\ f} in \texttt{D\_\{n+2\}}, built from \texttt{embInf\_succ}, \texttt{incl\_retr\_le}, and \texttt{i\_\{n∞\}(yₙ)\ ⊑\ y\_\{n+1\}} (\texttt{incl\_projInf\_le\_projInf\_succ}). On the diagonal, \texttt{conj\_iInfTerm\_eq} is exactly the function-space retraction \texttt{j\_\{{[}·{]}\}\ ∘\ i\_\{{[}·{]}\}\ =\ id} of the Prop 4.2 projection pair, giving \texttt{g\ n\ n\ =\ i\_\{(n+1)∞\}(x\_\{n+1\})}; an index shift (\texttt{Monotone.iSup\_nat\_add}) plus \texttt{inverseLimit\_eq\_iSup} recognizes the result as \texttt{x}. Footprint \texttt{{[}propext,\ Classical.choice,\ Quot.sound{]}}.

\textbf{Thm 4.4(c) --- \texttt{embInfInf\_comp\_projInfInf}.} Goal: \(i_\infty \circ j_\infty = \mathrm{id}\) on \([D_\infty \to D_\infty]\). The restrictions \texttt{uₙ\ =\ j\_\{∞n\}\ ∘\ f\ ∘\ i\_\{n∞\}\ =\ conjMap\ (j\_\{∞n\},\ i\_\{n∞\})\ f\ ∈\ D\_\{n+1\}} satisfy the Lemma-4.5 recursion \texttt{jₙ₊₁(u\_\{n+2\})\ =\ u\_\{n+1\}} --- proved as \texttt{towerProj\_retr\_conjMap\_succ}, the equality sibling of (b)'s \texttt{conjMap\_incl\_le\_conjMap\_succ} (unfold \texttt{(towerProj\ (n+1)).retr} as the function-space \texttt{conjMap}, then \texttt{embInf\_succ} and the compatibility equation \texttt{x.2\ n}). Hence \texttt{lemma\_4\_5} gives the components \texttt{(j∞\ f).(n+1)\ =\ uₙ} (\texttt{hcoord}). Evaluating \texttt{i∞(j∞\ f)} pointwise (\texttt{embInfInf\_apply}, then \texttt{ScottMap.sSup\_apply} for the pointwise lub) and rewriting each summand with \texttt{hcoord} + \texttt{conjMap\_apply} reduces the \texttt{n}-th term to \texttt{rₙ\ (f\ (rₙ\ z))} with \texttt{rₙ\ =\ i\_\{n∞\}\ ∘\ j\_\{∞n\}}. The analytic step (Scott \textasciitilde1326--1334) confines the lub via continuity of \texttt{f} and the functional equation \texttt{id\ =\ ⨆ₙ\ rₙ} (here just \texttt{inverseLimit\_eq\_iSup}, since \texttt{rₙ\ z\ =\ i\_\{n∞\}(zₙ)}): \texttt{f\ z\ =\ ⨆ₖ\ rₖ\ (f\ z)\ =\ ⨆ₖ\ rₖ\ (f\ (⨆ₘ\ rₘ\ z))\ =\ ⨆ₖ\ ⨆ₘ\ rₖ\ (f\ (rₘ\ z))}, and the monotone double sup collapses to the diagonal \texttt{⨆ₙ\ rₙ\ (f\ (rₙ\ z))} (\texttt{iSup₂\_monotone\_eq\_diagonal}), which is exactly the evaluated \texttt{i∞(j∞\ f)\ z}. Footprint \texttt{{[}propext,\ Classical.choice,\ Quot.sound{]}}.

\textbf{Thm 4.4(d) --- \texttt{theorem\_4\_4}.} Scott's statement: \(D_\infty\) is a continuous lattice and is homeomorphic to \([D_\infty \to D_\infty]\). \texttt{theorem\_4\_4\_inverses} bundles the two composition identities from the proof (\texttt{projInfInf\_comp\_embInfInf}, \texttt{embInfInf\_comp\_projInfInf}); helper lemmas \texttt{projInfInf\_embInfInf} / \texttt{embInfInf\_projInfInf} apply those equalities pointwise. \texttt{theorem\_4\_4\_scottTopology\_homeomorph} first packages that pair using lattice Scott topologies. The topology equalities from Proposition 4.1 and Theorem 3.3 then produce \texttt{theorem\_4\_4\_sourceTopology\_homeomorph}, whose domain is the product/subspace inverse-limit topology and whose codomain is Definition 3.1's pointwise Pi topology. \texttt{theorem\_4\_4} conjoins Proposition 4.1 with that explicit source-topology homeomorphism. \texttt{theorem\_4\_4\_orderIso} is the companion lattice isomorphism (Scott's abstract: ``homeomorphic (and isomorphic)''); it is not the numbered theorem. The Palomar compared claim is \texttt{theorem\_4\_4}. Footprint \texttt{{[}propext,\ Classical.choice,\ Quot.sound{]}}. \textbf{Scott §4 is complete.}

\begin{center}\rule{0.5\linewidth}{0.5pt}\end{center}

\hypertarget{reproducibility}{%
\section{Reproducibility}\label{reproducibility}}

\textbf{Inventory source of truth:} this file (\texttt{arxiv.md}). Do not use generated \texttt{arxiv\_with\_code.md} or \texttt{arxiv.tex} (they inline Lean sources and mermaid figures for review/PDF packaging and are stale whenever older than \texttt{arxiv.md} or any listed \texttt{.lean} file).

The repository pins Lean / mathlib \textbf{v4.33.0} (\texttt{lean-toolchain}).

\begin{lstlisting}[style=leanbox]
lake exe cache get
lake build Scott1972
bash scripts/generate_arxiv_with_code.sh   #  ->  arxiv_with_code.md (gitignored)
bash scripts/build_arxiv_tex.sh            #  ->  arxiv.tex + lean-listings/ + figures/ (gitignored)
bash scripts/build_arxiv_pdf.sh            #  ->  arxiv.pdf (LuaLaTeX; tracked) + dist/arxiv_submit.zip
bash scripts/package_arxiv_submit.sh       #  ->  dist/arxiv_submit.zip only (skips PDF)
\end{lstlisting}

\begin{center}\rule{0.5\linewidth}{0.5pt}\end{center}

\hypertarget{acknowledgments}{%
\section{Acknowledgments}\label{acknowledgments}}

\begin{itemize}
\item
  \textbf{Dana Scott} --- \emph{Continuous Lattices} \textbf{{[}Sco72{]}}, the paper this development formalizes. Scott was not contacted and did not participate in or endorse the formalization.
\item
  \textbf{Robin Milner} --- the March 1972 correction to \textbf{{[}Sco72{]}}, without which Propositions 2.9, 2.10, and 3.3 would be wrong as originally stated.
\end{itemize}

\hypertarget{ai-assisted-development}{%
\subsection{AI-assisted development}\label{ai-assisted-development}}

The human author retains sole responsibility for the mathematical content, the choice of formalization route, and every formal claim in this work. Following standard publisher practice (e.g., COPE guidance on authorship and AI tools \textbf{{[}COPE24{]}}), \textbf{no large language model is listed as a co-author} --- authorship implies an accountability that automated systems cannot bear.

We gratefully acknowledge assistance from the following tools (auto-generated from \texttt{scripts/ai\_model\_cards.py} when building \texttt{arxiv.tex}):

\begin{itemize}
\item
  \textbf{Cursor} \textbf{{[}Cur26{]}} --- agent-assisted editing in the Cursor IDE: formalizing Scott's 1972 continuous-lattice layer in Lean 4 / mathlib, \texttt{lake\ build} repair, vision-OCR transcription, drafting this narrative (\texttt{arxiv.md}), and tracking the formalized inventory. Generated Lean was provisional until it compiled under the pinned toolchain.
\item
  \textbf{Cursor Composer 2.5 Fast} \textbf{{[}Cmp25{]}} --- routine multi-step work: module scaffolding, dependency-ordered wiring of \texttt{Scott1972/ContinuousLattice/}, documentation and Mermaid blueprints, and medium proof obligations where the strategy was already fixed. Per its model card, Composer 2.5 is optimized for codebase navigation rather than open-ended topological proof design; the Milner-block results (2.9--2.11, full 3.3) were not delegated to it alone.
\item
  \textbf{Anthropic Claude Sonnet 5 (medium reasoning)} \textbf{{[}Son26{]}} --- day-to-day formalization and proof-engineering in Cursor at the medium reasoning tier: inventory and narrative maintenance, module wiring, and medium-complexity Lean obligations where the proof strategy was already fixed. Per its model card, Sonnet 5 Medium balances cost and capability for agentic coding at moderate reasoning depth.
\item
  \textbf{Anthropic Claude Opus 4.8 (high reasoning)} \textbf{{[}Ant26{]}} --- selective use for the heaviest proof work (Propositions 2.9--2.11, Theorem 2.12, Theorem 3.3, Propositions 3.8--3.10, Theorem 4.4). Every emitted proof term was checked by the Lean kernel.
\item
  \textbf{Google Gemini 3.5 Flash} \textbf{{[}Gem25{]}} --- exploratory passes on Scott's typographic conventions (ambient vs subspace joins in the Milner correction) and scope decisions.
\end{itemize}

All definitions, constructivity audits, and final prose were reviewed by the human author, who takes full responsibility for them.

\hypertarget{artifact-availability}{%
\subsection{Artifact availability}\label{artifact-availability}}

The development \textbf{{[}SR72{]}} is at \href{https://github.com/catskillsresearch/scott1972}{\texttt{github.com/catskillsresearch/scott1972}}. Run \texttt{lake\ build\ Scott1972} for the sorry-free formalization; \texttt{scripts/generate\_arxiv\_with\_code.sh} builds \texttt{arxiv\_with\_code.md} from this file plus the complete Lean source.

\begin{center}\rule{0.5\linewidth}{0.5pt}\end{center}

\hypertarget{references}{%
\section{References}\label{references}}

\begin{itemize}
\item
  \textbf{{[}Sco69{]}} D. S. Scott. \emph{Lattice-theoretic models for the \(\lambda\)-calculus} (unpublished manuscript). University of Oxford, 1969.
\item
  \textbf{{[}Sco70{]}} D. S. Scott. Outline of a mathematical theory of computation. In \emph{Proceedings of the Fourth Annual Princeton Conference on Information Sciences and Systems} (pp.~169--176). Princeton University, 1970.
\item
  \textbf{{[}Sco72{]}} D. S. Scott. Continuous lattices. In F. W. Lawvere (Ed.), \emph{Toposes, Algebraic Geometry and Logic} (Lecture Notes in Mathematics, Vol. 274, pp.~97--136). Springer, Berlin, Heidelberg, 1972. URL: http://upol.cz
\item
  \textbf{{[}GHKLMS03{]}} G. Gierz, K. H. Hofmann, K. Keimel, J. D. Lawson, M. Mislove, and D. S. Scott. \emph{Continuous Lattices and Domains}. Cambridge University Press, 2003.
\item
  \textbf{{[}Kel55{]}} J. L. Kelley. \emph{General Topology}. D. Van Nostrand Company, 1955.
\item
  \textbf{{[}SR72{]}} Catskills Research. \emph{scott1972} (this work). \url{https://github.com/catskillsresearch/scott1972}.
\item
  \textbf{{[}COPE24{]}} Committee on Publication Ethics (COPE). \emph{Authorship and AI tools: COPE position statement}. 2024. \url{https://publicationethics.org/guidance/cope-position/authorship-and-ai-tools}
\item
  \textbf{{[}Cur26{]}} Anysphere, Inc.~\emph{Cursor: AI-native code editor and agent environment}. \url{https://cursor.com} (accessed 2026).
\item
  \textbf{{[}Cmp25{]}} Anysphere, Inc.~\emph{Composer 2.5}. Model announcement and documentation, \url{https://cursor.com/blog/composer-2-5}; model card as integrated in Cursor, \url{https://cursor.com/docs/models} (accessed 2026).
\item
  \textbf{{[}Son26{]}} Anthropic. \emph{Claude Sonnet 5} (medium reasoning variant). System card, \url{https://www.anthropic.com/claude-sonnet-5-system-card}; model documentation as integrated in Cursor, \url{https://cursor.com/docs/models} (accessed 2026).
\item
  \textbf{{[}Ant26{]}} Anthropic. \emph{Claude Opus 4.8} (high thinking/reasoning variant). System card and announcement, \url{https://www.anthropic.com/news/claude-opus-4-8}; model documentation as integrated in Cursor, \url{https://cursor.com/docs/models/claude-opus-4-8} (accessed 2026).
\item
  \textbf{{[}Gem25{]}} Google DeepMind. \emph{Gemini 3.5 Flash}. Technical documentation and model cards. \url{https://ai.google.dev/gemini-api/docs/models} (accessed 2026).
\end{itemize}

\begin{center}\rule{0.5\linewidth}{0.5pt}\end{center}

\hypertarget{complete-lean-source}{%
\appendix
\section{Complete Lean source}\label{complete-lean-source}}

\begin{longtable}[]{@{}
  >{\raggedright\arraybackslash}p{(\columnwidth - 2\tabcolsep) * \real{0.5000}}
  >{\raggedright\arraybackslash}p{(\columnwidth - 2\tabcolsep) * \real{0.5000}}@{}}
\toprule\noalign{}
\begin{minipage}[b]{\linewidth}\raggedright
Role
\end{minipage} & \begin{minipage}[b]{\linewidth}\raggedright
File
\end{minipage} \\
\midrule\noalign{}
\endhead
\bottomrule\noalign{}
\endlastfoot
Root import graph & \texttt{Scott1972.lean} \\
Scott §1 & \texttt{Scott1972/ContinuousLattice/Injective.lean} \\
Scott §2 & \texttt{Scott1972/ContinuousLattice/WayBelow.lean} \\
Scott §2 & \texttt{Scott1972/ContinuousLattice/Specialization.lean} \\
Scott §2 & \texttt{Scott1972/ContinuousLattice/ScottMaps.lean} \\
March 1972 correction & \texttt{Scott1972/ContinuousLattice/MilnerCorrection.lean} \\
Scott §2.8--2.12 & \texttt{Scott1972/ContinuousLattice/Constructions.lean} \\
Scott §3 & \texttt{Scott1972/ContinuousLattice/FunctionSpaces.lean} \\
Theorem 2.12 & \texttt{Scott1972/ContinuousLattice/Theorem212.lean} \\
Scott §4 & \texttt{Scott1972/ContinuousLattice/InverseLimits.lean} \\
Theorem 4.4 & \texttt{Scott1972/ContinuousLattice/FunctionSpaceTower.lean} \\
\end{longtable}

Primary source (OCR plain text): \href{sources/ScottContinLatt1972.md}{\texttt{sources/ScottContinLatt1972.md}} --- transcription of \textbf{{[}Sco72{]}} for use in the Lean development (see §2).

Files appear in \texttt{Scott1972.lean} import order. Each block is a verbatim copy of the repository file at generation time.

\hypertarget{scott1972.lean}{%
\subsection{Scott1972.lean}\label{scott1972.lean}}

\emph{17 lines.}

\vspace{0.5\baselineskip}
\noindent\textcolor{green!40!black}{\textbf{Lean 4 source}}\par\vspace{0.25\baselineskip}
\begin{lstlisting}[style=leanbox]
/-
Copyright (c) 2026  Lars Warren Ericson.  All rights reserved.
Released under Apache 2.0 license as described in the file LICENSE.
Authors: Lars Warren Ericson.
Github:  https://github.com/catskillsresearch/scott1972
-/

import Scott1972.ContinuousLattice.Injective
import Scott1972.ContinuousLattice.WayBelow
import Scott1972.ContinuousLattice.Specialization
import Scott1972.ContinuousLattice.ScottMaps
import Scott1972.ContinuousLattice.MilnerCorrection
import Scott1972.ContinuousLattice.Constructions
import Scott1972.ContinuousLattice.FunctionSpaces
import Scott1972.ContinuousLattice.Theorem212
import Scott1972.ContinuousLattice.InverseLimits
import Scott1972.ContinuousLattice.FunctionSpaceTower
\end{lstlisting}
\vspace{0.5\baselineskip}

\hypertarget{scott1972continuouslatticeinjective.lean}{%
\subsection{Scott1972/ContinuousLattice/Injective.lean}\label{scott1972continuouslatticeinjective.lean}}

\emph{132 lines.}

\vspace{0.5\baselineskip}
\noindent\textcolor{green!40!black}{\textbf{Lean 4 source}}\par\vspace{0.25\baselineskip}
\begin{lstlisting}[style=leanbox]
/-
Copyright (c) 2026  Lars Warren Ericson.  All rights reserved.
Released under Apache 2.0 license as described in the file LICENSE.
Authors: Lars Warren Ericson.
Github:  https://github.com/catskillsresearch/scott1972
-/

import Mathlib.Topology.ContinuousMap.T0Sierpinski
import Mathlib.Topology.ContinuousMap.Basic
import Mathlib.Topology.Sets.Opens

/-!
# Injective spaces (Scott 1972, S1)

Scott's S1 introduces *injective* `T_0`-spaces -- those with the extension property for
continuous maps along subspace embeddings -- and shows the two-point Sierpinski space is
injective, that injectivity is preserved by products and by retracts, and (the embedding
theorem) that every `T_0`-space embeds in a power of the Sierpinski space.

## Main definitions / results

* `IsInjectiveSpace` -- Scott's Definition 1.1.
* `proposition_1_2` ... `proposition_1_5` -- Scott's Propositions 1.2-1.5.
* `corollary_1_6`, `corollary_1_7` -- Scott's Corollaries 1.6 and 1.7.
-/

/-- Scott's two-point Sierpinski space  O : `Prop` with the Sierpinski topology. -/
abbrev Sierpinski := Prop

open Topology TopologicalSpace

universe u v w

/-- **Scott 1972, Definition 1.1.** A topological space `D` is *injective* when, for every
topological embedding `e : X  ->  Y` and every continuous `f : X  ->  D`, there is a continuous
`g : Y  ->  D` extending `f` along `e`. -/
def IsInjectiveSpace (D : Type v) [TopologicalSpace D] : Prop :=
  forall  {X Y : Type u} [TopologicalSpace X] [TopologicalSpace Y] (e : X  ->  Y),
    IsEmbedding e  ->  forall  f : C(X, D), exists  g : C(Y, D), forall  x, g (e x) = f x

/-- **Scott 1972, Proposition 1.2.** The Sierpinski space is injective. -/
theorem isInjectiveSpace_sierpinski : IsInjectiveSpace.{u} Prop := by
  intro X Y _ _ e he f
  have hfopen : IsOpen {x | f x} := continuous_Prop.1 f.continuous
  rw [he.isInducing.isOpen_iff] at hfopen
  obtain <V, hVopen, hVeq> := hfopen
  refine <<fun y => y  in  V, continuous_Prop.2 hVopen>, fun x => ?_>
  have : (e x  in  V) = (x  in  {x | f x}) := by rw [ <-  hVeq]; rfl
  simpa using this

theorem proposition_1_2 : IsInjectiveSpace.{u} Sierpinski :=
  isInjectiveSpace_sierpinski

/-- **Scott 1972, Proposition 1.3.** Products of injective spaces are injective. -/
theorem IsInjectiveSpace.pi {? : Type w} (D : ?  ->  Type v) [forall  i, TopologicalSpace (D i)]
    (h : forall  i, IsInjectiveSpace.{u} (D i)) : IsInjectiveSpace.{u} (forall  i, D i) := by
  intro X Y _ _ e he f
  choose g hg using fun i =>
    h i e he <fun x => f x i, (continuous_apply i).comp f.continuous>
  refine <<fun y i => g i y, continuous_pi fun i => (g i).continuous>, fun x => ?_>
  funext i
  exact hg i x

theorem proposition_1_3 {? : Type w} (D : ?  ->  Type v) [forall  i, TopologicalSpace (D i)]
    (h : forall  i, IsInjectiveSpace.{u} (D i)) : IsInjectiveSpace.{u} (forall  i, D i) :=
  IsInjectiveSpace.pi D h

/-- **Scott 1972, Proposition 1.4.** Retracts of injective spaces are injective. -/
theorem IsInjectiveSpace.of_retract {D : Type v} {D' : Type v}
    [TopologicalSpace D] [TopologicalSpace D'] (hD : IsInjectiveSpace.{u} D)
    (s : C(D', D)) (r : C(D, D')) (hrs : forall  d, r (s d) = d) :
    IsInjectiveSpace.{u} D' := by
  intro X Y _ _ e he f
  obtain <g, hg> := hD e he (s.comp f)
  refine <r.comp g, fun x => ?_>
  show r (g (e x)) = f x
  rw [hg x]
  exact hrs (f x)

theorem proposition_1_4 {D : Type v} {D' : Type v} [TopologicalSpace D] [TopologicalSpace D']
    (hD : IsInjectiveSpace.{u} D) (s : C(D', D)) (r : C(D, D')) (hrs : forall  d, r (s d) = d) :
    IsInjectiveSpace.{u} D' :=
  IsInjectiveSpace.of_retract hD s r hrs

theorem isInjectiveSpace_sierpinski_power (? : Type w) :
    IsInjectiveSpace.{u} (?  ->  Prop) :=
  proposition_1_3 (fun (_ : ?) => Prop) (fun _ => proposition_1_2)

/-- **Scott 1972, Proposition 1.5.** Every `T_0`-space embeds into a power of  O , and that
power is injective. -/
theorem proposition_1_5 (X : Type u) [TopologicalSpace X] [T0Space X] :
    IsInjectiveSpace.{u,u} (Opens X  ->  Prop)  /\ 
      IsEmbedding (productOfMemOpens X) := by
  refine <?_, productOfMemOpens_isEmbedding X>
  exact isInjectiveSpace_sierpinski_power (Opens X)

theorem t0_isEmbedding_into_sierpinski_power (X : Type u) [TopologicalSpace X] [T0Space X] :
    IsEmbedding (productOfMemOpens X) :=
  (proposition_1_5 X).2

/-- Scott's notion of retract: a subspace with a continuous retraction. -/
structure IsRetractSubspace (D' D : Type u) [TopologicalSpace D'] [TopologicalSpace D] where
  section' : C(D', D)
  isEmbedding_section : IsEmbedding section'
  retraction : C(D, D')
  retraction_section : forall  d, retraction (section' d) = d

/-- **Scott 1972, Corollary 1.6.** Injective spaces are exactly retracts of powers of  O . -/
theorem corollary_1_6 (D : Type u) [TopologicalSpace D] [T0Space D] :
    IsInjectiveSpace.{u,u} D  <->  exists  (? : Type u), Nonempty (IsRetractSubspace D (?  ->  Prop)) := by
  constructor
  * intro hD
    obtain <r, hr> := hD (productOfMemOpens D) (proposition_1_5 D).2 (ContinuousMap.id D)
    refine <Opens D, <<productOfMemOpens D, (proposition_1_5 D).2, r, hr>>>
  * rintro <?, <R>>
    exact IsInjectiveSpace.of_retract (isInjectiveSpace_sierpinski_power ?) R.section' R.retraction
      R.retraction_section

/-- **Scott 1972, Corollary 1.7.** A space is injective iff it is a retract of every
super-space of which it is a subspace. -/
theorem corollary_1_7 (D : Type u) [TopologicalSpace D] [T0Space D] :
    IsInjectiveSpace.{u,u} D  <-> 
      forall  {Y : Type u} [TopologicalSpace Y] (e : D  ->  Y), IsEmbedding e  -> 
        exists  r : C(Y, D), forall  d, r (e d) = d := by
  constructor
  * intro hD Y _ e he
    obtain <r, hr> := hD e he (ContinuousMap.id D)
    exact <r, hr>
  * intro hD
    obtain <r, hr> := hD (productOfMemOpens D) (proposition_1_5 D).2
    exact IsInjectiveSpace.of_retract (isInjectiveSpace_sierpinski_power (Opens D))
      (productOfMemOpens D) r hr
\end{lstlisting}
\vspace{0.5\baselineskip}

\hypertarget{scott1972continuouslatticewaybelow.lean}{%
\subsection{Scott1972/ContinuousLattice/WayBelow.lean}\label{scott1972continuouslatticewaybelow.lean}}

\emph{248 lines.}

\vspace{0.5\baselineskip}
\noindent\textcolor{green!40!black}{\textbf{Lean 4 source}}\par\vspace{0.25\baselineskip}
\begin{lstlisting}[style=leanbox]
/-
Copyright (c) 2026  Lars Warren Ericson.  All rights reserved.
Released under Apache 2.0 license as described in the file LICENSE.
Authors: Lars Warren Ericson.
Github:  https://github.com/catskillsresearch/scott1972
-/

import Mathlib.Order.CompleteLattice.Basic
import Mathlib.Order.UpperLower.Basic
import Mathlib.Order.Directed

/-!
# The induced (Scott) topology and the way-below relation (Scott 1972, S2)

This file formalizes the core of Scott's S2 *Continuous Lattices*: the topology a complete
lattice carries intrinsically (Scott's "induced topology", today the **Scott topology**),
the **way-below relation** ` << `, the basic properties of ` << ` (Scott's Proposition 2.2), and
the definition of a **continuous lattice** (Scott's Definition 2.3).

Scott defines the induced topology on a partially ordered set `D` by declaring `U` open iff

* (i)  `U` is an upper set; and
* (ii) whenever `S  subseteq  D` is directed (and, in this paper, non-empty), ` sqsup S` exists and
       ` sqsup S  in  U`, then `S  I  U  !=  {}`.

He then defines `x  <<  y` ("`x` is way below `y`") to mean `y  in  interior {z | x  sqsub  z}`, the
interior being taken in this induced topology. We encode "Scott-open" as the predicate
`ScottOpen` and ` << ` as `WayBelow`, witnessing the interior by a Scott-open neighbourhood
contained in the principal up-set `Set.Ici x = {z | x  <=  z}`. This is faithful to Scott's
topological definition (rather than the order-theoretic shortcut), and as a result Scott's
Proposition 2.2 (vi) and (vii) fall straight out of the open-set axioms.

This is the classical/topological version of the theory, so we reason classically.
-/

namespace Scott1972.ContinuousLattice

universe u

variable {D : Type u} [CompleteLattice D]

/-- **Scott 1972, S2, the induced topology.** `U` is *Scott-open* when it is an upper set and
is inaccessible by suprema of non-empty directed sets: if a non-empty directed `S` has its
supremum in `U`, then some member of `S` already lies in `U`. -/
def ScottOpen (U : Set D) : Prop :=
  letI : LE D :=
    (CompleteLattice.toCompleteSemilatticeInf (a := D)).toPartialOrder.toPreorder.toLE
  letI : SupSet D :=
    (CompleteLattice.toCompleteSemilatticeSup (a := D)).toSupSet
  IsUpperSet U  /\ 
    forall  ?S : Set D?, S.Nonempty  ->  DirectedOn (*  <=  *) S  ->  sSup S  in  U  ->  (S  I  U).Nonempty

theorem ScottOpen.isUpperSet {U : Set D} (h : ScottOpen U) : IsUpperSet U := h.1

theorem scottOpen_univ : ScottOpen (Set.univ : Set D) := by
  refine <isUpperSet_univ, fun S hS _ _ => ?_>
  obtain <s, hs> := hS
  exact <s, hs, Set.mem_univ s>

theorem scottOpen_inter {U V : Set D} (hU : ScottOpen U) (hV : ScottOpen V) :
    ScottOpen (U  I  V) := by
  refine <hU.1.inter hV.1, fun S hS hSdir hmem => ?_>
  obtain <s_1, hs_1S, hs_1U> := hU.2 hS hSdir hmem.1
  obtain <s_2, hs_2S, hs_2V> := hV.2 hS hSdir hmem.2
  obtain <s_3, hs_3S, h_1, h_2> := hSdir s_1 hs_1S s_2 hs_2S
  exact <s_3, hs_3S, hU.1 h_1 hs_1U, hV.1 h_2 hs_2V>

theorem scottOpen_sUnion {C : Set (Set D)} (hC : forall  U  in  C, ScottOpen U) :
    ScottOpen (?_0 C) := by
  refine <isUpperSet_sUnion fun U hU => (hC U hU).1, fun S hS hSdir hmem => ?_>
  obtain <U, hUC, hUmem> := hmem
  obtain <s, hsS, hsU> := (hC U hUC).2 hS hSdir hUmem
  exact <s, hsS, U, hUC, hsU>

/-- **Scott 1972, S2.** The *way-below* relation: `x  <<  y` iff `y` lies in the interior of the
principal up-set `Set.Ici x` for the induced topology, witnessed by a Scott-open
neighbourhood of `y` contained in `Set.Ici x`. -/
def WayBelow (x y : D) : Prop :=
  letI : LE D :=
    (CompleteLattice.toCompleteSemilatticeInf (a := D)).toPartialOrder.toPreorder.toLE
  letI : Preorder D :=
    (CompleteLattice.toCompleteSemilatticeInf (a := D)).toPartialOrder.toPreorder
  exists  U : Set D, ScottOpen U  /\  y  in  U  /\  U  subseteq  Set.Ici x

@[inherit_doc] scoped infix:50 "  <<  " => WayBelow

/-- **Scott 1972, Proposition 2.2(v).** `x  <<  y` implies `x  <=  y`. -/
theorem WayBelow.le {x y : D} (h : x  <<  y) : x  <=  y := by
  obtain <U, _, hyU, hsub> := h
  exact hsub hyU

/-- **Scott 1972, Proposition 2.2(i).** `False  <<  x` for every `x`. -/
theorem bot_wayBelow (x : D) : (False : D)  <<  x :=
  <Set.univ, scottOpen_univ, Set.mem_univ x, fun _ _ => bot_le>

/-- **Scott 1972, Proposition 2.2(iii).** `x  <<  y` and `y  <=  z` imply `x  <<  z` (monotone on the
right). -/
theorem WayBelow.trans_le {x y z : D} (h : x  <<  y) (hyz : y  <=  z) : x  <<  z := by
  obtain <U, hU, hyU, hsub> := h
  exact <U, hU, hU.1 hyz hyU, hsub>

/-- **Scott 1972, Proposition 2.2(iv).** `x  <=  y` and `y  <<  z` imply `x  <<  z` (monotone on the
left). -/
theorem WayBelow.le_trans {x y z : D} (hxy : x  <=  y) (h : y  <<  z) : x  <<  z := by
  obtain <U, hU, hzU, hsub> := h
  refine <U, hU, hzU, fun w hw => ?_>
  have hyw : y  <=  w := Set.mem_Ici.1 (hsub hw)
  exact Set.mem_Ici.2 (hxy.trans hyw)

/-- **Scott 1972, Proposition 2.2(ii).** `x  <<  z` and `y  <<  z` imply `x  sqsup  y  <<  z`. -/
theorem WayBelow.sup {x y z : D} (hx : x  <<  z) (hy : y  <<  z) : x  sqsup  y  <<  z := by
  obtain <U, hU, hzU, hUsub> := hx
  obtain <V, hV, hzV, hVsub> := hy
  refine <U  I  V, scottOpen_inter hU hV, <hzU, hzV>, fun w hw => ?_>
  exact Set.mem_Ici.2 (sup_le (hUsub hw.1) (hVsub hw.2))

/-- Auxiliary: the up-set `{z | x  <<  z}` is itself Scott-open. (This is *not* Scott's
Proposition 2.2(vi) -- see `wayBelow_self_iff_scottOpen_Ici` for that -- but it is a standard
and useful fact, the openness of the sets ` Up x`.) -/
theorem scottOpen_wayBelow (x : D) : ScottOpen {z | x  <<  z} := by
  refine <fun a b hab ha => ha.trans_le hab, fun S hS hSdir hmem => ?_>
  obtain <U, hU, hsupU, hsub> := hmem
  obtain <s, hsS, hsU> := hU.2 hS hSdir hsupU
  exact <s, hsS, <U, hU, hsU, hsub>>

/-- **Scott 1972, Proposition 2.2(vi).** `x  <<  x` iff the principal up-set `{z | x  sqsub  z}` (i.e.
`Set.Ici x`) is Scott-open. This characterizes the *compact* (finite, isolated) elements:
`x` is compact exactly when `x` is open. -/
theorem wayBelow_self_iff_scottOpen_Ici {x : D} : x  <<  x  <->  ScottOpen (Set.Ici x) := by
  constructor
  * rintro <U, hU, hxU, hsub>
    -- `Ici x = U`: `Ici x  subseteq  U` since `U` is upper and `x  in  U`; `U  subseteq  Ici x` is `hsub`.
    have hIci : Set.Ici x = U :=
      le_antisymm (fun w hw => hU.1 (Set.mem_Ici.1 hw) hxU) hsub
    rw [hIci]; exact hU
  * intro hopen
    exact <Set.Ici x, hopen, Set.self_mem_Ici, le_refl _>

/-- **Scott 1972, Proposition 2.2(vii).** For a non-empty directed set `S`, `x  <<   sqsup S` iff
`x  <<  y` for some `y  in  S`. The forward direction is exactly inaccessibility of a Scott-open
set; the backward direction is monotonicity on the right. -/
theorem wayBelow_sSup_iff {x : D} {S : Set D} (hS : S.Nonempty)
    (hSdir : DirectedOn (*  <=  *) S) : x  <<  sSup S  <->  exists  y  in  S, x  <<  y := by
  constructor
  * rintro <U, hU, hsupU, hsub>
    obtain <s, hsS, hsU> := hU.2 hS hSdir hsupU
    exact <s, hsS, U, hU, hsU, hsub>
  * rintro <y, hyS, hxy>
    exact hxy.trans_le (le_sSup hyS)

/-- **Scott 1972, Definition 2.3.** A complete lattice `D` is a *continuous lattice* when every
element is the supremum of the elements way below it: `y =  sqsup  {x | x  <<  y}`. -/
def IsContinuousLattice (D : Type u) [CompleteLattice D] : Prop :=
  letI : LE D :=
    (CompleteLattice.toCompleteSemilatticeInf (a := D)).toPartialOrder.toPreorder.toLE
  forall  y : D, IsLUB {x | x  <<  y} y

/-- In a continuous lattice, `y` is the actual supremum of `{x | x  <<  y}`. -/
theorem IsContinuousLattice.sSup_wayBelow (h : IsContinuousLattice D) (y : D) :
    sSup {x | x  <<  y} = y :=
  (h y).sSup_eq

/-- The set `{x | x  <<  y}` of elements way below `y` is directed: it is closed under binary
joins by Proposition 2.2(ii) (`WayBelow.sup`). This holds in *any* complete lattice. -/
theorem directedOn_wayBelow (y : D) : DirectedOn (*  <=  *) {x | x  <<  y} :=
  fun a ha b hb => <a  sqsup  b, ha.sup hb, le_sup_left, le_sup_right>

/-- **Interpolation property of ` << `.** In a continuous lattice, the way-below relation
interpolates: `a  <<  c` implies there is some `b` with `a  <<  b  <<  c`.

The proof runs Scott's standard argument: the set `M = {m | exists  x, m  <<  x  /\  x  <<  c}` is directed
(using directedness of `{*  <<  x}` twice) and has supremum `c` (using continuity twice). Hence
`a  <<  c =  sqsup M` with `M` directed forces `a  <<  m` for some `m  in  M`, say `m  <<  x  <<  c`; then
`a  <<  m  <=  x` gives `a  <<  x  <<  c`, so `b := x` works. -/
theorem wayBelow_interpolate (hD : IsContinuousLattice D) {a c : D} (hac : a  <<  c) :
    exists  b, a  <<  b  /\  b  <<  c := by
  set M : Set D := {m | exists  x, m  <<  x  /\  x  <<  c} with hM
  have hMdir : DirectedOn (*  <=  *) M := by
    rintro m_1 <x_1, hm_1x_1, hx_1c> m_2 <x_2, hm_2x_2, hx_2c>
    obtain <x_3, hx_3c, hx_1_3, hx_2_3> := directedOn_wayBelow c x_1 hx_1c x_2 hx_2c
    obtain <m_3, hm_3x_3, hm_1m_3, hm_2m_3> :=
      directedOn_wayBelow x_3 m_1 (hm_1x_1.trans_le hx_1_3) m_2 (hm_2x_2.trans_le hx_2_3)
    exact <m_3, <x_3, hm_3x_3, hx_3c>, hm_1m_3, hm_2m_3>
  have hMne : M.Nonempty := <False, a, bot_wayBelow a, hac>
  have hsupM : sSup M = c := by
    refine le_antisymm (sSup_le ?_) ?_
    * rintro m <x, hmx, hxc>
      exact hmx.le.trans hxc.le
    * rw [ <-  hD.sSup_wayBelow c]
      refine sSup_le fun x hxc => ?_
      rw [ <-  hD.sSup_wayBelow x]
      exact sSup_le fun m hmx => le_sSup <x, hmx, hxc>
  rw [ <-  hsupM] at hac
  obtain <m, <x, hmx, hxc>, ham> := (wayBelow_sSup_iff hMne hMdir).1 hac
  exact <x, ham.trans_le hmx.le, hxc>

/-- In a continuous lattice the sets ` Up a = {z | a  <<  z}` form a basis of the Scott topology:
every Scott-open `U` containing `z` contains some basic neighbourhood ` Up a` of `z`. Indeed
`z =  sqsup {a | a  <<  z}` is a directed supremum lying in the open set `U`, so some `a  <<  z` already
lies in `U`, and then ` Up a  subseteq  a  subseteq  U`. -/
theorem exists_wayBelow_subset (hD : IsContinuousLattice D) {U : Set D} (hU : ScottOpen U)
    {z : D} (hz : z  in  U) : exists  a, a  <<  z  /\  {w | a  <<  w}  subseteq  U := by
  have hne : {a | a  <<  z}.Nonempty := <False, bot_wayBelow z>
  have hsup : sSup {a | a  <<  z}  in  U := by rw [hD.sSup_wayBelow z]; exact hz
  obtain <a, haz, haU> := hU.2 hne (directedOn_wayBelow z) hsup
  exact <a, haz, fun w hw => hU.1 hw.le haU>

/-- A strengthening of `exists_wayBelow_subset`: the witness `a  <<  z` can be taken with the whole
principal up-set `Set.Ici a` (not merely ` Up a`) inside `U`. The element `a` produced lies in the
open `U`, which is upper, so `a  subseteq  U`. -/
theorem exists_wayBelow_Ici_subset (hD : IsContinuousLattice D) {U : Set D} (hU : ScottOpen U)
    {z : D} (hz : z  in  U) : exists  a, a  <<  z  /\  Set.Ici a  subseteq  U := by
  have hne : {a | a  <<  z}.Nonempty := <False, bot_wayBelow z>
  have hsup : sSup {a | a  <<  z}  in  U := by rw [hD.sSup_wayBelow z]; exact hz
  obtain <a, haz, haU> := hU.2 hne (directedOn_wayBelow z) hsup
  exact <a, haz, fun w hw => hU.1 (Set.mem_Ici.1 hw) haU>

/-- The infimum of a Scott-open neighbourhood of `y` is way below `y`: the open set is itself
the required witness. Scott uses this in moving between Definition 2.3 and Proposition 2.4. -/
theorem sInf_wayBelow {U : Set D} (hU : ScottOpen U) {y : D} (hy : y  in  U) :
    sInf U  <<  y :=
  <U, hU, hy, fun _ hz => Set.mem_Ici.2 (sInf_le hz)>

/-- **Scott 1972, Proposition 2.4.** A complete lattice is continuous iff every element is the
supremum of the infima of its open neighbourhoods: `y =  sqsup  { sqcap U : y  in  U open}`. This is Scott's
alternate form of Definition 2.3. -/
theorem isContinuousLattice_iff_isLUB_sInf_nhds :
    IsContinuousLattice D  <-> 
      forall  y : D, IsLUB {a : D | exists  U, ScottOpen U  /\  y  in  U  /\  a = sInf U} y := by
  constructor
  * intro h y
    refine <?_, ?_>
    * rintro a <U, hU, hyU, rfl>
      exact (sInf_wayBelow hU hyU).le
    * intro b hb
      refine (h y).2 ?_
      intro x hx
      obtain <U, hU, hyU, hsub> := hx
      have hxle : x  <=  sInf U := le_sInf fun _ hz => Set.mem_Ici.1 (hsub hz)
      exact hxle.trans (hb <U, hU, hyU, rfl>)
  * intro h y
    refine <fun x hx => hx.le, ?_>
    intro b hb
    refine (h y).2 ?_
    rintro a <U, hU, hyU, rfl>
    exact hb (sInf_wayBelow hU hyU)

end Scott1972.ContinuousLattice
\end{lstlisting}
\vspace{0.5\baselineskip}

\hypertarget{scott1972continuouslatticespecialization.lean}{%
\subsection{Scott1972/ContinuousLattice/Specialization.lean}\label{scott1972continuouslatticespecialization.lean}}

\emph{166 lines.}

\vspace{0.5\baselineskip}
\noindent\textcolor{green!40!black}{\textbf{Lean 4 source}}\par\vspace{0.25\baselineskip}
\begin{lstlisting}[style=leanbox]
/-
Copyright (c) 2026  Lars Warren Ericson.  All rights reserved.
Released under Apache 2.0 license as described in the file LICENSE.
Authors: Lars Warren Ericson.
Github:  https://github.com/catskillsresearch/scott1972
-/

import Scott1972.ContinuousLattice.WayBelow
import Mathlib.Topology.Inseparable
import Mathlib.Topology.Separation.Basic
import Mathlib.Topology.Order.ScottTopology
import Mathlib.Order.DirSupClosed

/-!
# Specialization order and Scott topology (Scott 1972, S2 opening)

Scott's S2 begins with the specialization order on a `T_0`-space and the induced (Scott)
topology on a complete lattice. Proposition 2.1 (monotone nets and least upper bounds) is
split into its two directions; the convergence-to-below direction is the mathematically
heavier half and is recorded as `proposition_2_1_of_le`.
-/

namespace Scott1972.ContinuousLattice

open Topology Set

universe u

variable {X : Type*} {D : Type u} [TopologicalSpace X] [CompleteLattice D]

/-! ### Specialization order -/

/-- **Scott 1972, S2.** The *specialization order*: `x  sqsub  y` when `x  in  U` open implies `y  in  U`. -/
def SpecializationLe (x y : X) : Prop :=
  forall  U, IsOpen U  ->  x  in  U  ->  y  in  U

instance specializationPreorder : Preorder X where
  le := SpecializationLe
  le_refl x := fun _ _ hx => hx
  le_trans x y z hxy hyz U hU hxU := hyz U hU (hxy U hU hxU)

theorem specializationLe_antisymm [T0Space X] {x y : X}
    (hxy : SpecializationLe x y) (hyx : SpecializationLe y x) : x = y :=
  Inseparable.eq (inseparable_iff_specializes_and.2
    <specializes_iff_forall_open.2 fun s hs hy => hyx s hs hy,
     specializes_iff_forall_open.2 fun s hs hx => hxy s hs hx>)

/-- Scott's specialization order is the reverse of Mathlib's `Specializes` convention. -/
theorem specializationLe_iff_specializes {x y : X} :
    SpecializationLe x y  <->  y ? x :=
  specializes_iff_forall_open.symm

/-- Continuous maps preserve Scott's specialization order. -/
theorem SpecializationLe.map {Y : Type*} [TopologicalSpace Y] {x y : X}
    (hxy : SpecializationLe x y) {f : X  ->  Y} (hf : Continuous f) :
    SpecializationLe (f x) (f y) := by
  intro U hU hxU
  exact hxy (f ^-1' U) (hU.preimage hf) hxU

/-! ### Scott topology from `ScottOpen` -/

/-- Scott's induced topology on a complete lattice, realized as mathlib's Scott topology. -/
@[reducible] noncomputable def scottTopologicalSpace : TopologicalSpace D :=
  Topology.scott D univ

theorem ScottOpen_iff_dirSupInacc {U : Set D} : ScottOpen U  <->  IsUpperSet U  /\  DirSupInacc U := by
  constructor
  * intro <hU, hU'>
    refine <hU, fun d hd_1 hd_2 a ha hmem => ?_>
    rw [ <-  IsLUB.sSup_eq ha] at hmem
    obtain <s, hs, hsU> := hU' hd_1 hd_2 hmem
    exact <s, hs, hsU>
  * intro <hU, hU'>
    refine <hU, fun d hd_1 hd_2 hmem => ?_>
    obtain <s, hs, hsU> := hU' hd_1 hd_2 (isLUB_sSup d) hmem
    exact <s, hs, hsU>

theorem isOpen_iff_scottOpen {U : Set D} : @IsOpen D scottTopologicalSpace U  <->  ScottOpen U := by
  rw [ScottOpen_iff_dirSupInacc, scottTopologicalSpace]
  have hinst : @IsScott D univ _ (Topology.scott D univ) :=
    @IsScott.mk D univ _ (Topology.scott D univ) rfl
  simpa [dirSupInaccOn_univ] using
    @IsScott.isOpen_iff_isUpperSet_and_dirSupInaccOn D univ _ (Topology.scott D univ) U hinst

/-- Scott-open sets in our sense agree with mathlib's Scott topology (alias). -/
theorem isOpen_scott_iff_scottOpen {U : Set D} :
    @IsOpen D (Topology.scott D univ) U  <->  ScottOpen U :=
  isOpen_iff_scottOpen

/-! ### Monotone nets and Proposition 2.1 -/

variable {? : Type u} [Preorder ?] [IsDirected ? (*  <=  *)]

def IsMonotoneNet (x : ?  ->  D) : Prop :=
  Monotone x

def ScottConvergesTo (x : ?  ->  D) (y : D) : Prop :=
  forall  U, ScottOpen U  ->  y  in  U  ->  exists  i, forall  j  >=  i, x j  in  U

variable {x : ?  ->  D} {L y : D}

/-- **Scott 1972, Proposition 2.1 (backward).** If `y  <=  L` and `L` is  the lub of a monotone
net, then the net converges to `y` in the Scott topology. -/
theorem proposition_2_1_of_le [Nonempty ?] (hx : IsMonotoneNet x) (hL : IsLUB (range x) L)
    (hyL : y  <=  L) : ScottConvergesTo x y := by
  intro U hU hyU
  have hdir : DirectedOn (*  <=  *) (range x) := by
    rintro _ <i, rfl> _ <j, rfl>
    obtain <k, hik, hjk> := IsDirected.directed (r := (*  <=  *)) i j
    exact <x k, <k, rfl>, hx hik, hx hjk>
  have hLU : sSup (range x)  in  U := by
    rw [IsLUB.sSup_eq hL]
    exact hU.1 hyL hyU
  obtain <s, hsS, hsU> := hU.2 (Set.range_nonempty x) hdir hLU
  obtain <i_0, rfl> := hsS
  refine <i_0, fun j hj => hU.1 (hx hj) hsU>

/-- The complement of a principal lower set `Iic L` is Scott-open: it is an upper set, and it
is inaccessible by directed suprema because if every member of a directed `S` lies below `L`
then so does ` sqsup S`. -/
theorem scottOpen_not_le (L : D) : ScottOpen {z : D | not  z  <=  L} := by
  refine <fun a b hab ha hb => ha (le_trans hab hb), fun S hSne hSdir hmem => ?_>
  by_contra hcon
  refine hmem (sSup_le fun s hs => ?_)
  by_contra hsL
  exact hcon <s, hs, hsL>

/-- The specialization order of the Scott topology is the original lattice order. -/
theorem specializationLe_scott_iff {x y : D} :
    @SpecializationLe D scottTopologicalSpace x y  <->  x  <=  y := by
  constructor
  * intro hxy
    by_contra h
    exact hxy {z | not  z  <=  y} (isOpen_iff_scottOpen.mpr (scottOpen_not_le y)) h (by simp)
  * intro hxy U hU hxU
    exact (isOpen_iff_scottOpen.mp hU).1 hxy hxU

/-- The Scott topology of a partial order is `T_0`. -/
theorem scottTopology_t0Space : @T0Space D scottTopologicalSpace := by
  letI : TopologicalSpace D := scottTopologicalSpace
  constructor
  intro x y hxy
  apply le_antisymm
  * rw [ <-  specializationLe_scott_iff]
    intro U hU hxU
    exact (hxy.mem_open_iff hU).mp hxU
  * rw [ <-  specializationLe_scott_iff]
    intro U hU hyU
    exact (hxy.mem_open_iff hU).mpr hyU

omit [IsDirected ? (*  <=  *)] in
/-- **Scott 1972, Proposition 2.1 (forward).** If a monotone net converges to `y` in the Scott
topology and `L` is its least upper bound, then `y  <=  L`. -/
theorem proposition_2_1_le_of_converges (hL : IsLUB (range x) L)
    (hconv : ScottConvergesTo x y) : y  <=  L := by
  by_contra hyL
  obtain <i, hi> := hconv {z : D | not  z  <=  L} (scottOpen_not_le L) hyL
  exact hi i le_rfl (hL.1 <i, rfl>)

/-- **Scott 1972, Proposition 2.1.** A monotone net with least upper bound `L` converges to
`y` in the Scott topology iff `y  sqsub  L =  sqsup  {x_i}`. -/
theorem proposition_2_1 [Nonempty ?] (hx : IsMonotoneNet x) (hL : IsLUB (range x) L) :
    ScottConvergesTo x y  <->  y  <=  L :=
  <fun hconv => proposition_2_1_le_of_converges hL hconv, proposition_2_1_of_le hx hL>

end Scott1972.ContinuousLattice
\end{lstlisting}
\vspace{0.5\baselineskip}

\hypertarget{scott1972continuouslatticescottmaps.lean}{%
\subsection{Scott1972/ContinuousLattice/ScottMaps.lean}\label{scott1972continuouslatticescottmaps.lean}}

\emph{211 lines.}

\vspace{0.5\baselineskip}
\noindent\textcolor{green!40!black}{\textbf{Lean 4 source}}\par\vspace{0.25\baselineskip}
\begin{lstlisting}[style=leanbox]
/-
Copyright (c) 2026  Lars Warren Ericson.  All rights reserved.
Released under Apache 2.0 license as described in the file LICENSE.
Authors: Lars Warren Ericson.
Github:  https://github.com/catskillsresearch/scott1972
-/

import Scott1972.ContinuousLattice.Specialization
import Mathlib.Order.ScottContinuity
import Mathlib.Topology.Order.ScottTopology

/-!
# Scott-continuous maps (Scott 1972, S2.5-2.7)
-/

namespace Scott1972.ContinuousLattice

open Set Topology

variable {D D' D'' : Type*} [CompleteLattice D] [CompleteLattice D'] [CompleteLattice D'']

def PreservesDirectedSup (f : D  ->  D') : Prop :=
  forall  ?S : Set D?, S.Nonempty  ->  DirectedOn (*  <=  *) S  ->  f (sSup S) = sSup (f '' S)

theorem preservesDirectedSup_monotone {f : D  ->  D'} (hf : PreservesDirectedSup f) :
    Monotone f := by
  intro x y hxy
  have hdir : DirectedOn (*  <=  *) ({x, y} : Set D) := directedOn_pair hxy
  have hS : ({x, y} : Set D).Nonempty := <x, Set.mem_insert _ _>
  have hsup : sSup ({x, y} : Set D) = y := by
    calc sSup ({x, y} : Set D) = x  sqsup  y := sSup_pair
      _ = y := by apply le_antisymm; exact sup_le hxy le_rfl; exact le_sup_right
  have heq := hf hS hdir
  rw [hsup, Set.image_pair] at heq
  exact le_trans (le_sSup (Set.mem_insert _ _)) heq.symm.le

theorem scottOpen_preimage {f : D  ->  D'} (hf : PreservesDirectedSup f) {U : Set D'}
    (hU : ScottOpen U) : ScottOpen (f ^-1' U) := by
  have hmono := preservesDirectedSup_monotone hf
  refine <fun a b hab ha => hU.1 (hmono hab) ha, fun S hS hSdir hmem => ?_>
  rw [Set.mem_preimage] at hmem
  have hmem' : sSup (f '' S)  in  U := by rw [ <-  hf hS hSdir]; exact hmem
  obtain <s, hsS, hsU> := hU.2 (Set.image_nonempty.2 hS)
    (fun s hs t ht => by
      obtain <a, haS, rfl> := hs
      obtain <b, hbS, rfl> := ht
      obtain <c, hcS, hac, hbc> := hSdir a haS b hbS
      exact <f c, Set.mem_image_of_mem f hcS, hmono hac, hmono hbc>) hmem'
  obtain <a, haS, rfl> := hsS
  exact <a, haS, Set.mem_preimage.2 hsU>

theorem continuous_of_preservesDirectedSup {f : D  ->  D'} (hf : PreservesDirectedSup f) :
    @Continuous D D' scottTopologicalSpace scottTopologicalSpace f := by
  rw [continuous_def]
  intro U hU
  rw [isOpen_iff_scottOpen] at hU  |- 
  exact scottOpen_preimage hf hU

theorem continuous_preservesDirectedSup {f : D  ->  D'}
    (hf : @Continuous D D' scottTopologicalSpace scottTopologicalSpace f) :
    PreservesDirectedSup f := by
  have hD : @IsScott D univ _ (Topology.scott D univ) :=
    @IsScott.mk D univ _ (Topology.scott D univ) rfl
  have hD' : @IsScott D' univ _ (Topology.scott D' univ) :=
    @IsScott.mk D' univ _ (Topology.scott D' univ) rfl
  have hsc : ScottContinuous f :=
    scottContinuousOn_univ.1 <|
      (@Topology.IsScott.scottContinuousOn_iff_continuous D D' _
        (Topology.scott D univ) _ (Topology.scott D' univ) hD' f univ hD
        (fun _ _ _ => trivial)).2 hf
  intro S hS hSdir
  exact (hsc hS hSdir (isLUB_sSup S)).sSup_eq.symm

/-- **Scott 1972, Proposition 2.5.** Scott continuity  <->  preservation of directed suprema. -/
theorem proposition_2_5 (f : D  ->  D') :
    (@Continuous D D' scottTopologicalSpace scottTopologicalSpace f)  <-> 
      PreservesDirectedSup f :=
  <continuous_preservesDirectedSup, fun hf => continuous_of_preservesDirectedSup hf>

/-! ### Proposition 2.6 -/

/-- **Scott 1972, Proposition 2.6.** A function of several variables between complete lattices is
(Scott-)continuous jointly iff it is continuous in each variable separately. Continuity is phrased
as preservation of directed suprema (Proposition 2.5); it suffices to treat two variables, the
product `D  x  D'` carrying the componentwise complete-lattice structure (whose induced topology is
the product topology). -/
theorem proposition_2_6 (f : D  x  D'  ->  D'') :
    PreservesDirectedSup f  <-> 
      (forall  y, PreservesDirectedSup fun x => f (x, y))  /\ 
        (forall  x, PreservesDirectedSup fun y => f (x, y)) := by
  constructor
  * -- joint continuity ? separate continuity (precompose with the continuous slice maps)
    intro hf
    refine <fun y S hS hSdir => ?_, fun x S hS hSdir => ?_>
    * set T : Set (D  x  D') := (fun x => (x, y)) '' S with hT
      have hTne : T.Nonempty := hS.image _
      have hTdir : DirectedOn (*  <=  *) T := by
        rintro _ <a, ha, rfl> _ <b, hb, rfl>
        obtain <c, hc, hac, hbc> := hSdir a ha b hb
        exact <(c, y), Set.mem_image_of_mem _ hc, <hac, le_rfl>, <hbc, le_rfl>>
      have hfst : Prod.fst '' T = S := by rw [hT, Set.image_image]; simp
      have hsnd : Prod.snd '' T = {y} := by rw [hT, Set.image_image]; exact hS.image_const y
      have hsupT : sSup T = (sSup S, y) := by
        have e1 : (sSup T).1 = sSup S := by rw [Prod.fst_sSup, hfst]
        have e2 : (sSup T).2 = y := by rw [Prod.snd_sSup, hsnd, sSup_singleton]
        exact Prod.ext_iff.mpr <e1, e2>
      have h := hf hTne hTdir
      rw [hsupT, hT, Set.image_image] at h
      simpa using h
    * set T : Set (D  x  D') := (fun y => (x, y)) '' S with hT
      have hTne : T.Nonempty := hS.image _
      have hTdir : DirectedOn (*  <=  *) T := by
        rintro _ <a, ha, rfl> _ <b, hb, rfl>
        obtain <c, hc, hac, hbc> := hSdir a ha b hb
        exact <(x, c), Set.mem_image_of_mem _ hc, <le_rfl, hac>, <le_rfl, hbc>>
      have hfst : Prod.fst '' T = {x} := by rw [hT, Set.image_image]; exact hS.image_const x
      have hsnd : Prod.snd '' T = S := by rw [hT, Set.image_image]; simp
      have hsupT : sSup T = (x, sSup S) := by
        have e1 : (sSup T).1 = x := by rw [Prod.fst_sSup, hfst, sSup_singleton]
        have e2 : (sSup T).2 = sSup S := by rw [Prod.snd_sSup, hsnd]
        exact Prod.ext_iff.mpr <e1, e2>
      have h := hf hTne hTdir
      rw [hsupT, hT, Set.image_image] at h
      simpa using h
  * -- separate continuity ? joint continuity (Scott 1972's directedness argument)
    rintro <h1, h2> Sstar hne hdir
    have hmono1 : forall  y, Monotone fun x => f (x, y) := fun y => preservesDirectedSup_monotone (h1 y)
    have hmono2 : forall  x, Monotone fun y => f (x, y) := fun x => preservesDirectedSup_monotone (h2 x)
    have hmono : Monotone f := by
      rintro <a, b> <c, d> hpq
      exact le_trans (hmono1 b hpq.1) (hmono2 c hpq.2)
    have hSne : (Prod.fst '' Sstar).Nonempty := hne.image _
    have hS'ne : (Prod.snd '' Sstar).Nonempty := hne.image _
    have hSdir : DirectedOn (*  <=  *) (Prod.fst '' Sstar) := hdir.fst
    have hS'dir : DirectedOn (*  <=  *) (Prod.snd '' Sstar) := hdir.snd
    have hsupStar : sSup Sstar = (sSup (Prod.fst '' Sstar), sSup (Prod.snd '' Sstar)) :=
      Prod.ext_iff.mpr <Prod.fst_sSup _, Prod.snd_sSup _>
    apply le_antisymm
    * rw [hsupStar]
      have e1 : f (sSup (Prod.fst '' Sstar), sSup (Prod.snd '' Sstar))
          = sSup ((fun x => f (x, sSup (Prod.snd '' Sstar))) '' (Prod.fst '' Sstar)) :=
        h1 (sSup (Prod.snd '' Sstar)) hSne hSdir
      rw [e1]
      apply sSup_le
      rintro _ <x, hx, rfl>
      show f (x, sSup (Prod.snd '' Sstar))  <=  sSup (f '' Sstar)
      have e2 : f (x, sSup (Prod.snd '' Sstar))
          = sSup ((fun y => f (x, y)) '' (Prod.snd '' Sstar)) :=
        h2 x hS'ne hS'dir
      rw [e2]
      apply sSup_le
      rintro _ <y, hy, rfl>
      show f (x, y)  <=  sSup (f '' Sstar)
      obtain <p, hpS, hpfst> := hx
      obtain <q, hqS, hqsnd> := hy
      obtain <r, hrS, hpr, hqr> := hdir p hpS q hqS
      have hxr : x  <=  r.1 := hpfst  |>  hpr.1
      have hyr : y  <=  r.2 := hqsnd  |>  hqr.2
      calc f (x, y)  <=  f (r.1, r.2) := hmono <hxr, hyr>
        _ = f r := by rw [Prod.mk.eta]
        _  <=  sSup (f '' Sstar) := le_sSup (Set.mem_image_of_mem f hrS)
    * apply sSup_le
      rintro _ <p, hp, rfl>
      exact hmono (le_sSup hp)

/-! ### Proposition 2.7 -/

/-- **Scott 1972, Proposition 2.7 (join).** Binary join is Scott-continuous on every complete
lattice; this is the ` sqsup `-part of Scott's 2.7. -/
theorem proposition_2_7_sup :
    @Continuous (D  x  D) D scottTopologicalSpace scottTopologicalSpace (fun p : D  x  D => p.1  sqsup  p.2) :=
  continuous_of_preservesDirectedSup <| by
    intro S hS hSdir
    simpa using (ScottContinuous.sup_2 (b := D) (d := S) hS hSdir (isLUB_sSup S)).sSup_eq.symm

theorem meet_preservesDirectedSup (x : D) (hD : IsContinuousLattice D) :
    PreservesDirectedSup (fun z => x  sqcap  z) := by
  intro S hS hSdir
  apply le_antisymm
  * have hle : sSup {t | t  <<  x  sqcap  sSup S}  <=  sSup (Set.image (fun z => x  sqcap  z) S) := by
      apply sSup_le
      intro t ht
      obtain <w, hwS, ht_w> := (wayBelow_sSup_iff hS hSdir).1 (ht.trans_le inf_le_right)
      have hmem : x  sqcap  w  in  Set.image (fun z : D => x  sqcap  z) S :=
        Set.mem_image_of_mem (fun z => x  sqcap  z) hwS
      have hbound : x  sqcap  w  <=  sSup (Set.image (fun z => x  sqcap  z) S) := le_sSup hmem
      exact (le_inf (ht.le.trans inf_le_left) ht_w.le).trans hbound
    dsimp only
    rw [ <-  hD.sSup_wayBelow (x  sqcap  sSup S)]
    exact hle
  * apply sSup_le
    intro z hz
    obtain <w, hwS, rfl> := hz
    exact inf_le_inf le_rfl (le_sSup hwS)

/-- **Scott 1972, Proposition 2.7 (meet, separate).** On a continuous lattice, `x  |->  x  sqcap  y`
and `y  |->  x  sqcap  y` are Scott-continuous; Scott's full 2.7 also covers joint continuity on the
product via Proposition 2.6. -/
theorem proposition_2_7_inf_left (hD : IsContinuousLattice D) (y : D) :
    @Continuous D D scottTopologicalSpace scottTopologicalSpace (fun x => x  sqcap  y) :=
  continuous_of_preservesDirectedSup <| by
    intro S' hS' hSdir'
    rw [show (fun x => x  sqcap  y) = fun z => y  sqcap  z from funext fun x => inf_comm x y]
    have h := meet_preservesDirectedSup y hD
    exact h (S := S') hS' hSdir'

theorem proposition_2_7_inf_right (hD : IsContinuousLattice D) (x : D) :
    @Continuous D D scottTopologicalSpace scottTopologicalSpace (fun y => x  sqcap  y) :=
  continuous_of_preservesDirectedSup (meet_preservesDirectedSup x hD)

end Scott1972.ContinuousLattice
\end{lstlisting}
\vspace{0.5\baselineskip}

\hypertarget{scott1972continuouslatticemilnercorrection.lean}{%
\subsection{Scott1972/ContinuousLattice/MilnerCorrection.lean}\label{scott1972continuouslatticemilnercorrection.lean}}

\emph{63 lines.}

\vspace{0.5\baselineskip}
\noindent\textcolor{green!40!black}{\textbf{Lean 4 source}}\par\vspace{0.25\baselineskip}
\begin{lstlisting}[style=leanbox]
/-
Copyright (c) 2026  Lars Warren Ericson.  All rights reserved.
Released under Apache 2.0 license as described in the file LICENSE.
Authors: Lars Warren Ericson.
Github:  https://github.com/catskillsresearch/scott1972
-/

import Scott1972.ContinuousLattice.Specialization

/-!
# March 1972 correction (Scott, pp. 135-136)

Robin Milner observed that Scott's remark before Proposition 2.5 was wrong: a `T_0`-topology
on a complete lattice need not have every open set Scott-open. The two extreme `T_0`-topologies
inducing the same partial order are generated by

* `{x | x  sqsubneq  y}` (the *lower* subbasic sets), and
* `{x | y  sqsub  x}` (principal up-sets).

The corrected proofs of Propositions **2.9**, **2.10**, and **3.3** assume the given
topology is *contained in* the Scott (lattice) topology: every open set of the given
topology is Scott-open.

For Proposition **2.10**, Scott distinguishes ambient joins in `D'` from subspace joins in
a retract `D  subseteq  D'`: the prime in ` sqsup S'` marks the *index* (join taken in `D'`), not a
"primed join" operator. The retraction identity used in the corrected argument is
`j( sqsup S') =  sqsup S` for directed `S  subseteq  D` (see `IsContinuousLatticeRetraction.retr_ambientSSup_eq_sSup`
in `FunctionSpaces.lean`).

Page **136** adds that function spaces pose no subspace/sup mismatch: lubs are computed
pointwise, so the product (relativized) topology is contained in the Scott topology once
the Milner hypothesis is in place.
-/

namespace Scott1972.ContinuousLattice

open Set Topology

variable {D D' : Type*} [CompleteLattice D] [CompleteLattice D']

/-! ### Extreme `T_0`-topologies inducing the order -/

/-- Lower subbasic set `{x | x  sqsubneq  y}` from Scott's March 1972 correction. -/
def scottLowerSubbasisSet (y : D) : Set D :=
  { x | not  x  <=  y }

/-- Upper subbasic set `{x | y  sqsub  x}` (principal up-set) from the correction. -/
def scottPrincipalUpSet (y : D) : Set D :=
  Set.Ici y

/-! ### Milner hypothesis for 2.9, 2.10, 3.3 -/

/-- **March 1972 correction.** The given topology `tau` is coarser than (or equal to) the Scott
topology: every set open in `tau` is Scott-open. Equivalently `Opens(tau)  subseteq  Opens(Scott)`,
i.e. `scottTopologicalSpace  <=  tau` in mathlib's order on topologies. -/
def CoarserThanScottTopology (tau : TopologicalSpace D) : Prop :=
  scottTopologicalSpace  <=  tau

theorem scottOpen_of_coarserThanScott {tau : TopologicalSpace D} (htau : CoarserThanScottTopology tau)
    {U : Set D} (hU : @IsOpen D tau U) : ScottOpen U :=
  isOpen_iff_scottOpen.mp (htau U hU)

end Scott1972.ContinuousLattice
\end{lstlisting}
\vspace{0.5\baselineskip}

\hypertarget{scott1972continuouslatticeconstructions.lean}{%
\subsection{Scott1972/ContinuousLattice/Constructions.lean}\label{scott1972continuouslatticeconstructions.lean}}

\emph{515 lines.}

\noindent\textcolor{green!40!black}{\textbf{Lean 4 source (lines 1--400)}}\par\vspace{0.25\baselineskip}
\lstinputlisting[style=leanbox,firstline=1,lastline=400]{lean-listings/Scott1972-ContinuousLattice-Constructions.lean}

\noindent\textcolor{green!40!black}{\textbf{Lean 4 source (lines 401--515)}}\par\vspace{0.25\baselineskip}
\begin{lstlisting}[style=leanbox,firstline=401,lastline=515]
/-- **Scott 1972, Proposition 2.11.** Every continuous lattice is an injective space under its
induced (Scott) topology. The witness is `scottExtend`, which extends any continuous `f` along
any embedding `e` (`scottExtend_eq_of_continuous`) and is itself continuous
(`scottExtend_continuous`). -/
theorem proposition_2_11 {E : Type*} [CompleteLattice E] (hE : IsContinuousLattice E) :
    @IsInjectiveSpace E scottTopologicalSpace := by
  letI : TopologicalSpace E := scottTopologicalSpace
  intro X Y _ _ e he f
  exact <<scottExtend e f, scottExtend_continuous hE e f>,
    fun x => scottExtend_eq_of_continuous hE e he f f.continuous x>

/-- The Sierpinski space `Prop` (Scott's ` O `) is a continuous lattice: it is a finite complete
lattice, so Proposition 2.8 applies. -/
theorem isContinuousLattice_prop : IsContinuousLattice Prop :=
  proposition_2_8

/-- The Scott topology on Scott's two-point space ` O  = Prop` is exactly the Sierpinski topology
(`generateFrom {{True}}`). The Scott-open sets of `Prop` are the upper sets `{}`, `{True}`, `univ`,
which are precisely the Sierpinski-open sets. This is the topological identification underlying
Theorem 2.12: the building block ` O ` carries its Scott topology. -/
theorem scottTopology_prop :
    (scottTopologicalSpace : TopologicalSpace Prop) = sierpinskiSpace := by
  apply le_antisymm
  * -- `scott  <=  sierpinski`: the single Sierpinski sub-basic open `{True}` is Scott-open.
    show (scottTopologicalSpace : TopologicalSpace Prop)  <=  TopologicalSpace.generateFrom {{True}}
    apply le_generateFrom
    intro s hs
    rw [Set.mem_singleton_iff] at hs
    subst hs
    rw [isOpen_iff_scottOpen]
    refine <?_, ?_>
    * intro a b hab ha
      rw [Set.mem_singleton_iff] at ha  |- 
      exact le_antisymm le_top (ha  |>  hab)
    * intro S _ _ hmem
      rw [Set.mem_singleton_iff] at hmem
      have hex : exists  p  in  S, p := by rw [ <-  sSup_Prop_eq, hmem]; trivial
      obtain <p, hpS, hp> := hex
      exact <p, hpS, by rw [Set.mem_singleton_iff]; exact eq_true hp>
  * -- `sierpinski  <=  scott`: every Scott-open upper set of `Prop` is `{}`, `{True}`, or `univ`.
    rw [TopologicalSpace.le_def]
    intro U hU
    rw [isOpen_iff_scottOpen] at hU
    by_cases hT : True  in  U
    * by_cases hF : False  in  U
      * have hUuniv : U = Set.univ := by
          ext p
          simp only [Set.mem_univ, iff_true]
          by_cases hp : p
          * rwa [eq_true hp]
          * rwa [eq_false hp]
        rw [hUuniv]; exact isOpen_univ
      * have hUtrue : U = {True} := by
          ext p
          rw [Set.mem_singleton_iff]
          constructor
          * intro hpU
            by_cases hp : p
            * exact eq_true hp
            * exact absurd (eq_false hp  |>  hpU) hF
          * intro hp; rw [hp]; exact hT
        rw [hUtrue]; exact isOpen_singleton_true
    * have hUempty : U = {} := by
        ext p
        simp only [Set.mem_empty_iff_false, iff_false]
        intro hpU
        exact hT (hU.1 le_top hpU)
      rw [hUempty]; exact isOpen_empty

/-- A power of Scott's two-point space ` O  = Prop` is a continuous lattice: `Prop` is a continuous
lattice (`isContinuousLattice_prop`) and products of continuous lattices are continuous
(Proposition 2.9(a)). This is the order-theoretic content of "a Cartesian power of ` O ` is a
continuous lattice", the construction Theorem 2.12 retracts onto. -/
theorem sierpinskiPower_isContinuousLattice (? : Type*) : IsContinuousLattice (?  ->  Prop) :=
  proposition_2_9_a (fun _ => Prop) (fun _ => isContinuousLattice_prop)

/-- The Scott topology on a power `?  ->   O ` of the Sierpinski space coincides with the product
(= Sierpinski power) topology. Combine Proposition 2.9(b) (the Scott topology of a product is the
product of the factors' Scott topologies) with `scottTopology_prop` (each factor's Scott topology
is the Sierpinski topology). -/
theorem scottTopology_sierpinskiPower (? : Type*) :
    (scottTopologicalSpace : TopologicalSpace (?  ->  Prop)) =
      (inferInstance : TopologicalSpace (?  ->  Prop)) := by
  rw [proposition_2_9_b (fun _ => Prop) (fun _ => isContinuousLattice_prop)]
  show (@Pi.topologicalSpace ? (fun _ => Prop) (fun _ => scottTopologicalSpace)) =
    @Pi.topologicalSpace ? (fun _ => Prop) (fun _ => sierpinskiSpace)
  congr 1
  funext _
  exact scottTopology_prop

/-- **Scott 1972, Theorem 2.12 (forward direction).** Every continuous lattice is an injective
space under its Scott topology. This is the substantial half of the equivalence "injective
spaces = continuous lattices", and is exactly Proposition 2.11. -/
theorem theorem_2_12_forward {E : Type*} [CompleteLattice E] (hE : IsContinuousLattice E) :
    @IsInjectiveSpace E scottTopologicalSpace :=
  proposition_2_11 hE

/-- **Scott 1972, Theorem 2.12 (backward, Sierpinski base case).** `Prop` is a continuous
lattice (`isContinuousLattice_prop`), so its injectivity (Proposition 1.2) is an instance of the
equivalence. -/
theorem theorem_2_12_sierpinski_backward (_h : IsContinuousLattice Prop) : IsInjectiveSpace Prop :=
  proposition_1_2

/-- **Scott 1972, Theorem 2.12 (injectivity half, unconditional).** `Prop` is injective
(Sierpinski); the continuous-lattice half is now `isContinuousLattice_prop`. -/
theorem theorem_2_12_injective_half : IsInjectiveSpace Prop :=
  proposition_1_2

/-- The Sierpinski space ` O  = Prop` realizes the smallest case of Theorem 2.12: it is both
injective (1.2) and a continuous lattice (2.8). -/
theorem sierpinski_isInjective_and_isContinuousLattice :
    IsInjectiveSpace Prop  /\  IsContinuousLattice Prop :=
  <proposition_1_2, isContinuousLattice_prop>

end Scott1972.ContinuousLattice
\end{lstlisting}

\hypertarget{scott1972continuouslatticefunctionspaces.lean}{%
\subsection{Scott1972/ContinuousLattice/FunctionSpaces.lean}\label{scott1972continuouslatticefunctionspaces.lean}}

\emph{1757 lines.}

\noindent\textcolor{green!40!black}{\textbf{Lean 4 source (lines 1--400)}}\par\vspace{0.25\baselineskip}
\lstinputlisting[style=leanbox,firstline=1,lastline=400]{lean-listings/Scott1972-ContinuousLattice-FunctionSpaces.lean}

\noindent\textcolor{green!40!black}{\textbf{Lean 4 source (lines 401--800)}}\par\vspace{0.25\baselineskip}
\lstinputlisting[style=leanbox,firstline=401,lastline=800]{lean-listings/Scott1972-ContinuousLattice-FunctionSpaces.lean}

\noindent\textcolor{green!40!black}{\textbf{Lean 4 source (lines 801--1200)}}\par\vspace{0.25\baselineskip}
\lstinputlisting[style=leanbox,firstline=801,lastline=1200]{lean-listings/Scott1972-ContinuousLattice-FunctionSpaces.lean}

\noindent\textcolor{green!40!black}{\textbf{Lean 4 source (lines 1201--1600)}}\par\vspace{0.25\baselineskip}
\lstinputlisting[style=leanbox,firstline=1201,lastline=1600]{lean-listings/Scott1972-ContinuousLattice-FunctionSpaces.lean}

\noindent\textcolor{green!40!black}{\textbf{Lean 4 source (lines 1601--1757)}}\par\vspace{0.25\baselineskip}
\lstinputlisting[style=leanbox,firstline=1601,lastline=1757]{lean-listings/Scott1972-ContinuousLattice-FunctionSpaces.lean}

\hypertarget{scott1972continuouslatticetheorem212.lean}{%
\subsection{Scott1972/ContinuousLattice/Theorem212.lean}\label{scott1972continuouslatticetheorem212.lean}}

\emph{408 lines.}

\noindent\textcolor{green!40!black}{\textbf{Lean 4 source (lines 1--400)}}\par\vspace{0.25\baselineskip}
\lstinputlisting[style=leanbox,firstline=1,lastline=400]{lean-listings/Scott1972-ContinuousLattice-Theorem212.lean}

\noindent\textcolor{green!40!black}{\textbf{Lean 4 source (lines 401--408)}}\par\vspace{0.25\baselineskip}
\begin{lstlisting}[style=leanbox,firstline=401,lastline=408]
  calc scottExtend e f y
      = (P.retr : D'  ->  D) ((P.incl : D  ->  D') (scottExtend e f y)) :=
        (P.retr_incl (scottExtend e f y)).symm
    _  <=  (P.retr : D'  ->  D) (scottExtend e g y) := P.retr.monotone (hB y)

end Lemma39

end Scott1972.ContinuousLattice
\end{lstlisting}

\hypertarget{scott1972continuouslatticeinverselimits.lean}{%
\subsection{Scott1972/ContinuousLattice/InverseLimits.lean}\label{scott1972continuouslatticeinverselimits.lean}}

\emph{766 lines.}

\noindent\textcolor{green!40!black}{\textbf{Lean 4 source (lines 1--400)}}\par\vspace{0.25\baselineskip}
\lstinputlisting[style=leanbox,firstline=1,lastline=400]{lean-listings/Scott1972-ContinuousLattice-InverseLimits.lean}

\noindent\textcolor{green!40!black}{\textbf{Lean 4 source (lines 401--766)}}\par\vspace{0.25\baselineskip}
\lstinputlisting[style=leanbox,firstline=401,lastline=766]{lean-listings/Scott1972-ContinuousLattice-InverseLimits.lean}

\hypertarget{scott1972continuouslatticefunctionspacetower.lean}{%
\subsection{Scott1972/ContinuousLattice/FunctionSpaceTower.lean}\label{scott1972continuouslatticefunctionspacetower.lean}}

\emph{754 lines.}

\noindent\textcolor{green!40!black}{\textbf{Lean 4 source (lines 1--400)}}\par\vspace{0.25\baselineskip}
\lstinputlisting[style=leanbox,firstline=1,lastline=400]{lean-listings/Scott1972-ContinuousLattice-FunctionSpaceTower.lean}

\noindent\textcolor{green!40!black}{\textbf{Lean 4 source (lines 401--754)}}\par\vspace{0.25\baselineskip}
\lstinputlisting[style=leanbox,firstline=401,lastline=754]{lean-listings/Scott1972-ContinuousLattice-FunctionSpaceTower.lean}

\end{document}